\documentclass[twocolumn]{aastex63}
\usepackage{comment}

\newcommand{\flx}{erg~cm$^{-2}$~s$^{-1}$}
\newcommand{\lum}{erg~s$^{-1}~$}

\def\approxgt{\ifmmode \rlap{$>$}{}_{{}_{{}_{\textstyle\sim}}} \else%
$\rlap{$>$}{}_{{}_{{}_{\textstyle\sim}}}$\fi} 
\def\approxlt{\ifmmode \rlap{$<$}{}_{{}_{{}_{\textstyle\sim}}} \else%
$\rlap{$<$}{}_{{}_{{}_{\textstyle\sim}}}$\fi}

\received{June 1, 2019}
\revised{January 10, 2019}
\accepted{\today}
\submitjournal{ApJ}

\shorttitle{Black hole mass bias}
\shortauthors{Jonker et al.}
\graphicspath{{./}{figures/}}

\begin{document}

\title{The observed mass distribution of Galactic black hole LMXBs is biased against massive black holes}

\correspondingauthor{Peter G.~Jonker}
\email{p.jonker@astro.ru.nl}

\author[0000-0001-5679-0695]{Peter G.~Jonker}
\affiliation{Department of Astrophysics/IMAPP, Radboud University\\
P.O. Box 9010, 6500 GL Nijmegen, the Netherlands}
\affiliation{SRON, Netherlands Institute for Space Research\\
Sorbonnelaan, 2, 3584CA Utrecht, the Netherlands}

\author{Karamveer Kaur}
\affiliation{Racah Institute of Physics, The Hebrew University, Jerusalem, 91904, Israel}

\author{Nicholas Stone}
\affiliation{Racah Institute of Physics, The Hebrew University, Jerusalem, 91904, Israel}

\author{Manuel A.~P.~Torres}
\affiliation{Instituto de Astrof\'{i}sica de Canarias, E-38205 La Laguna, S/C de Tenerife, Spain}
\affiliation{Departamento de Astrof\'{i}sica, Universidad de La Laguna, E-38206
La Laguna, S/C de Tenerife, Spain}

\begin{abstract}
The discovery of gravitational wave radiation from merging black holes
(BHs) also uncovered BHs with masses in the range of $\approx$20--160
M$_\odot$. In contrast, the most massive Galactic stellar-mass BH
currently known has a mass $\approx21$ M$_\odot$. While low-mass X-ray
binaries (LMXBs) will never independently evolve into a binary BH
system, and binary evolution effects can play an important role
explaining the different BH masses found through studies of X-ray
binaries and gravitational wave events, (electromagnetic) selection
effects may also play a role in this discrepancy. Assuming BH LMXBs
originate in the Galactic Plane, we show that the spatial distribution
of the current sample of confirmed and candidate BH LMXBs are both
biased to sources that lie at a large distance from the
Plane.  Specifically, most of the confirmed and candidate BH LMXBs are
found at a Galactic height larger than 3 times the scale height for
massive star formation. In addition, the confirmed BH LMXBs are found
at larger distances to the Galactic Center than the candidate BH
LMXBs.  Interstellar absorption makes candidate BH LMXBs in the Plane
and those in the Bulge too faint for a dynamical mass measurement
using current instrumentation. Given the observed and theoretical
evidence for BH natal and/or Blaauw kicks, their relation with BH mass
and binary orbital period, and the relation between outburst
recurrence time and BH mass, the observational selection effects imply
that the current sample of confirmed BH LMXBs is biased against the
most massive BHs.

\end{abstract}


\section{Introduction} \label{sec:intro}

Black hole (BH) X--ray binaries are systems in which a BH accretes
mass from a companion star. Typically, a distinction is made between low- and high-mass X-ray binaries on the basis of the mass of the donor star in the binary. In this paper we focus on low-mass X-ray binaries (LMXBs), excluding BH X-ray binaries with O- and early B-type donor stars. When the accretion rate through
the disc is large, these systems show up as bright X--ray sources. When
the mass flow rate through the disc decreases the systems go (back) to
quiescence. During the quiescent phase the mass donor or companion star can be detected in the optical and/or the near-infrared
(NIR). Such observations can be used to determine the mass of the BH
(for a review and a detailed explanation of the methods involved see \citealt{2014SSRv..183..223C}). 

The observed BH mass distribution in these LMXBs
(\citealt{2010ApJ...725.1918O}; \citealt{2011ApJ...741..103F}) has an
apparent lack of BHs in the mass range of 2--5
M$_\odot$ (the so called mass gap; \citealt{1998ApJ...499..367B}) and BHs more massive than $\approx$15~M$_\odot$ are also not observed. Recently, the mass of the BH in the high-mass X-ray binary Cyg~X-1 has been adjusted upwards to 21.2$\pm$2.2 M$_\odot$ (\citealt{2021arXiv210209091M}), making this the heaviest stellar-mass BH with an {\it electromagnetically-measured} mass. In some mass determinations a systematic error is introduced by assuming the accretion disk light is not contributing to the optical light. Properly accounting for this may remove the apparent lack of BHs in LMXBs in the 2--5 M$_\odot$ range (e.g.,~\citealt{2012ApJ...757...36K}). Depending on the assumed BH kick properties micro-lensing mass determinations of single BH lenses may also cast doubt on the presence of the mass gap  (\citealt{2020A&A...636A..20W}).

The detection of gravitational waves (GWs) by the Laser
Interferometer Gravitational Wave Observatory (LIGO; \citealt{2015CQGra..32k5012A}) and the Virgo interferometer (\citealt{2015CQGra..32b4001A}) from binary BH mergers (e.g., \citealt{2016PhRvL.116f1102A};
\citealt{2016PhRvX...6d1015A}; \citealt{2019ApJ...882L..24A};  \citealt{2020arXiv201014527A}) revealed  
the existence of BHs more massive than the stellar--mass BHs previously detected in X-ray binaries. The current record holder is the detection of a BH merger product with a mass of $\approx$150 M$_\odot$, where the BH component masses were $85_{-14}^{+21} M_\odot$ and $66_{-18}^{+17} M_\odot$ (\citealt{2020PhRvL.125j1102A}).
While some of the BHs involved in binary BH mergers are inferred to have masses in the range electromagnetically observed for stellar--mass BHs, many are considerably larger (\citealt{2020arXiv201014527A}). It is likely that the different binary evolution histories of binary BH merger progenitors and the BH X-ray binaries (especially for the LMXB systems) are an important factor in these different mass distributions (e.g., \citealt{2019ApJ...878L...1P, 2021A&A...645A...5S}), however, some aspects of the BH formation process may be relevant to both groups of systems.

Stellar-mass BHs can form out of massive stars in two different ways (\citealt{2001ApJ...554..548F}), hence their \textbf{\textit{initial}} spatial distribution is expected to be correlated to that of massive star formation. For a sufficiently massive progenitor, BHs may form by direct collapse. Observational evidence
for this comes from the disappearance of red-supergiant stars without
evidence for a supernova (\citealt{2015MNRAS.453.2885R}) -- though a
faint (NIR) transient may appear (\citealt{2017MNRAS.468.4968A}) -- and also
from the zero peculiar velocity of some BH X-ray binaries, such as
Cyg~X-1 (\citealt{2003Sci...300.1119M};
\citealt{2011ApJ...742...83R}), although we note that the peculiar velocity is not a quantity conserved with time (see e.g., \citealt{2009MNRAS.394.1440M}).
Alternatively, if a supernova
explosion is not sufficiently energetic to unbind the complete stellar
envelope, fallback of material onto the proto-neutron star formed in
the explosion can create a BH (\citealt{1989ApJ...346..847C}; \citealt{2014arXiv1401.3032W}; \citealt{Chan_2018}). 

Given that most O stars are formed in binaries (\citealt{2012Sci...337..444S}) or higher-order multiples \citep{MoeDiStefano17}, and given that both the accreting BHs and those found through GW merger events are in binaries, we summarize here some important aspects of compact object formation on the space velocities of ensuing BHs, as that influences their observed spatial distribution. First, if one star in a binary explodes in a supernova, leading to impulsive mass loss from the binary system, a Blaauw kick will be imparted on the system, regardless of the type of compact object formed during the supernova event (\citealt{1961BAN....15..265B}). This Blaauw kick is directed in the orbital plane of the binary.

BH formation may also be accompanied by natal kicks, powered by the mechanisms that have been proposed to explain the high peculiar velocities of some neutron stars (e.g., \citealt{2017A&A...608A..57V}).  Natal kicks from anisotropic neutrino emission are thought to occur regardless of BH formation mechanism, while kicks related to asymmetric mass ejection (or, relatedly, asymmetric mass fallback) can only occur for the fallback channel of BH formation.  In principle, hydrodynamical kicks from
asymmetries in the supernova ejecta can accelerate a nascent BH to
similarly high {\it velocities} as observed in neutron stars (\citealt{2013MNRAS.434.1355J}).  Kicks from neutrino asymmetries, conversely, impart roughly the same {\it momentum} to the BH as they do in cases of neutron star formation \citep{2013MNRAS.434.1355J}, and therefore produce kick velocities that are sensitive to the final compact remnant mass.

Because of the lack of (significant) mass loss, direct collapse BHs are not subject to Blaauw or ejecta kicks, though in principle they might still undergo kicks from neutrino anisotropy; as direct collapse BHs are usually thought to be larger in mass than fallback BHs (e.g., \citealt{Fryer99}), this may introduce a mass dependence into the natal kick distribution, with the largest kicks going to the lowest-mass BHs (e.g., \citealt{2012ApJ...749...91F}).

LMXBs have long been recognized as a valuable probe of the BH mass distribution (if dynamical mass measurements are obtained), because the mass of the BH will not change appreciably due to gas accretion (as the mass of the secondary is low). LMXBs can also be used to study natal kick physics.  Many BH LMXB formation models involve a supernova and hence the Blaauw and/or natal kick will be imparted on the system and/or BH in the binary (see \citealt{2006csxs.book..623T} for a review)\footnote{The evolutionary model of \citet{1986MNRAS.220P..13E}, involving triple star evolution, provides a formation channel that does not impart a large kick velocity upon the system. }.  As a result, the distribution of LMXB altitudes, above and below the Galactic Plane, will encode their natal kick distribution, so long as they formed in the disk of the Milky Way. Note that the space velocity of old LMXB systems could well have received some component of peculiar velocity through non-axisymmetric forces experienced on their orbits through the Galaxy by scattering from the potentials of spiral arms or interstellar molecular clouds during the time between the BH formation and the system becoming active as an X-ray binary (\citealt{1977A&A....60..263W}). Estimates of the velocity dispersion of the thin disc population show that it is approximately 40--43 km s$^{-1}$ for K0-M5 late-type stars (Dehnen \& Binney 1998; \citealt{2000A&A...354..522M}), indicating that this effect needs to be taken into account (e.g., such as was done to determine the peculiar space velocity of V404~Cyg; \citealt{2009MNRAS.394.1440M}).

Initially it was thought that most BHs would quickly be kicked out of globular clusters, although a few could remain (\citealt{1993Natur.364..421K}). However, recent findings of actively accreting and quiescent candidate BH X-ray binaries in both Galactic and extra-galactic globular clusters, as well as the detection of binary BH mergers through GW radiation, has led to a renewed interest in the possibility of clusters retaining BHs.  \citet{2007Natur.445..183M} provide evidence for an accreting BH in an extra-galactic globular cluster associated with the Virgo elliptical galaxy NGC~4472. Quiescent candidate BH LMXBs were identified in globular clusters around the Milky Way. Typically radio and X-ray luminosity and radio spectral index measurements were used to argue for a candidate quiescent BH LMXB (in M~22, \citealt{2012Natur.490...71S}; in M~62, \citealt{2013ApJ...777...69C}; in 47~Tuc, \citealt{2015MNRAS.453.3918M}; in M~10, \citealt{2018ApJ...855...55S}; in Terzan~5, \citealt{2020ApJ...904..147U}).

If a sizable number of BHs are indeed retained by globular clusters, the high stellar density and large interaction rate could lead to the formation of BHs with a binary companion. A significant fraction of these systems will be ejected from the globular cluster through interactions with stars. The evolution of such a binary could imply it becomes a BH LMXB. This can occur both for systems that are retained or ejected by globular clusters  (e.g.,~\citealt{2018MNRAS.477.1853G}; \citealt{2018ApJ...852...29K}). Dynamical evidence for the presence of two or even three BHs in the globular cluster NGC~3201 was presented by \citet{2018MNRAS.475L..15G, 2019A&A...632A...3G}.

In this paper we investigate if the samples of confirmed and candidate BH LMXBs suffer from selection effects by studying their spatial distributions. We exclude the Galactic BH high-mass X-ray binaries, Cyg~X$-$1 (\citealt{2021arXiv210209091M}) and MWC~656 (\citealt{2014Natur.505..378C}) from our analysis, because the BHs in those systems formed recently (as the young age of the mass donor star testifies). The current high metallicity in the Galaxy probably precludes the formation of the most massive stellar-mass BH (e.g., see \citealt{2020arXiv201011730V} for the role of stellar metallicity and remnant mass). However, the BH LMXBs that have late-type mass donors  may have formed billions of years ago, when the metallicity was lower. Hence, we cannot exclude that the precursors of the BHs in LMXBs could have formed massive stellar-mass BHs such as found through GW events by the LIGO-Virgo collaboration.  In \S 2, we discuss the existing Galactic BH LMXB sample, and analyze its spatial distribution.  In \S 3, we discuss the implications of this distribution for source selection effects and BH LMXB formation channels.  We conclude in \S 4.

\section{Source sample and results}
\label{sample} 
    \begin{figure*}
        \includegraphics[width=9cm]{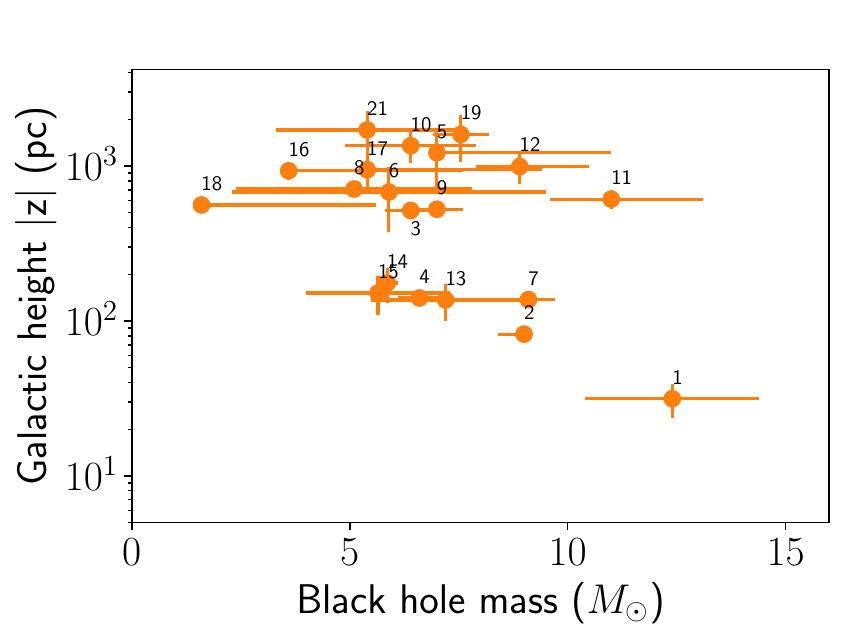} 
        \includegraphics[width=9cm]{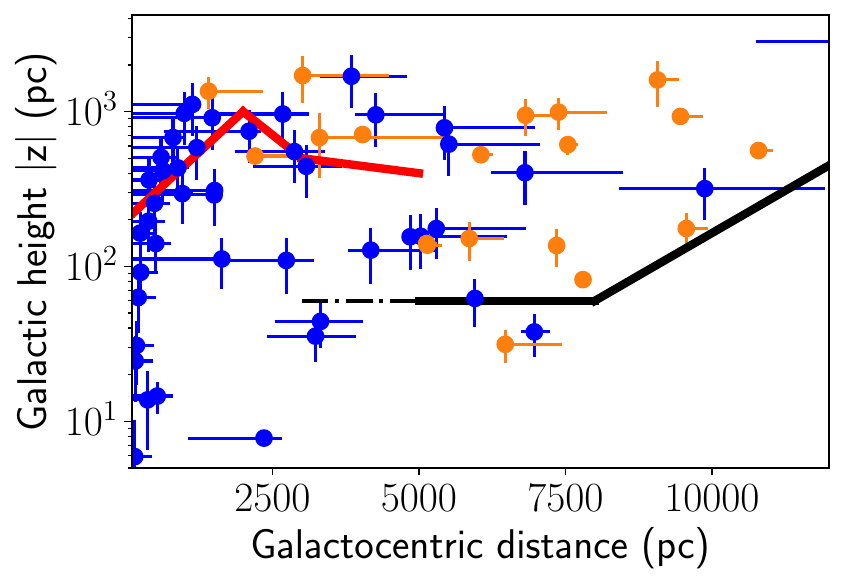}
        \caption{{\it Left panel:} The $|z|$ distance to the Galactic Plane (in pc) as a function of BH mass for the sample of Galactic BH LMXBs with dynamical mass measurements. The number refers to the source number in Table~\ref{tab:sam}. {\it Right panel:} The Galacto-centric distance, $d_{GC}$, as a function of $|z|$ for the sample of BHs with dynamical mass measurements in LMXBs (orange bullets). In addition, we show the candidate BH LMXBs (sources without a dynamical mass measurement) as blue bullets. The thick horizontal black line indicates 3$\times$ the Galactic scale height of 20 pc for massive stars from \citet{urquhart2014} and \citet{reid2019} for massive stars outside the Bulge with a Galacto-centric distance $5<d_{GC}<8$~kpc. For smaller $d_{GC}$ we extend this by a dash-dot line down to 3 kpc. For larger $d_{GC}$ the scale height increases rapidly to $\approx$450 pc (3~$\sigma$) at 12 kpc as shown by the thick drawn black line. All Galactic BH LMXBs with dynamical mass measurements, except GRS~1915$+$105, lie at larger distances above the Plane than 3$\times$ the scale height of massive star formation, implying that these BHs all received some sort of kick moving them to this location, unless they were formed outside the Galactic Plane. The sources \#4 BW~Cir and \#19 Swift~J1357.2$-$0933 fall outside this figure (see Table~\ref{tab:sam}). The red drawn line denotes the approximate height of the box-peanut shaped Galactic Bulge and bar (\citealt{2015MNRAS.448..713P, 2015MNRAS.450.4050W}). }
        \label{fig:mgcz}   
    \end{figure*}

We use the sample of confirmed BH LMXBs listed in \citet{2014SSRv..183..223C}. Table~\ref{tab:sam} provides basic information on the sources, including distances  and mass measurements which are updated with respect to the above mentioned work. In addition to those sources we included four additional systems with recent (dynamical) mass measurements; Swift~J1357.2$-$0933, MAXI~J1820$+$070, MAXI~J1659$-$152, and MAXI~J1305$-$704. Note that a few of these dynamically confirmed BH LMXBs have mass determinations that formally do not rule out a neutron star nature of the compact object, i.e., the mass determination is consistent with a compact object mass of $<3$~M$_\odot$. This includes GX~339$-$4, 4U~1543$-$47 and perhaps GRO~J0422$+$32. Nevertheless, we classify these systems as a BH here because the radio and X-ray spectral and timing properties of these sources are best described if they host a BH. In this respect it is interesting to note that none of the candidate BH LMXBs where a dynamical mass was determined turned out to have a best-fit mass of 1.4--2~M$_\odot$ (all confirmed BH LMXBs used to be candidate BH LMXBs before their dynamical mass was determined, of course). \\

For Swift~J1357.2$-$0933, \citet{2015MNRAS.454.2199M} constrained the BH mass using optical observations of the full-width at half-maximum (FWHM) of the H$\alpha$ emission line and the correlation between the FWHM and the semi-amplitude of the radial velocity of the mass donor star (\citealt{2015ApJ...808...80C}). The BH mass determination was further refined by \citet{2016ApJ...822...99C}.
For MAXI~J1820$+$070 we used the mass measurement of \citet{torres-1820-2019,torres-1820-2020} and the distance determination of 2.96$\pm$0.33 kpc obtained through the radio parallax measurement by \citet{atri2020}. For MAXI~J1659$-$152 we use the (approximate) dynamical mass measurement from \citet{torres-1659-2020}. For the distance to this source we adopt the value of $6\pm2$ kpc from \citet{jonker2012}. However, our conclusions are unchanged if we would have used the value of 8.6$\pm 3.7$~kpc reported by \citet{2013A&A...552A..32K}. For MAXI~J1305$-$704 we use the recently derived mass and distance estimate from \citet{2021MNRAS.tmp.1510M}.

For all sources, the equatorial coordinates and the distance $d$ are used to calculate the absolute value of the distance to the Plane, $|z|$, and the distance to the Galactic Center, $d\rm _{GC}$, using the python {\sc astropy:SkyCoord} routine. For the calculation we took 8.15 kpc for the Sun -- Galactic Center distance and 5.5 pc for the height of the Sun above the Plane following \citet{reid2019}.

\begin{table*}
  \caption{The name, orbital period P$_{\rm orb}$, BH mass, the distance $d$, the Galacto-centric distance $d\rm _{GC}$, and the absolute distance to the Galactic plane $|z|$ for the sample of Galactic BH LMXBs with a dynamical mass measurement sorted on decreasing orbital period from top to bottom. 
  }
\label{tab:sam}
\begin{center}
\begin{tabular}{lccccccc}
\hline
\# & Name & P$_{\rm orb}$ & Mass  &  $d$ & $d\rm _{GC}$ & $|z|$ & mass / distance\\
 &     & d &  M$_\odot$ &  kpc & kpc & pc & reference\\
\hline
1&   GRS~1915$+$105   & 33.85(0.16)    & 12.4$\pm$2.0  &8.6$^{+2}_{-1.6}$ & 6.5$^{+1.0}_{-0.6}$ & 30$\pm$8 & [1,2]/[1]\\
2&   V404~Cyg    & 6.47129(7)  &9$^{+0.2}_{-0.6}$&   2.39$\pm$0.14  & 7.80$\pm$0.01 & 82$\pm$5 & [3]/[4]\\
3 & XTE~J1819$-$2525$^{\dagger\dagger\dagger\dagger}$ & 2.81730(1)  & 6.4$\pm$0.6  & 6.2$\pm$0.7 & 2.2$^{+0.55}_{-0.62}$  & 515$\pm$60 & [30]/[30]\\
4 &  GRO~J1655$-$40  &  2.62168(14)  &   6.6$\pm$0.5   &  3.2 $\pm$0.2  & 5.13$\pm 0.18$ & 140$\pm$10 & [5]/[6]\\
5 &  Ginga~1354$-$645$^{\dagger\dagger\dagger}$  & 2.54451(8)   &$>7^\dagger$  &  25$^{+10}_{0}$ & 21$\pm$9.5 & 1220$^{+500}_{-0}$ & [7]/[7]\\
6 &   GX~339$-$4    & 1.7587(5)    &   5.9$\pm$3.6&  9$\pm$4  &  3.95$^{+2.2}_{-0.6}$ & 680$\pm$300 & [8]/[8]\\
7 &   XTE~J1550 $-$564 & 1.542033(2)   &  9.1$\pm$0.6    &  4.4$\pm$0.5 & 5.14$^{+0.25}_{-0.21}$    & 140$\pm$15  & [9]/[9]\\
8 & 4U~1543$-$47 &  1.123(8)   &  5.1$\pm$2.7 &7.5$\pm$0.5 &  4.04$^{+0.08}_{-0.02}$  & 710$\pm$45  & [21]/[11] \\
9 &  MAXI~J1820$+$070 & 0.68549(1) & 7.0$\pm$0.6  &   2.96$\pm$0.33 & 6.1$\pm$0.2    &  525$\pm$60 & [23]/[24] \\
10 & H~1705$-$250 &  0.5213(13)  &   6.4$\pm$1.5     &  8.6$\pm$2.0 &  1.94$^{+0.9}_{-0.5}$  & 1400$\pm$300 & [10]/[11]\\
11 &  Ginga~1124$-$684 &  0.432606(3)  &11.0$^{+2.1}_{-1.4}$ &  5.0$\pm$0.7 & 7.54$^{+0.17}_{-0.11}$   & 600$\pm$100 & [12]/[12]\\
12 & MAXI~J1305$-$704 & 0.394(4) & 8.9$^{+1.5}_{-1.0}$ & 7.5$^{+1.8}_{-1.4}$ & 7.4$^{+0.8}_{-0.5}$ & 1000$\pm225$ & [27]/[27] \\
13 &   Ginga~2000$+$250 & 0.3440873(2)   &   7.2$\pm$1.7    &  2.7$\pm$0.7 &  7.35$^{+0.12}_{-0.06}$  & 135$\pm$35  & [10]/[11]\\
14 &  1A~0620$-$00  & 0.32301405(1)      &   5.86$\pm$0.24   &  1.6$\pm$0.4 & 9.56$\pm$0.36   & 175$\pm$45  & [13]/[14]\\
15 &  XTE~J1650$-$500  & 0.3205(7)  &  5.65$\pm$1.65   &  2.6$\pm$0.7 & 5.86$\pm$0.58   & 150$\pm$40  & [15]/[16] \\
16 &  GRS~1009$-$45  &  0.285206(14)   &  $>3.6^\dagger$&    5.7$\pm$0.7 & 9.46$\pm$0.36   & 930$\pm$110 &  [17]/[11]\\
17 &   XTE~J1859$+$226  & 0.274(2)  &  $>5.4^\dagger$&     6.3$\pm$1.7 &  6.82$^{+0.58}_{-0.18}$  & 950$\pm$250 & [18]/[19]\\
18 &   GRO~J0422$+$32  &  0.2121600(2) &  $4\pm 1 ^{\dagger\dagger}$ & 2.75$\pm$0.25& 10.8$\pm$0.2  & 560$\pm$50  & [20]/[11] \\
19 &    XTE~J1118$+$480  & 0.1699339(2)  & 7.55$\pm$0.65  &  1.8$\pm$0.6 &  9.1$\pm$0.35   &  1600$\pm$500 & [22]/[11] \\
20 &    Swift~J1357.2$-$0933 & 0.11(4)  & 12.4$\pm$3.6  &  $8\pm1^{\dagger\dagger\dagger\dagger\dagger}$ &  7.67$^{+0.51}_{-0.41}$   &  6130$\pm$770 & [28,29]/[29] \\
21 &   MAXI~J1659$-$152  & 0.10058(19)  & 5.4$\pm$2.1  &  6$\pm$2 &    3$^{+1.5}_{-0.6}$ & 1700$\pm$570  & [25]/[26] \\
\end{tabular}
\end{center}
\footnotesize{$^\dagger$ Lower limit to the BH mass.$^{\dagger\dagger}$ BH mass uncertain see \citet{2014SSRv..183..223C} for a discussion.
$^{\dagger\dagger\dagger}$ Also referred to as BW~Cir.
$^{\dagger\dagger\dagger\dagger}$ Also referred to as V4641~Sgr.
$^{\dagger\dagger\dagger\dagger\dagger}$ The formal distance estimate is $>$6.7 kpc, we took 8$\pm$1 kpc for the calculation of $d_{GC}$ and $|z|$.\\ References: [1]=\cite{2014ApJ...796....2R},[2]=\citealt{2013ApJ...768..185S},[3]=\cite{2010ApJ...716.1105K},[4]=\cite{2009ApJ...706L.230M},[5]=\cite{2003MNRAS.339.1031S},[6]=\cite{1995Natur.375..464H},[7]=\cite{2009ApJS..181..238C},[8]=\cite{2017ApJ...846..132H},[9]=\cite{2011ApJ...730...75O},[10]=\cite{2014SSRv..183..223C}, [11]=\cite{2004MNRAS.354..355J}, [12]=\cite{2016ApJ...825...46W}, [13]=\cite{2017MNRAS.472.1907V}, [14]=\cite{poshak2019}, [15]=\cite{2004ApJ...616..376O} , [16]=\cite{2006MNRAS.366..235H}, [17]=\cite{1999PASP..111..969F}, [18]=\cite{2011MNRAS.413L..15C},[19]=\cite{2002MNRAS.331..169H},[20]=\cite{2003ApJ...599.1254G},[21]=\cite{1998ApJ...499..375O},[22]=\cite{2013AJ....145...21K},[23]=\cite{torres-1820-2020},[24]=\cite{atri2020},[25]=\cite{torres-1659-2020},[26]=\cite{jonker2012},[27]=\cite{2021MNRAS.tmp.1510M},
[28]=\cite{2016ApJ...822...99C},[29]=\cite{2015MNRAS.454.2199M}, [30]=\cite{2014ApJ...784....2M}.}
\end{table*}

In the left panel of Fig.~\ref{fig:mgcz} we show the BH mass versus its $|z|$ and in the right panel the BH $d_{GC}$ versus $|z|$ for both the dynamically confirmed BH LMXBs as well as known candidate BH  LMXBs-- those systems that display characteristics during transient outbursts typically also seen in dynamically confirmed BH X-ray transients (cf., \citealt{blackcat}; see Table~\ref{tab:candi}). The thick drawn black line in this figure is at three times the scale height of 19--20 pc for massive stars observed in current star forming regions with a $d_{GC}\approxlt$8  kpc (\citealt{urquhart2014,reid2019}). As shown by these authors, the Galactic scale height for massive star formation sites increases rapidly for sources with $d_{GC}\approxgt 8$~kpc to a 1~$\sigma$ value of $\approx$150 pc at a $d_{GC}=12$~kpc. While we draw a straight line  at three times the scale height, reaching 450 pc for 12 kpc starting at 8 kpc, the exact way the scale height increases with Galacto-centric distance for $d_{GC}\approxgt$8 kpc is somewhat uncertain and might be slightly different for regions below and above the Plane.

For many of the candidate BH LMXBs the distance is often not well known due to the fact that a reliable distance measurement in BH LMXBs often (though not exclusively; see e.g.~\citealt{2004MNRAS.354..355J} for details) comes from the spectroscopic detection of the mass donor star. For all BH candidates except MAXI~J1348$-$630 we take $d=8\pm$3 kpc as this is virtually always consistent with the (uncertain) distance estimates present in the literature for these systems. For MAXI~J1348$-$630 an accurate distance of 2.2$\pm0.6$ kpc has been derived by \citet{2020MNRAS.tmpL.236C} using H~I absorption line measurements.

Using as a null hypothesis that all BHs in LMXBs originate in regions where massive stars form, which we generously take to lie within three times the scale height of the massive star-forming regions, the main conclusion from these two plots is that only the confirmed BH LMXB GRS~1915$+$105 is found at a location close to its origin, i.e., close to the Galactic Plane. Poisson statistics shows that to observe such a configuration where one source is found in the region of origin out of the 21 sources by chance is negligible. Similarly, only about 4 candidate BH LMXBs are at $|z|<60$~pc, excluding the sources in the Bulge region with $d_{GC}<3 - 5$~kpc. Here the 3 or 5 kpc depends on what we take as limit for the boxy/peanut shaped Bulge (\citealt{2015MNRAS.448..713P, 2015MNRAS.450.4050W}, and see \citealt{2020RAA....20..159S} for a recent review). 

Next, we compare the $|z|$ distribution of the confirmed BH LMXBs with that of the candidate BH LMXBs using a two-sample Kolmogorov-Smirnov (K-S) test.  For both the confirmed BH LMXBs, as well as the candidate BH LMXBs, we simulated 10$^4$ distributions where we take a random $|z|$ value from the range of possible $|z|$ values in the 1~$\sigma$ uncertainty range, for each source. While the hypothesis that the two $|z|$ distributions are drawn from the same parent population has a low mean probability of being true (the mean p-value of the 10$^4$ samples is 0.05), this p-value does not rule out this possibility at a high confidence level. Similarly, we also compare the two distributions in $d_{GC}$. A two-sample K-S test shows that the hypothesis that the distributions in $d_{GC}$ are drawn from the same parent population has a low mean probability of being true (p-value =2$\times 10^{-4}$). It can be seen from the right panel of Fig.~\ref{fig:zGC} that there are much more candidate BH LMXBs (32) at $d_{GC}\approxlt 3$~kpc than confirmed BH LMXBs (there are only two such sources, \#3 XTE~J1819--2525 and \#10, H~1705$-$250).

We combine the sets of dynamically confirmed BH LMXBs and candidate BH LMXBs and require $d_{GC}>3$~kpc for both BH and BH candidate sources to avoid systems located in the Bulge (see the right panel of Fig.~\ref{fig:mgcz}). This combined data set has 30 systems. We compared the observed $|z|$ distribution of all (candidate) BH LMXBs with a null-hypothesis distribution that is continuous in $60<|z|<2000$ pc.  To do so, we compare the $|z|$ distribution of the combined systems with the distributions of the same number of systems randomly drawn from a distribution with uniform probability of having a value between $60<|z|<2000$~pc. As described above, we draw from the range of possible $|z|$ values afforded by the 1~$\sigma$ uncertainties in the $|z|$ determination for each source, BH and BH candidate alike. We repeat the whole procedure $10^4$ times and subsequently determine the mean value of the two-sample K-S (\citealt{smirnov1948}) p-value and the mean value of the two-sample Anderson-Darling (A-D) p-value (\citealt{andersondarling}). Because we take the 1~$\sigma$ uncertainty ranges in $|z|$ and $d_{GC}$ into account in these tests, the number of systems varies by 1 or 2 for different draws out of the 10$^4$. The mean K-S p-value is 4$\times 10^{-2}$ whereas that of the A-D test is $1\times 10^{-2}$. The latter number represents a lower limit because in about 3850 out of the 10$^4$ cases the result was capped at the lowest value allowed by the routine (0.001).

Besides these statistical tests, we also calculate the Mann-Whitney (\citealt{mann1947}) U-statistic. Here, the null hypothesis is that for two randomly selected values from the BH $|z|$ distribution (X) and the simulated distribution of uniform probability for $60<|z|<2000$ (Y), the probability of X being greater than Y is equal to the probability of Y being greater than X. This null-hypothesis would be true if the two distributions are drawn from the same parent population. The U-statistic has a mean p-value of $6\times 10^{-3}$. Overall, we conclude that the observed $|z|$ distribution is inconsistent with being uniform over the 60--2000 pc range.

From the left panel of Fig.~\ref{fig:mgcz} there is suggestive evidence for the absence of dynamically confirmed BH LMXBs with $|z|$ between $\approx$200--400 pc. 
To investigate if this apparent gap in the $|z|$ distribution of BH and candidate BH LMXBs is significant 
we repeat the statistical tests done above comparing the observed distribution with a distribution that is uniform between $60<|z|<2000$~pc, except now we enforce the presence of a gap between 200--400~pc. The mean K-S p-value is 0.01, the mean A-D p-value is 4$\times 10^{-3}$ (although again the latter number represents a lower limit because in about 6850 out of the 10$^4$ times it was capped at the lowest value allowed by the routine of 0.001). The mean U-statistic p-value is 2$\times 10^{-3}$. Comparing these statistics with those assuming a homogeneous distribution in the $|z|$-direction reveals that a hypothetical distribution with a gap does not lead to higher probabilities that the two distributions are consistent with being the same. Therefore, we conclude that there is no evidence for a gap in the $|z|$-distribution of the BH and candidate BH LMXBs. Finally, we checked if this conclusion was altered if we changed the upper bound of the distribution from 2000 pc to 1200 pc, but it does not.

\section{Discussion}

This section is structured as follows: we will first discuss our main findings on the observed spatial distribution of confirmed and candidate BH LMXBs comparing their distributions with each other and with the expected distribution assuming BH LMXBs are formed in the Galactic Plane. Next, in \S\ref{disc:disc} and \S\ref{disc:rec} we investigate if the prevalent discovery method of (candidate) BH LMXBs implies the presence of potential biases in BH mass. In \S\ref{disc:mass} we show that the bias affecting the chance for discovery, together with a mass measurement bias for the confirmed BH LXMBs implies that a selection effect on BH mass exists, assuming they are formed in the Galactic Plane. We investigate this assumption in \S\ref{disc:gc} by turning to an alternative view where BH LMXBs are formed in globular clusters. We end the Discussion by listing additional (proposed) ways of finding BHs in our Milky Way which may (in the future) alleviate some of the existing selection effects (\S\ref{disc:alt}).

\subsection{The observed spatial distributions of confirmed and candidate BH LMXBs}
\label{disc:spat}
Dynamical BH mass measurements are known for only $\approx$30\% of the total sample of BH candidates among the LMXBs (21 out of the list of 67 available on BlackCAT; \citealt{blackcat}.)\footnote{As of Spring 2021.} Comparing the observed $\{ d_{\rm GC}, |z|\}$ distributions of confirmed BH LMXBs with the theoretical expectation that these sources originate in the Galactic Plane, we note that confirmed BH LMXBs are nearly exclusively found outside their region of origin (Fig.~\ref{fig:mgcz}). All of the dynamically confirmed BHs in LMXBs, except GRS~1915$+$105, currently lie outside the region associated with the formation of massive stars (and hence BHs). Similarly, only 4 of the 17 candidate BH LMXBs with $d_{GC}>3$~kpc are at $|z|<60$~pc, implying that for discovery of (candidate) BH LMXBs a location outside the Plane is also favored. We also compared the $\{ d_{\rm GC}, |z|\}$ distributions of confirmed BH LMXBs with that of the candidate BH LMXBs and we find that the latter have, on average, a significantly lower $d_{\rm GC}$ than the confirmed BH sources. This spatial configuration could imply that all BHs in LMXBs get a significant kick upon formation moving them away from their origin, or there are selection effects at play. To end up with a confirmed BH LMXB it first has to be discovered as a candidate BH LMXB, therefore, we first investigate potential selection effects against discovering sources in the Plane.

\subsection{Biases in discovering candidate BH LMXBs at low $|z|$}
\label{disc:disc}
The overall low number of confirmed and candidate BH LMXBs to lie within three times the root mean square scale height of $\approx$20~pc for massive stars observed in current star forming regions with a Galacto-centric distance of less than $\approx$8~kpc (\citealt{urquhart2014,reid2019}) raises the question if selection effects are important in discovering (candidate) BH LMXBs. They are nearly always discovered as new X-ray sources by scanning or large field-of-view X-ray satellites when they go into outburst triggered by an accretion disc instability (e.g., \citealt{2001A&A...373..251D}). 

Example X-ray satellites include(d) the Rossi X-ray Timing Explorer with its All Sky Monitor (2--10 keV; \citealt{1996ApJ...469L..33L}), BeppoSAX which its Wide Field Camera (2--30 keV; \citealt[reporting on its Galactic Center monitoring program]{2004NuPhS.132..486I}), and the currently operational {\it Swift} Burst Alert Telescope (15--50 keV; \citealt{Krimm_2013}), INTEGRAL IBIS (15--10 MeV; \citealt{2003A&A...411L...1W, 2003A&A...411L.131U, 2007A&A...466..595K}), and the Monitor of All-sky X-ray Image (MAXI, it has two detectors with nominal sensitivity ranges 0.5--12 keV and 2--30 keV; \citealt{10.1093/pasj/61.5.999}) on the International Space Station. The sensitivity of each of these instruments to X-rays with energies above 5 keV implies that they should detect outbursts from sources even if their soft X-ray emission is attenuated by a very large interstellar column density. If we take the MAXI satellite as an example, its one-orbit (90 minute) flux limit in the 5--12 keV band is $\approx 2\times 10^{-10}$\flx. Here we assumed a source with a power law spectrum with photon index of 1.7 -- typical for the X-ray spectrum of a low-hard state source (cf.~\citealt{2006csxs.book..157M}) -- and we scaled the values reported in \citet{10.1093/pasj/61.5.999} for 2--30 keV to the 5--12 keV energy band. If we assume all sources transit to the soft state when reaching a luminosity of a few percent of the Eddington limit (for a $\approx$10 M$_\odot$ BH; \citealt{2003A&A...409..697M}) where their hard X-ray luminosity drops, the peak hard X-ray luminosity at that Eddington fraction is $\approx 5\times 10^{37}$\lum\,implying that such sources could be detected by MAXI even if located at a distance of 20 kpc.\footnote{Because we took a flux limit in the 5--12 keV band extinction should not play an important role in this distance limit, if the neutral hydrogen column density N$_H\approxlt 10^{23}$~cm$^{-2}$, since such an N$_H$ value only reduces the observed flux in that band at the 10\% level.} 

However, some outburst (candidate) BH LMXBs remain in the low-hard state and do not become very bright (e.g., faint X-ray transients; \citealt{2006A&A...449.1117W}). These sources are suggested to have a short P$_{\rm orb}$ (\citealt{2004A&A...423..321M}). Sources with outbursts with a peak X-ray luminosity of only $\approx 2\times 10^{36}$\lum\,would still be detected out to a distance of 8.5~kpc. However, the observed Galactic distribution of both BH and neutron star LMXBs (e.g., see figures 4 in both \citealt[]{2004MNRAS.354..355J} and \citealt{2021arXiv210504549C}) suggests that a significant number of BH LMXBs outbursts, notably those far away such as those on the other side of the Galactic Center, are not detected by X-ray all-sky monitors. This discovery-bias can possibly be explained by a combination of confusion in coded mask cameras due to the simultaneous presence of a large number of sources in its field-of-view, the limited sensitivity, and a reduced brightness due to Galactic absorption for N$_H$ values larger than $10^{23}$~cm$^{-2}$. 

\subsection{Discovery-bias against long BH LMXB outburst recurrence times}
\label{disc:rec}
Besides the aforementioned observational X-ray selection effects operating for sources at low $|z|$, a significant fraction of the sources with long recurrence times (i.e., those that have been in quiescence for at least the duration of the operations of the aforementioned and other X-ray satellites) may not have been discovered yet, and these preferentially also reside at low $|z|$ as we will show below. It is expected that BH LMXBs with a long P$_{\rm orb}$ -- such as those hosting a giant mass donor star like GRS~1915$+$105 -- have long recurrence times (\citealt{1997ApJ...484..844K,2009MNRAS.400.1337D}) implying that many will not have undergone an outburst in the last 50--60 years.

In addition, higher mass BH LMXBs are thought to have longer recurrence times than lower mass BH LMXBs under the disc instability model (\citealt{2001A&A...373..251D}).
If true, this would imply that a lower fraction of the total population of systems with more massive BHs in LMXBs has been discovered. Such systems may preferentially exist in the Galactic Plane, as they are more likely to include direct collapse BHs that receive at most weak kicks at birth. In conclusion, it seems likely that a population of (candidate) BH LMXBs is still hidden in the Plane because the sources have not yet shown a bright outburst or they have not yet shown an outburst at all during the period when sensitive all-sky X-ray satellites were operational. Together, these effects could help explain the low number of (candidate) BH LMXBs with $|z|<60$~pc. Furthermore, these effects suggest that there is a bias against discovering the most massive BH LMXBs in outburst. 

\subsection{Discovery-bias aggravated by a mass measurement bias}
\label{disc:mass}
The bias against discovering LMXBs with more massive BHs that might reside in the Plane can possibly be circumvented by observing sources in the Bulge, since the Bulge is thought to be formed through bar instabilities that might scatter old, massive, BH LMXBs to higher $|z|$ (see \citealt{2017MNRAS.469.1587D}, and \citealt{2020RAA....20..159S} for a recent review). However, the extinction and crowding that prevent the optical and ground based NIR observations necessary for BH mass measurements for candidate BH LMXBs in the Plane, are likewise precluding such measurements to be obtained for sources in the Bulge. Evidence for this can be seen comparing the $\{ d_{\rm GC}, |z|\}$ spatial distributions of confirmed and candidate BH LMXBs. Whereas there are many candidate BH LMXBs in the Bulge region, the confirmed BH LMXBs are all outside the Bulge. The current sample of confirmed BH LMXBs avoids the extinction (and crowding) that plagues sources that reside in the Plane and sources that lie in the Bulge at small $d_{GC}$, i.e., near the Galactic Center. 

Therefore, the observed $\{ d_{\rm GC}, |z|\}$ distribution of dynamically confirmed BH LMXBs implies that the current selection of BH masses from these systems is biased towards those systems that obtained a kick velocity that moved them out of the Plane (assuming, for now, that these LMXBs are formed in the Plane). This observed bias towards large $\{d_{GC},|z|$\} sources among the dynamically confirmed BHs favours those with lower BH masses. For example, for a fixed impulse
kick at birth (as is roughly expected for e.g., neutrino anisotropy kicks) the low BH-mass systems will travel further from the
Plane than would the high BH-mass systems. 
The likely lower-mass BHs forming through mass fallback receive additional kicks from asymmetric mass ejection, but the likely higher-mass BHs forming from direct collapse do not.  These direct collapse BHs are thought to be more massive as virtually no mass is lost in the absence of a supernova. The absence of a kick would keep them in the plane of the Galaxy where their massive stellar predecessors originate (\citealt{2012ApJ...749...91F}).

Supportive evidence for the scenario described above can be found in the work of \citet{poshak2020}. These authors provided evidence for a bias against long P$_{\rm orb}$ systems at large $|z|$, as the natal or Blaauw kick necessary to move the BH plus its companion star to a large $|z|$ will more quickly result in the disruption of long P$_{\rm orb}$ binaries when compared to short period binaries. Under the direct collapse formation scenario for BHs involving red-supergiants, their P$_{\rm orb}$ before collapse is necessarily long (if not, mass transfer to the initially less massive star in the binary would ensue). Interestingly, while there is a bias against discovering long P$_{\rm orb}$ systems in the Plane, as soon as their location is known when they do go into an X-ray outburst, in principle the (sub)-giant mass donor required to fill the Roche lobe in the long P$_{\rm orb}$ will be more amenable to the (NIR) detection and further study required for a mass measurement. However, deep searches for the NIR counterpart in quiescence, as has been done in the case of GRS~1915$+$105, are necessary to find the NIR counterparts and while this has been done for several candidate BH LMXB (e.g., \citealt{2002MNRAS.331.1065C, 2019MNRAS.482.2149L}), it has not been done systematically for all. This and the difficulty inherent to sensitive multi-epoch NIR observations in crowded regions like the Bulge and Plane explain the general lack of mass determinations for the few candidate BH LMXBs with $|z|<60$~pc. 

Next, we investigate the crucial assumption in the work above, namely, that BH LMXBs originate in the Galactic Plane by assuming a globular cluster origin for BH LMXBs. 

\subsection{Testing the assumption that BH LMXBs form in the Plane: BH LMXBs originating from globular clusters?}
\label{disc:gc}
If a large fraction of the BH LMXBs are formed in globular clusters, then the observed discovery-bias and the spatial selection effect in determining the BH mass do not necessarily imply a selection in BH mass. 

\subsubsection{BH LMXB orbital period distribution  constraints on a globular cluster origin}
Theoretical modeling suggests that only BH LMXBs with P$_{\rm orb}\approxlt6.5$~h could have originated in globular clusters (see figure 14 in \citealt{2018MNRAS.477.1853G}). Comparing the P$_{\rm orb}$ of the dynamically confirmed BHs in LMXBs with the theoretical limits on the P$_{\rm orb}$ of BH LMXBs originating in clusters only a small number (6 out of the 20) of the current population of confirmed BH LMXBs could potentially have been formed in globular clusters. Note that the P$_{\rm orb}\approxlt 6.5$~h is not necessarily a strict limit, as this value depends on the globular cluster properties and on the details of the (time of) the ejection of the binary containing a BH.

\subsubsection{BH LMXB metallicity constraints on a globular cluster origin}
The metallicity of the companion stars in several of the systems that have P$_{\rm orb}<6.5$~h seems to rule out a globular cluster origin (e.g., XTE~J1118$+$480; \citealt{2006ApJ...644L..49G}\footnote{We note that although the work of \citet{2001ApJ...561.1006F} is consistent with a low-metallicity object, the method they employed is not very sensitive to the metallicity, and perhaps more importantly, is much more sensitive to other parameters in the modeling of the X-ray reflection spectrum such as the geometry of the medium responsible for the incident spectrum, the emissivity profile of the accretion disk, and the spin of the black hole. Therefore, we deem their work not a contradiction of the much more direct metallicity determination of \citet{2006ApJ...644L..49G}.}). However, metallicity determinations use the current elemental abundances of the companion star, and these are likely polluted by the supernova explosion, and therefore, not an accurate proxy for the metallicity of the (putative) host.\footnote{The spectrum of a BH-main sequence binary formed dynamically in a globular cluster would not show pollution products even if the BH was formed in a supernova explosion as long as the binary was formed after the BH (\citealt{2020PASJ...72...45S}).}
Furthermore, the effect that in core-collapse supernovae the metallicity of the ejected material can be strongly direction-dependent has not been taken into account. For instance, a strong bi-polar distribution of Fe synthesized in the core collapse was found in the supernova remnant G11.2$-$0.3 (\citealt{2009ApJ...703L..81M}). If this situation is typical, the amount of Fe captured by the companion star will strongly depend on the inclination of the bi-polar structure with respect to the binary orbital plane prior to the supernova.  

Finally, the pollution itself is not well known; the observed abundances in clusters of galaxies are difficult to (re)produce with (linear) combinations of existing supernova nucleosynthesis models (\citealt{2019MNRAS.483.1701S}). The inclusion of neutrino physics in the core-collapse supernova models will be crucial to predict the elemental yields (see e.g., \citealt{2021Natur.589...29B} for a recent review), but this still leaves the directional uncertainty in the pollution of the companion star. Consequently, the current observed metallicity can be strongly influenced by the supernova pollution in a way that is difficult to reconstruct. Therefore, the current metallicity determinations do not provide stringent constraints on the metallicity of the (host) environment of the (BH) LMXB. We conclude that the metallicity determinations alone do not preclude that short P$_{\rm orb}$ BH LMXBs originated in a globular cluster. 

\subsubsection{BH LMXB $|z|$-distribution constraints on a globular cluster origin}
To investigate the resulting $|z|$ distribution of BH LMXBs originating in globular clusters,
we integrated the orbits of 158 Galactic globular clusters backwards in time for 1 Gyr. The current spatial and velocity coordinates are taken from a publicly available globular cluster catalogue\footnote{\url{https://people.smp.uq.edu.au/HolgerBaumgardt/globular/}; e.g., see \citet{2019MNRAS.482.5138B}.}.  A subset of 94 globular clusters with orbits tracing the relevant portion of the Galactic disk are chosen; we select globular clusters whose radial distances of apocenter $d_{GC, apo} > 4$ kpc and pericenter $d_{GC,peri} < 12$ kpc. 
We next investigate if the current $|z|$ distribution of globular clusters is reflected in the $|z|$ distribution of LMXBs.

We assume that dynamically formed BH LMXBs will trace the time-averaged positions of their parent globulars, motivated by the low speed at which binaries are dynamically ejected from globular clusters (comparable to the globular cluster escape speed, $\sim 10~{\rm km~s}^{-1}$). We compute the resulting probability distribution $P(z)$ in the $z$-coordinate for the globular clusters, and their descendant BH LMXBs, in the following way. Firstly, the orbit of each globular is sampled at a time interval of 0.1 Myr (up to 1 Gyr) and the instantaneous $z$ value is recorded. Next, the normalized histogram of these $z$ values for each globular cluster is taken to be the probability distribution $P_i(z)$ of binary ejecta for a single globular cluster. The final probability distribution $P(z) = \sum_{i=1}^{89} 10^{c_i} P_i(z)/\sum_{i=1}^{89}10^{c_i}$ for globulars is a normalized sum of these individual single-globular probability distributions, weighted by a factor of $10^c$, where $c$ is the concentration parameter for globular clusters\footnote{We take concentration parameters from the catalogue \url{https://www.physics.mcmaster.ca/~harris/mwgc.dat}; \citep{1996AJ....112.1487H}}. Note that here we restrict ourselves further to the subset of 89 globular clusters which have $c$ values in the Harris dataset. This measure of weights is motivated by the analysis of \citet{2018MNRAS.477.1853G} (their figure 4), which suggests that the number of binaries ejected from a globular cluster is roughly proportional to $10^c$. 

\begin{figure}
    \includegraphics[width=9cm]{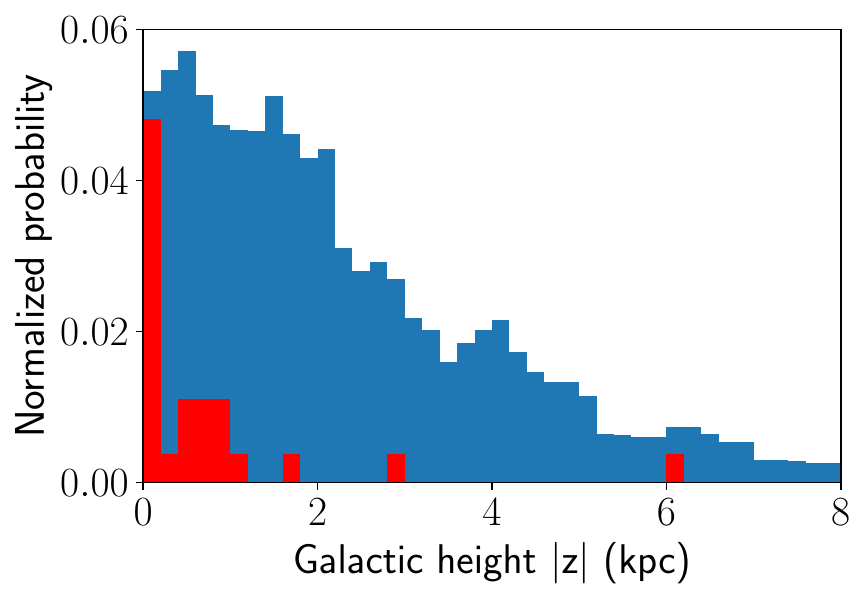} 
    \caption{ The probability distribution $P(z)$ of  globular clusters (blue histogram with bin size of 200 pc). The arbitrarily normalized histogram of the 27 observed BH LMXBs with $4 < d_{GC} < 12$~kpc is over-plotted in red. We caution that these 27 sources are comprised of confirmed BHs, including those with P$_{\rm orb}>6.5$~h, and candidate BH LMXBs for which the P$_{\rm orb}$ may not be known at present. 
    }
    \label{fig:zGC}   
\end{figure}

The probability distribution $P(z)$ of globular clusters is shown in  Fig.~\ref{fig:zGC} (as a blue histogram with bin size of 200 pc), while an arbitrarily normalized histogram of the $|z|$ coordinates of 27 LMXBs tracing the disk ($4 < d_{GC} < 12$~kpc) is plotted in red. We conclude that 
a population of BH LMXBs escaping from globular clusters will have a much larger high-$|z|$ tail than is observed. 
It is in principle possible that the observed lack of very high-$|z|$ binaries could be related to other selection effects. Studying these possibilities goes beyond the scope of this paper, however, so we conclude this analysis by noting that a simple, empirically motivated approach to BH LMXB populations originating from globular clusters does not appear to be broadly compatible with the observed candidate and confirmed BH LMXB $|z|$ distribution. 

Note that by utilizing the current orbits of globular clusters we ignore a potential contribution to the existing BH LMXB population from globular clusters that have been disrupted in the time period after their birth at redshifts greater than $\approx$1 but before the current epoch. Including the contribution of such disrupted globular clusters will probably increase the number of LMXBs at $|z|\approxlt 1$ kpc, leading to a steeper profile of the probability distribution P(z) (shown in blue in Figure 2).

In addition, in this simple analysis we have assumed that most of the (BH) LMXBs are ejected with small relative velocities with respect to the globular cluster, so that they would roughly track the orbit of the globular and hence, the $z$-distribution of BH LMXBs would roughly trace the globular cluster probability distribution, the former being only slightly broader than the later. The assumption of small ejection velocities for binaries is valid owing to the small escape velocities of globulars (roughly 90\% of globular clusters have escape velocities $<$ 40 km s$^{-1}$ evaluated at half light radii from the Baumgardt data). Figure 5 of \citet{2018MNRAS.477.1853G} shows the distribution of ejection velocities of binaries as a function of globular cluster mass in their simulations; the distribution of ejection velocities should peak below 10 km s$^{-1}$, because the mode of the mass distribution of globular clusters lies around $10^{5-5.5} M_\odot$ (here the mass determinations from the Baumgardt data are used).

\subsubsection{Bias against finding BH LMXBs originating in globular clusters }
While there is no strong evidence that a significant fraction of the known (dynamically confirmed) BH LMXBs should have formed in globular clusters, this current lack of evidence should not be considered as proof for the hypothesis that none of the BH LMXBs is formed in clusters. There is a significant bias against determining the mass of a BH dynamically for short orbital period systems (such as those originating in globular clusters).\footnote{Incidentally, often the orbital period of a candidate BH LMXBs is only determined during the BH mass measurement procedure. Therefore, a significant fraction of the candidate BH LMXBs could, in principle, have a P$_{{\rm orb}}\approxlt$6.5~h.} This is both caused by the fact that the shorter the orbital period, the fainter the mass of the donor star in the optical bands, and by the requirement to limit the spectroscopic integration time to roughly 1/20th of the orbital period. This latter limitation is to avoid significant source motion during the integration time. Such movement would cause variable Doppler shifts in the absorption lines and thereby smear out the (weak) stellar absorption features, making their detection more difficult.

Overall, we conclude that the observed low number of BH LMXBs with P$_{{\rm orb}}\approxlt$6.5~h and the difference between the observed and predicted BH LMXB $|z|$ distribution for a globular cluster origin suggests that a globular cluster origin for the majority of the known candidate and confirmed BH LMXBs is unlikely.

\subsection{Alternative ways to detect Galactic BHs}
\label{disc:alt}
Given that the population of BH LMXBs discovered through their X-ray outbursts suffers from selection effects, we below discuss alternative known methods to detect Galactic stellar-mass BHs. Other proposed and sometimes successfully employed ways of finding single BHs, non-interacting BHs in binaries, and BH LMXBs in quiescence -- those that have not been detected in outburst so far -- include micro-lensing, searches focusing on the large FWHM of the Balmer H$\alpha$ emission line caused by the accretion disk in quiescence, and searches focusing on the faint X-ray emission that is present even in quiescence. We briefly discuss some results and (potential) biases of these methods below. 

Using the OGLE micro-lensing survey \citet{2016MNRAS.458.3012W} and \citet{2020A&A...636A..20W} reported median dark lens masses of $\approx 2-12$~M$_\odot$, i.e., in the BH range. However, the OGLE surveys avoid to a large extent the Galactic Plane region $|b|<1^\circ$ (\citealt[their figure 1]{2019ApJS..244...29M}), especially the latest OGLE~IV incarnation that covers the largest area on the sky. Therefore, this survey will not find the most massive (single) BHs if they indeed reside in the Galactic Plane (a Galactic latitude $b=1^\circ$ corresponds to $d\approx3.4$~kpc for $|z|<$60~pc and therefore, the disk volume probed is small).

\citet{2018MNRAS.473.5195C} put forward a new method using a combination of optical photometric filter observations to extract the FWHM and equivalent width of the Balmer H$\alpha$ emission line. The method has been tested successfully (\citealt{2018MNRAS.481.4372C}). This, combined with a (photometric) orbital period determination of the binary system, can be used to single out interacting BH LMXBs from among the large population of H$\alpha$ emission line objects. The method is using two correlations: 
\begin{itemize} 
\item 1) a correlation between the H$\alpha$ FWHM and the radial velocity semi-amplitude of the mass donor star ($K_2$; \citealt{2015ApJ...808...80C}) 
\item 2) a correlation between $A$ and $B$ where $A$ is the ratio between the peak--separation of the two H$\alpha$ disk emission lines and the H$\alpha$ FWHM and $B$ is the ratio between the mass of the two binary components in confirmed BH LMXBs ($q$; \citealt{2016ApJ...822...99C}). 
\end{itemize}

\citet{2018MNRAS.473.5195C} suggests a lower limit on the FWHM value of $\approx 2200$~km~s$^{-1}$ to avoid the detection of other much more numerous Galactic H$\alpha$ emission line sources. This limits the projected radial velocity amplitude to values of $K_2\approxgt 500$~km s$^{-1}$, implying that $\frac{{\rm P}_{\rm orb}}{M_{BH}}\approxlt 1.9\times F(q,i)$, with P$_{\rm orb}$ expressed in hours and $M_{BH}$ in units of a solar mass. Here, $F(q,i)=\frac{\sin^3 i}{1+q^2}$, with $i$ the binary inclination and $q$ the binary mass ratio -- defined here as M$_2$/M$_{BH}$ with M$_2$ the donor star mass -- implying $F(q,i)$ is at $\mathcal{O}(1)$ for systems viewed under a high inclination. The positive correlation between orbital period and BH mass for this method suggests that if long orbital period systems with massive stellar-mass BHs exists they can be found with this tool. A small caveat: here we assume that the correlations underlying this method can be extrapolated to the values associated with more massive BHs in long orbital period LMXBs. 

Alternatively, one can try to use the faint X-ray emission that is emitted even if the system is in quiescence (cf.~\citealt{2013ApJ...773...59P}). This is for instance one of the goals behind the Galactic Bulge Survey (\citealt[][]{2011ApJS..194...18J,2014ApJS..210...18J}), and the faint X-ray emission combined with the (faint) radio emission has been used to argue for the detection of BH LMXBs in quiescence in globular clusters (see \citealt{2012Natur.490...71S} and other references in the Introduction). This method might favour long-period orbital systems given the observed positive correlation between quiescent X-ray luminosity and orbital period (e.g., \citealt{2001ApJ...553L..47G, 2013ApJ...775....9H}), assuming that systems that are currently in quiescence and have not undergone a recent outburst have the same quiescent X-ray luminosity as the systems that recently showed an outburst (which populate the orbital period and quiescent X-ray luminosity correlation). No dependence on BH mass is known for this method.

A final method to detect BHs in binaries involves the astrometric detection of the (projected) orbital motion of the binary stellar companion. The enormous potential of the {\it Gaia} satellite for this has been realized by several authors already (e.g., \citealt{2002ApJ...572..944G, 2017ApJ...850L..13B, 2018ApJ...861...21Y, 2019MNRAS.487.5610S, 2019MNRAS.486.4098I, 2020PASJ...72...45S, 2020ApJ...905..134W, 2020MNRAS.496.1922B, 2020arXiv200907277G}). However, for this potential to be realized we need to await the {\it Gaia} data release that includes the  astrometric measurements for individual epochs and not only their time average as has been the case for the data releases so far. In general, the larger the mass of the BH, the smaller the mass ratio, and the larger the orbital period the larger the potential astrometric signal. Although, a larger  orbital period also requires a longer astrometric monitoring sequence before the signal can be detected. Similarly, there is a linear dependence between the size of the projected orbit on the sky and the distance to the binary.

\section{Conclusions}
Since the first direct GW detections in 2015, astronomers have been puzzled by the apparent discrepancy between BH mass distributions measured (i) electromagnetically, in LMXBs and (ii) gravitationally, via binary BH mergers.  For the most extreme cases, such as GW-selected BHs in the pair instability mass gap (e.g., GW190521), there is likely a true physical difference in formation channels between these two populations.  However, outside the pair instability mass gap, it remains unclear whether the EM-selected and GW-selected samples reflect physically different populations or simply different selection effects, for two different messengers, applied to the {\it same} underlying population.  This question is of more than academic interest given its connection to longstanding problems in binary evolution, core collapse supernova physics, and dynamical channels for binary BH assembly, all of which are plausible candidates for creating a real astrophysical difference between BH LMXB and BH-BH mass distributions.  The observed difference between these mass distributions could in principle be used to probe these problems, {\it but only if said difference is not due to selection effects.}

In this paper we have explored possible selection effects in electromagnetic determination of BH LMXB masses.  In general, examining the spatial distribution of confirmed BH LMXBs and candidate BH LMXBs suggests multiple new selection effects that may be shaping their mass distribution.  Our main conclusions are the following: 
\begin{itemize}
    \item The Galactic height $|z|$ distribution of confirmed {\it and} candidate BH LMXBs shows a paucity of sources in the region where they are expected to form (i.e., in the regions of massive star formation in the Plane). The paucity of even candidate BH LMXBs in the Milky Way Plane suggests a strong selection effect in the X-ray discovery of BH LMXBs, likely due to a combination of crowding and absorption at low Galactic heights, combined with an intrinsic lower outburst probability of LMXBs hosting more massive BHs that reside in the Plane.
    \item The confirmed BH LMXBs have a significantly 
    greater distance to the Galactic Center than the population of candidate BH LMXBs.  This bias can be explained by a {\it different} selection effect acting on the optical spectroscopy needed to dynamically confirm a candidate BH LMXB: high levels of dust extinction in the Plane and towards the Bulge precluding optical BH mass measurements. 
    \item Taken together, both of the above selection effects combine to give the observed distribution of $|z|$ for confirmed BH LMXBs with dynamical mass measurements.  The combination of these two selection effects favors sources that obtained a significant (natal) kick that moved them out of the Plane of their origin. 
    \item Although the magnitudes and mechanisms of BH natal kicks are debated, it is generally believed on theoretical grounds that kicks arising from anisotropic neutrino emission will be of ``fixed momentum,'' i.e., will impart larger velocity kicks to compact remnants of smaller mass (e.g., \citealt{2012ApJ...749...91F}).  Natal kicks due to asymmetric mass ejection may instead be in the ``fixed velocity'' regime (i.e., similar kick velocities for remnants of different mass), but most models for core collapse supernovae predict that above a certain BH remnant mass, BHs form through direct collapse without any natal kick from asymmetric ejection (direct collapse BHs will also not be subject to Blaauw kicks).  The selection effects discussed above, which favor X-ray outburst occurence and probably also detection, and electromagnetic mass measurements in high-$|z|$ BH LMXBs, therefore disfavor mass measurement of high-mass BH LMXBs.
    \item The observed BH LMXB orbital periods disfavor a globular cluster origin for a majority of dynamically confirmed BH LXMB systems.  This constraint does not apply to the $\approx$6 shortest orbital period BH LMXBs, for which there is little decisive evidence for or against a globular cluster origin. In addition, the predicted $|z|$-distribution of BH LMXBs originating in globular clusters is inconsistent with the observed $|z|$ distribution of the combined population of candidate and confirmed BH LMXBs (although, here we ignore the potential contribution of an initial population of globular clusters that has been disrupted).
\end{itemize}

Current electromagnetic detection and mass measurement techniques appear biased against massive Galactic BH LMXBs, and these selection effects may help explain the discrepancy between BH mass distributions in LMXB and gravitational wave samples. However, as we have explored in \S ~\ref{disc:alt}, future detection or mass determination strategies, focused on BH LMXBs residing in the Plane of the Galaxy (for instance, mass measurement using the James Webb Space Telescope) may allow the identification of Galactic LMXBs as massive as those found through LIGO/Virgo's GW detections.

\section*{Acknowledgments} \noindent PGJ acknowledges the Caltech Kingsley distinguished-visitor program and the hospitality at Caltech where this work was initiated. KK and NCS gratefully acknowledge support from the Israel Science Foundation (Individual Research Grant 2565/19). KK also thanks the Israel Academy of Sciences and Humanities for supporting her work with an IASH Postdoctoral Fellowship.  MAPT acknowledges support via a Ram\'on y Cajal Fellowship RYC-2015-17854 and  MINECO grant AYA2017-83216-P. We would like to thank the anonymous referee for her/his comments which improved the paper.


\begin{thebibliography}{}
\expandafter\ifx\csname natexlab\endcsname\relax\def\natexlab#1{#1}\fi
\providecommand{\url}[1]{\href{#1}{#1}}
\providecommand{\dodoi}[1]{doi:~\href{http://doi.org/#1}{\nolinkurl{#1}}}
\providecommand{\doeprint}[1]{\href{http://ascl.net/#1}{\nolinkurl{http://ascl.net/#1}}}
\providecommand{\doarXiv}[1]{\href{https://arxiv.org/abs/#1}{\nolinkurl{https://arxiv.org/abs/#1}}}

\bibitem[{{Aasi} {et~al.}(2015){Aasi}, {Abadie}, {Abbott}, {Abbott}, {Abbott},
  {Abernathy}, {Accadia}, {Acernese}, {Adams}, {Adams}, \&
  et~al.}]{2015CQGra..32k5012A}
{Aasi}, J., {Abadie}, J., {Abbott}, B.~P., {et~al.} 2015, Classical and Quantum
  Gravity, 32, 115012, \dodoi{10.1088/0264-9381/32/11/115012}

\bibitem[{{Abbott} {et~al.}(2016{\natexlab{a}}){Abbott}, {Abbott}, {Abbott},
  {Abernathy}, {Acernese}, {Ackley}, {Adams}, {Adams}, {Addesso}, {Adhikari},
  \& et~al.}]{2016PhRvL.116f1102A}
{Abbott}, B.~P., {Abbott}, R., {Abbott}, T.~D., {et~al.} 2016{\natexlab{a}},
  Physical Review Letters, 116, 061102, \dodoi{10.1103/PhysRevLett.116.061102}

\bibitem[{{Abbott} {et~al.}(2016{\natexlab{b}}){Abbott}, {Abbott}, {Abbott},
  {Abernathy}, {Acernese}, {Ackley}, {Adams}, {Adams}, {Addesso}, {Adhikari},
  {Adya}, {Affeldt}, {Agathos}, {Agatsuma}, {Aggarwal}, {Aguiar}, {Aiello},
  {Ain}, {Ajith}, {Allen}, {Allocca}, {Altin}, {Anderson}, {Anderson}, {Arai},
  {Araya}, {Arceneaux}, {Areeda}, {Arnaud}, {Arun}, {Ascenzi}, {Ashton}, {Ast},
  {Aston}, {Astone}, {Aufmuth}, {Aulbert}, {Babak}, {Bacon}, {Bader}, {Baker},
  {Baldaccini}, {Ballardin}, {Ballmer}, {Barayoga}, {Barclay}, {Barish},
  {Barker}, {Barone}, {Barr}, {Barsotti}, {Barsuglia}, {Barta}, {Bartlett},
  {Bartos}, {Bassiri}, {Basti}, {Batch}, {Baune}, {Bavigadda}, {Bazzan},
  {Bejger}, {Bell}, {Berger}, {Bergmann}, {Berry}, {Bersanetti}, {Bertolini},
  {Betzwieser}, {Bhagwat}, {Bhandare}, {Bilenko}, {Billingsley}, {Birch},
  {Birney}, {Birnholtz}, {Biscans}, {Bisht}, {Bitossi}, {Biwer}, {Bizouard},
  {Blackburn}, {Blair}, {Blair}, {Blair}, {Bloemen}, {Bock}, {Boer}, {Bogaert},
  {Bogan}, {Bohe}, {Bond}, {Bondu}, {Bonnand}, {Boom}, {Bork}, {Boschi},
  {Bose}, {Bouffanais}, {Bozzi}, {Bradaschia}, {Brady}, {Braginsky},
  {Branchesi}, {Brau}, {Briant}, {Brillet}, {Brinkmann}, {Brisson}, {Brockill},
  {Broida}, {Brooks}, {Brown}, {Brown}, {Brown}, {Brunett}, {Buchanan},
  {Buikema}, {Bulik}, {Bulten}, {Buonanno}, {Buskulic}, {Buy}, {Byer},
  {Cabero}, {Cadonati}, {Cagnoli}, {Cahillane}, {Calder{\'o}n Bustillo},
  {Callister}, {Calloni}, {Camp}, {Cannon}, {Cao}, {Capano}, {Capocasa},
  {Carbognani}, {Caride}, {Casanueva Diaz}, {Casentini}, {Caudill},
  {Cavagli{\`a}}, {Cavalier}, {Cavalieri}, {Cella}, {Cepeda}, {Cerboni
  Baiardi}, {Cerretani}, {Cesarini}, {Chamberlin}, {Chan}, {Chao}, {Charlton},
  {Chassande-Mottin}, {Cheeseboro}, {Chen}, {Chen}, {Cheng}, {Chincarini},
  {Chiummo}, {Cho}, {Cho}, {Chow}, {Christensen}, {Chu}, {Chua}, {Chung},
  {Ciani}, {Clara}, {Clark}, {Cleva}, {Coccia}, {Cohadon}, {Colla}, {Collette},
  {Cominsky}, {Constancio}, {Conte}, {Conti}, {Cook}, {Corbitt}, {Cornish},
  {Corsi}, {Cortese}, {Costa}, {Coughlin}, {Coughlin}, {Coulon}, {Countryman},
  {Couvares}, {Cowan}, {Coward}, {Cowart}, {Coyne}, {Coyne}, {Craig},
  {Creighton}, {Cripe}, {Crowder}, {Cumming}, {Cunningham}, {Cuoco}, {Dal
  Canton}, {Danilishin}, {D'Antonio}, {Danzmann}, {Darman}, {Dasgupta}, {Da
  Silva Costa}, {Dattilo}, {Dave}, {Davier}, {Davies}, {Daw}, {Day}, {De},
  {DeBra}, {Debreczeni}, {Degallaix}, {De Laurentis}, {Del{\'e}glise}, {Del
  Pozzo}, {Denker}, {Dent}, {Dergachev}, {De Rosa}, {DeRosa}, {DeSalvo},
  {Devine}, {Dhurand har}, {D{\'\i}az}, {Di Fiore}, {Di Giovanni}, {Di
  Girolamo}, {Di Lieto}, {Di Pace}, {Di Palma}, {Di Virgilio}, {Dolique},
  {Donovan}, {Dooley}, {Doravari}, {Douglas}, {Downes}, {Drago}, {Drever},
  {Driggers}, {Ducrot}, {Dwyer}, {Edo}, {Edwards}, {Effler}, {Eggenstein},
  {Ehrens}, {Eichholz}, {Eikenberry}, {Engels}, {Essick}, {Etzel}, {Evans},
  {Evans}, {Everett}, {Factourovich}, {Fafone}, {Fair}, {Fairhurst}, {Fan},
  {Fang}, {Farinon}, {Farr}, {Farr}, {Favata}, {Fays}, {Fehrmann}, {Fejer},
  {Fenyvesi}, {Ferrante}, {Ferreira}, {Ferrini}, {Fidecaro}, {Fiori},
  {Fiorucci}, {Fisher}, {Flaminio}, {Fletcher}, {Fong}, {Fournier}, {Frasca},
  {Frasconi}, {Frei}, {Freise}, {Frey}, {Frey}, {Fritschel}, {Frolov}, {Fulda},
  {Fyffe}, {Gabbard}, {Gaebel}, {Gair}, {Gammaitoni}, {Gaonkar}, {Garufi},
  {Gaur}, {Gehrels}, {Gemme}, {Geng}, {Genin}, {Gennai}, {George}, {Gergely},
  {Germain}, {Ghosh}, {Ghosh}, {Ghosh}, {Giaime}, {Giardina}, {Giazotto},
  {Gill}, {Glaefke}, {Goetz}, {Goetz}, {Gondan}, {Gonz{\'a}lez}, {Gonzalez
  Castro}, {Gopakumar}, {Gordon}, {Gorodetsky}, {Gossan}, {Gosselin}, {Gouaty},
  {Grado}, {Graef}, {Graff}, {Granata}, {Grant}, {Gras}, {Gray}, {Greco},
  {Green}, {Groot}, {Grote}, {Grunewald}, {Guidi}, {Guo}, {Gupta}, {Gupta},
  {Gushwa}, {Gustafson}, {Gustafson}, {Hacker}, {Hall}, {Hall}, {Hamilton},
  {Hammond}, {Haney}, {Hanke}, {Hanks}, {Hanna}, {Hannam}, {Hanson},
  {Hardwick}, {Harms}, {Harry}, {Harry}, {Hart}, {Hartman}, {Haster},
  {Haughian}, {Healy}, {Heidmann}, {Heintze}, {Heitmann}, {Hello}, {Hemming},
  {Hendry}, {Heng}, {Hennig}, {Henry}, {Heptonstall}, {Heurs}, {Hild}, {Hoak},
  {Hofman}, {Holt}, {Holz}, {Hopkins}, {Hough}, {Houston}, {Howell}, {Hu},
  {Huang}, {Huerta}, {Huet}, {Hughey}, {Husa}, {Huttner}, {Huynh-Dinh},
  {Indik}, {Ingram}, {Inta}, {Isa}, {Isac}, {Isi}, {Isogai}, {Iyer}, {Izumi},
  {Jacqmin}, {Jang}, {Jani}, {Jaranowski}, {Jawahar}, {Jian},
  {Jim{\'e}nez-Forteza}, {Johnson}, {Johnson-McDaniel}, {Jones}, {Jones},
  {Jonker}, {Ju}, {K}, {Kalaghatgi}, {Kalogera}, {Kandhasamy}, {Kang},
  {Kanner}, {Kapadia}, {Karki}, {Karvinen}, {Kasprzack}, {Katsavounidis},
  {Katzman}, {Kaufer}, {Kaur}, {Kawabe}, {K{\'e}f{\'e}lian}, {Kehl}, {Keitel},
  {Kelley}, {Kells}, {Kennedy}, {Key}, {Khalili}, {Khan}, {Khan}, {Khan},
  {Khazanov}, {Kijbunchoo}, {Kim}, {Kim}, {Kim}, {Kim}, {Kim}, {Kim}, {Kim},
  {Kimbrell}, {King}, {King}, {Kissel}, {Klein}, {Kleybolte}, {Klimenko},
  {Koehlenbeck}, {Koley}, {Kondrashov}, {Kontos}, {Korobko}, {Korth},
  {Kowalska}, {Kozak}, {Kringel}, {Krishnan}, {Kr{\'o}lak}, {Krueger}, {Kuehn},
  {Kumar}, {Kumar}, {Kuo}, {Kutynia}, {Lackey}, {Land ry}, {Lange}, {Lantz},
  {Lasky}, {Laxen}, {Lazzarini}, {Lazzaro}, {Leaci}, {Leavey}, {Lebigot},
  {Lee}, {Lee}, {Lee}, {Lee}, {Lenon}, {Leonardi}, {Leong}, {Leroy},
  {Letendre}, {Levin}, {Lewis}, {Li}, {Libson}, {Littenberg}, {Lockerbie},
  {Lombardi}, {London}, {Lord}, {Lorenzini}, {Loriette}, {Lormand}, {Losurdo},
  {Lough}, {Lousto}, {L{\"u}ck}, {Lundgren}, {Lynch}, {Ma}, {Machenschalk},
  {MacInnis}, {Macleod}, {Maga{\~n}a-Sandoval}, {Maga{\~n}a Zertuche}, {Magee},
  {Majorana}, {Maksimovic}, {Malvezzi}, {Man}, {Mandel}, {Mandic}, {Mangano},
  {Mansell}, {Manske}, {Mantovani}, {Marchesoni}, {Marion}, {M{\'a}rka},
  {M{\'a}rka}, {Markosyan}, {Maros}, {Martelli}, {Martellini}, {Martin},
  {Martynov}, {Marx}, {Mason}, {Masserot}, {Massinger}, {Masso-Reid},
  {Mastrogiovanni}, {Matichard}, {Matone}, {Mavalvala}, {Mazumder}, {McCarthy},
  {McClelland}, {McCormick}, {McGuire}, {McIntyre}, {McIver}, {McManus},
  {McRae}, {McWilliams}, {Meacher}, {Meadors}, {Meidam}, {Melatos}, {Mendell},
  {Mercer}, {Merilh}, {Merzougui}, {Meshkov}, {Messenger}, {Messick},
  {Metzdorff}, {Meyers}, {Mezzani}, {Miao}, {Michel}, {Middleton}, {Mikhailov},
  {Milano}, {Miller}, {Miller}, {Miller}, {Miller}, {Millhouse}, {Minenkov},
  {Ming}, {Mirshekari}, {Mishra}, {Mitra}, {Mitrofanov}, {Mitselmakher},
  {Mittleman}, {Moggi}, {Mohan}, {Mohapatra}, {Montani}, {Moore}, {Moore},
  {Moraru}, {Moreno}, {Morriss}, {Mossavi}, {Mours}, {Mow-Lowry}, {Mueller},
  {Muir}, {Mukherjee}, {Mukherjee}, {Mukherjee}, {Mukund}, {Mullavey}, {Munch},
  {Murphy}, {Murray}, {Mytidis}, {Nardecchia}, {Naticchioni}, {Nayak},
  {Nedkova}, {Nelemans}, {Nelson}, {Neri}, {Neunzert}, {Newton}, {Nguyen},
  {Nielsen}, {Nissanke}, {Nitz}, {Nocera}, {Nolting}, {Normandin}, {Nuttall},
  {Oberling}, {Ochsner}, {O'Dell}, {Oelker}, {Ogin}, {Oh}, {Oh}, {Ohme},
  {Oliver}, {Oppermann}, {Oram}, {O'Reilly}, {O'Shaughnessy}, {Ottaway},
  {Overmier}, {Owen}, {Pai}, {Pai}, {Palamos}, {Palashov}, {Palomba},
  {Pal-Singh}, {Pan}, {Pan}, {Pankow}, {Pannarale}, {Pant}, {Paoletti},
  {Paoli}, {Papa}, {Paris}, {Parker}, {Pascucci}, {Pasqualetti}, {Passaquieti},
  {Passuello}, {Patricelli}, {Patrick}, {Pearlstone}, {Pedraza}, {Pedurand},
  {Pekowsky}, {Pele}, {Penn}, {Perreca}, {Perri}, {Pfeiffer}, {Phelps},
  {Piccinni}, {Pichot}, {Piergiovanni}, {Pierro}, {Pillant}, {Pinard}, {Pinto},
  {Pitkin}, {Poe}, {Poggiani}, {Popolizio}, {Porter}, {Post}, {Powell},
  {Prasad}, {Predoi}, {Prestegard}, {Price}, {Prijatelj}, {Principe},
  {Privitera}, {Prix}, {Prodi}, {Prokhorov}, {Puncken}, {Punturo}, {Puppo},
  {P{\"u}rrer}, {Qi}, {Qin}, {Qiu}, {Quetschke}, {Quintero}, {Quitzow-James},
  {Raab}, {Rabeling}, {Radkins}, {Raffai}, {Raja}, {Rajan}, {Rakhmanov},
  {Rapagnani}, {Raymond}, {Razzano}, {Re}, {Read}, {Reed}, {Regimbau}, {Rei},
  {Reid}, {Reitze}, {Rew}, {Reyes}, {Ricci}, {Riles}, {Rizzo}, {Robertson},
  {Robie}, {Robinet}, {Rocchi}, {Rolland}, {Rollins}, {Roma}, {Romano},
  {Romano}, {Romanov}, {Romie}, {Rosi{\'n}ska}, {Rowan}, {R{\"u}diger},
  {Ruggi}, {Ryan}, {Sachdev}, {Sadecki}, {Sadeghian}, {Sakellariadou},
  {Salconi}, {Saleem}, {Salemi}, {Samajdar}, {Sammut}, {Sanchez}, {Sandberg},
  {Sandeen}, {Sand ers}, {Sassolas}, {Sathyaprakash}, {Saulson}, {Sauter},
  {Savage}, {Sawadsky}, {Schale}, {Schilling}, {Schmidt}, {Schmidt},
  {Schnabel}, {Schofield}, {Sch{\"o}nbeck}, {Schreiber}, {Schuette}, {Schutz},
  {Scott}, {Scott}, {Sellers}, {Sengupta}, {Sentenac}, {Sequino}, {Sergeev},
  {Setyawati}, {Shaddock}, {Shaffer}, {Shahriar}, {Shaltev}, {Shapiro},
  {Shawhan}, {Sheperd}, {Shoemaker}, {Shoemaker}, {Siellez}, {Siemens},
  {Sieniawska}, {Sigg}, {Silva}, {Singer}, {Singer}, {Singh}, {Singh},
  {Singhal}, {Sintes}, {Slagmolen}, {Smith}, {Smith}, {Smith}, {Son}, {Sorazu},
  {Sorrentino}, {Souradeep}, {Srivastava}, {Staley}, {Steinke}, {Steinlechner},
  {Steinlechner}, {Steinmeyer}, {Stephens}, {Stevenson}, {Stone}, {Strain},
  {Straniero}, {Stratta}, {Strauss}, {Strigin}, {Sturani}, {Stuver},
  {Summerscales}, {Sun}, {Sunil}, {Sutton}, {Swinkels}, {Szczepa{\'n}czyk},
  {Tacca}, {Talukder}, {Tanner}, {T{\'a}pai}, {Tarabrin}, {Taracchini},
  {Taylor}, {Theeg}, {Thirugnanasamband am}, {Thomas}, {Thomas}, {Thomas},
  {Thorne}, {Thrane}, {Tiwari}, {Tiwari}, {Tokmakov}, {Toland}, {Tomlinson},
  {Tonelli}, {Tornasi}, {Torres}, {Torrie}, {T{\"o}yr{\"a}}, {Travasso},
  {Traylor}, {Trifir{\`o}}, {Tringali}, {Trozzo}, {Tse}, {Turconi},
  {Tuyenbayev}, {Ugolini}, {Unnikrishnan}, {Urban}, {Usman}, {Vahlbruch},
  {Vajente}, {Valdes}, {Vallisneri}, {van Bakel}, {van Beuzekom}, {van den
  Brand}, {Van Den Broeck}, {Vand er-Hyde}, {van der Schaaf}, {van Heijningen},
  {van Veggel}, {Vardaro}, {Vass}, {Vas{\'u}th}, {Vaulin}, {Vecchio},
  {Vedovato}, {Veitch}, {Veitch}, {Venkateswara}, {Verkindt}, {Vetrano},
  {Vicer{\'e}}, {Vinciguerra}, {Vine}, {Vinet}, {Vitale}, {Vo}, {Vocca},
  {Vorvick}, {Voss}, {Vousden}, {Vyatchanin}, {Wade}, {Wade}, {Wade}, {Walker},
  {Wallace}, {Walsh}, {Wang}, {Wang}, {Wang}, {Wang}, {Wang}, {Ward}, {Warner},
  {Was}, {Weaver}, {Wei}, {Weinert}, {Weinstein}, {Weiss}, {Wen}, {We{\ss}els},
  {Westphal}, {Wette}, {Whelan}, {Whitcomb}, {Whiting}, {Williams},
  {Williamson}, {Willis}, {Willke}, {Wimmer}, {Winkler}, {Wipf}, {Wittel},
  {Woan}, {Woehler}, {Worden}, {Wright}, {Wu}, {Wu}, {Yablon}, {Yam},
  {Yamamoto}, {Yancey}, {Yu}, {Yvert}, {Zadro{\.Z}ny}, {Zangrando}, {Zanolin},
  {Zendri}, {Zevin}, {Zhang}, {Zhang}, {Zhang}, {Zhao}, {Zhou}, {Zhou}, {Zhu},
  {Zucker}, {Zuraw}, {Zweizig}, {LIGO Scientific Collaboration}, \& {Virgo
  Collaboration}}]{2016PhRvX...6d1015A}
---. 2016{\natexlab{b}}, Physical Review X, 6, 041015,
  \dodoi{10.1103/PhysRevX.6.041015}

\bibitem[{{Abbott} {et~al.}(2019){Abbott}, {Abbott}, {Abbott}, {Abraham},
  {Acernese}, {Ackley}, {Adams}, {Adhikari}, {Adya}, {Affeldt}, {Agathos},
  {Agatsuma}, {Aggarwal}, {Aguiar}, {Aiello}, {Ain}, {Ajith}, {Allen},
  {Allocca}, {Aloy}, {Altin}, {Amato}, {Ananyeva}, {Anderson}, {Anderson},
  {Angelova}, {Antier}, {Appert}, {Arai}, {Araya}, {Areeda}, {Ar{\`e}ne},
  {Arnaud}, {Arun}, {Ascenzi}, {Ashton}, {Aston}, {Astone}, {Aubin}, {Aufmuth},
  {AultONeal}, {Austin}, {Avendano}, {Avila-Alvarez}, {Babak}, {Bacon},
  {Badaracco}, {Bader}, {Bae}, {Baker}, {Baldaccini}, {Ballardin}, {Ballmer},
  {Banagiri}, {Barayoga}, {Barclay}, {Barish}, {Barker}, {Barkett}, {Barnum},
  {Barone}, {Barr}, {Barsotti}, {Barsuglia}, {Barta}, {Bartlett}, {Bartos},
  {Bassiri}, {Basti}, {Bawaj}, {Bayley}, {Bazzan}, {B{\'e}csy}, {Bejger},
  {Belahcene}, {Bell}, {Beniwal}, {Berger}, {Bergmann}, {Bernuzzi}, {Bero},
  {Berry}, {Bersanetti}, {Bertolini}, {Betzwieser}, {Bhandare}, {Bidler},
  {Bilenko}, {Bilgili}, {Billingsley}, {Birch}, {Birney}, {Birnholtz},
  {Biscans}, {Biscoveanu}, {Bisht}, {Bitossi}, {Bizouard}, {Blackburn},
  {Blair}, {Blair}, {Blair}, {Bloemen}, {Bode}, {Boer}, {Boetzel}, {Bogaert},
  {Bondu}, {Bonilla}, {Bonnand}, {Booker}, {Boom}, {Booth}, {Bork}, {Boschi},
  {Bose}, {Bossie}, {Bossilkov}, {Bosveld}, {Bouffanais}, {Bozzi},
  {Bradaschia}, {Brady}, {Bramley}, {Branchesi}, {Brau}, {Briant}, {Briggs},
  {Brighenti}, {Brillet}, {Brinkmann}, {Brisson}, {Brockill}, {Brooks},
  {Brown}, {Brunett}, {Buikema}, {Bulik}, {Bulten}, {Buonanno}, {Buscicchio},
  {Buskulic}, {Buy}, {Byer}, {Cabero}, {Cadonati}, {Cagnoli}, {Cahillane},
  {Calder{\'o}n Bustillo}, {Callister}, {Calloni}, {Camp}, {Campbell},
  {Canepa}, {Cannon}, {Cao}, {Cao}, {Capocasa}, {Carbognani}, {Caride},
  {Carney}, {Carullo}, {Casanueva Diaz}, {Casentini}, {Caudill},
  {Cavagli{\`a}}, {Cavalier}, {Cavalieri}, {Cella}, {Cerd{\'a}-Dur{\'a}n},
  {Cerretani}, {Cesarini}, {Chaibi}, {Chakravarti}, {Chamberlin}, {Chan},
  {Chao}, {Charlton}, {Chase}, {Chassande-Mottin}, {Chatterjee}, {Chaturvedi},
  {Chatziioannou}, {Cheeseboro}, {Chen}, {Chen}, {Chen}, {Cheng}, {Cheong},
  {Chia}, {Chincarini}, {Chiummo}, {Cho}, {Cho}, {Cho}, {Christensen}, {Chu},
  {Chua}, {Chung}, {Chung}, {Ciani}, {Ciobanu}, {Ciolfi}, {Cipriano}, {Cirone},
  {Clara}, {Clark}, {Clearwater}, {Cleva}, {Cocchieri}, {Coccia}, {Cohadon},
  {Cohen}, {Colgan}, {Colleoni}, {Collette}, {Collins}, {Cominsky},
  {Constancio}, {Conti}, {Cooper}, {Corban}, {Corbitt}, {Cordero-Carri{\'o}n},
  {Corley}, {Cornish}, {Corsi}, {Cortese}, {Costa}, {Cotesta}, {Coughlin},
  {Coughlin}, {Coulon}, {Countryman}, {Couvares}, {Covas}, {Cowan}, {Coward},
  {Cowart}, {Coyne}, {Coyne}, {Creighton}, {Creighton}, {Cripe}, {Croquette},
  {Crowder}, {Cullen}, {Cumming}, {Cunningham}, {Cuoco}, {Dal Canton},
  {D{\'a}lya}, {Danilishin}, {D'Antonio}, {Danzmann}, {Dasgupta}, {Da Silva
  Costa}, {Datrier}, {Dattilo}, {Dave}, {Davier}, {Davis}, {Daw}, {DeBra},
  {Deenadayalan}, {Degallaix}, {De Laurentis}, {Del{\'e}glise}, {Del Pozzo},
  {DeMarchi}, {Demos}, {Dent}, {De Pietri}, {Derby}, {De Rosa}, {De Rossi},
  {DeSalvo}, {de Varona}, {Dhurandhar}, {D{\'\i}az}, {Dietrich}, {Di Fiore},
  {Di Giovanni}, {Di Girolamo}, {Di Lieto}, {Ding}, {Di Pace}, {Di Palma}, {Di
  Renzo}, {Dmitriev}, {Doctor}, {Donovan}, {Dooley}, {Doravari}, {Dorrington},
  {Downes}, {Drago}, {Driggers}, {Du}, {Ducoin}, {Dupej}, {Dwyer}, {Easter},
  {Edo}, {Edwards}, {Effler}, {Ehrens}, {Eichholz}, {Eikenberry}, {Eisenmann},
  {Eisenstein}, {Essick}, {Estelles}, {Estevez}, {Etienne}, {Etzel}, {Evans},
  {Evans}, {Fafone}, {Fair}, {Fairhurst}, {Fan}, {Farinon}, {Farr}, {Farr},
  {Fauchon-Jones}, {Favata}, {Fays}, {Fazio}, {Fee}, {Feicht}, {Fejer}, {Feng},
  {Fernandez-Galiana}, {Ferrante}, {Ferreira}, {Ferreira}, {Ferrini},
  {Fidecaro}, {Fiori}, {Fiorucci}, {Fishbach}, {Fisher}, {Fishner},
  {Fitz-Axen}, {Flaminio}, {Fletcher}, {Flynn}, {Fong}, {Font}, {Forsyth},
  {Fournier}, {Frasca}, {Frasconi}, {Frei}, {Freise}, {Frey}, {Frey},
  {Fritschel}, {Frolov}, {Fulda}, {Fyffe}, {Gabbard}, {Gadre}, {Gaebel},
  {Gair}, {Gammaitoni}, {Ganija}, {Gaonkar}, {Garcia},
  {Garc{\'\i}a-Quir{\'o}s}, {Garufi}, {Gateley}, {Gaudio}, {Gaur}, {Gayathri},
  {Gemme}, {Genin}, {Gennai}, {George}, {George}, {Gergely}, {Germain},
  {Ghonge}, {Ghosh}, {Ghosh}, {Ghosh}, {Giacomazzo}, {Giaime}, {Giardina},
  {Giazotto}, {Gill}, {Giordano}, {Glover}, {Godwin}, {Goetz}, {Goetz},
  {Goncharov}, {Gonz{\'a}lez}, {Gonzalez Castro}, {Gopakumar}, {Gorodetsky},
  {Gossan}, {Gosselin}, {Gouaty}, {Grado}, {Graef}, {Granata}, {Grant}, {Gras},
  {Grassia}, {Gray}, {Gray}, {Greco}, {Green}, {Green}, {Gretarsson}, {Groot},
  {Grote}, {Grunewald}, {Gruning}, {Guidi}, {Gulati}, {Guo}, {Gupta}, {Gupta},
  {Gustafson}, {Gustafson}, {Haegel}, {Halim}, {Hall}, {Hall}, {Hamilton},
  {Hammond}, {Haney}, {Hanke}, {Hanks}, {Hanna}, {Hannam}, {Hannuksela},
  {Hanson}, {Hardwick}, {Haris}, {Harms}, {Harry}, {Harry}, {Haster},
  {Haughian}, {Hayes}, {Healy}, {Heidmann}, {Heintze}, {Heitmann}, {Hello},
  {Hemming}, {Hendry}, {Heng}, {Hennig}, {Heptonstall}, {Hernandez Vivanco},
  {Heurs}, {Hild}, {Hinderer}, {Hoak}, {Hochheim}, {Hofman}, {Holgado},
  {Holland}, {Holt}, {Holz}, {Hopkins}, {Horst}, {Hough}, {Howell}, {Hoy},
  {Hreibi}, {Huerta}, {Huet}, {Hughey}, {Hulko}, {Husa}, {Huttner},
  {Huynh-Dinh}, {Idzkowski}, {Iess}, {Ingram}, {Inta}, {Intini}, {Irwin},
  {Isa}, {Isac}, {Isi}, {Iyer}, {Izumi}, {Jacqmin}, {Jadhav}, {Jani},
  {Janthalur}, {Jaranowski}, {Jenkins}, {Jiang}, {Johnson}, {Jones}, {Jones},
  {Jones}, {Jonker}, {Ju}, {Junker}, {Kalaghatgi}, {Kalogera}, {Kamai},
  {Kandhasamy}, {Kang}, {Kanner}, {Kapadia}, {Karki}, {Karvinen}, {Kashyap},
  {Kasprzack}, {Katsanevas}, {Katsavounidis}, {Katzman}, {Kaufer}, {Kawabe},
  {Keerthana}, {K{\'e}f{\'e}lian}, {Keitel}, {Kennedy}, {Key}, {Khalili},
  {Khan}, {Khan}, {Khan}, {Khan}, {Khazanov}, {Khursheed}, {Kijbunchoo}, {Kim},
  {Kim}, {Kim}, {Kim}, {Kim}, {Kim}, {Kimball}, {King}, {King},
  {Kinley-Hanlon}, {Kirchhoff}, {Kissel}, {Kleybolte}, {Klika}, {Klimenko},
  {Knowles}, {Koch}, {Koehlenbeck}, {Koekoek}, {Koley}, {Kondrashov}, {Kontos},
  {Koper}, {Korobko}, {Korth}, {Kowalska}, {Kozak}, {Kringel}, {Krishnendu},
  {Kr{\'o}lak}, {Kuehn}, {Kumar}, {Kumar}, {Kumar}, {Kumar}, {Kuo}, {Kutynia},
  {Kwang}, {Lackey}, {Lai}, {Lam}, {Landry}, {Lane}, {Lang}, {Lange}, {Lantz},
  {Lanza}, {Lartaux-Vollard}, {Lasky}, {Laxen}, {Lazzarini}, {Lazzaro},
  {Leaci}, {Leavey}, {Lecoeuche}, {Lee}, {Lee}, {Lee}, {Lee}, {Lee}, {Lee},
  {Lehmann}, {Lenon}, {Leroy}, {Letendre}, {Levin}, {Li}, {Li}, {Li}, {Li},
  {Lin}, {Linde}, {Linker}, {Littenberg}, {Liu}, {Liu}, {Lo}, {Lockerbie},
  {London}, {Longo}, {Lorenzini}, {Loriette}, {Lormand}, {Losurdo}, {Lough},
  {Lousto}, {Lovelace}, {Lower}, {L{\"u}ck}, {Lumaca}, {Lundgren}, {Lynch},
  {Ma}, {Macas}, {Macfoy}, {MacInnis}, {Macleod}, {Macquet},
  {Maga{\~n}a-Sandoval}, {Maga{\~n}a Zertuche}, {Magee}, {Majorana},
  {Maksimovic}, {Malik}, {Man}, {Mandic}, {Mangano}, {Mansell}, {Manske},
  {Mantovani}, {Mapelli}, {Marchesoni}, {Marion}, {M{\'a}rka}, {M{\'a}rka},
  {Markakis}, {Markosyan}, {Markowitz}, {Maros}, {Marquina}, {Marsat},
  {Martelli}, {Martin}, {Martin}, {Martynov}, {Mason}, {Massera}, {Masserot},
  {Massinger}, {Masso-Reid}, {Mastrogiovanni}, {Matas}, {Matichard}, {Matone},
  {Mavalvala}, {Mazumder}, {McCann}, {McCarthy}, {McClelland}, {McCormick},
  {McCuller}, {McGuire}, {McIver}, {McManus}, {McRae}, {McWilliams}, {Meacher},
  {Meadors}, {Mehmet}, {Mehta}, {Meidam}, {Melatos}, {Mendell}, {Mercer},
  {Mereni}, {Merilh}, {Merzougui}, {Meshkov}, {Messenger}, {Messick},
  {Metzdorff}, {Meyers}, {Miao}, {Michel}, {Middleton}, {Mikhailov}, {Milano},
  {Miller}, {Miller}, {Millhouse}, {Mills}, {Milovich-Goff}, {Minazzoli},
  {Minenkov}, {Mishkin}, {Mishra}, {Mistry}, {Mitra}, {Mitrofanov},
  {Mitselmakher}, {Mittleman}, {Mo}, {Moffa}, {Mogushi}, {Mohapatra},
  {Montani}, {Moore}, {Moraru}, {Moreno}, {Morisaki}, {Mours}, {Mow-Lowry},
  {Mukherjee}, {Mukherjee}, {Mukherjee}, {Mukund}, {Mullavey}, {Munch},
  {Mu{\~n}iz}, {Muratore}, {Murray}, {Nagar}, {Nardecchia}, {Naticchioni},
  {Nayak}, {Neilson}, {Nelemans}, {Nelson}, {Nery}, {Neunzert}, {Ng}, {Ng},
  {Nguyen}, {Nichols}, {Nissanke}, {Nocera}, {North}, {Nuttall},
  {Obergaulinger}, {Oberling}, {O'Brien}, {O'Dea}, {Ogin}, {Oh}, {Oh}, {Ohme},
  {Ohta}, {Okada}, {Oliver}, {Oppermann}, {Oram}, {O'Reilly}, {Ormiston},
  {Ortega}, {O'Shaughnessy}, {Ossokine}, {Ottaway}, {Overmier}, {Owen}, {Pace},
  {Pagano}, {Page}, {Pai}, {Pai}, {Palamos}, {Palashov}, {Palomba},
  {Pal-Singh}, {Pan}, {Pang}, {Pang}, {Pankow}, {Pannarale}, {Pant},
  {Paoletti}, {Paoli}, {Parida}, {Parker}, {Pascucci}, {Pasqualetti},
  {Passaquieti}, {Passuello}, {Patil}, {Patricelli}, {Pearlstone}, {Pedersen},
  {Pedraza}, {Pedurand}, {Pele}, {Penn}, {Perez}, {Perreca}, {Pfeiffer},
  {Phelps}, {Phukon}, {Piccinni}, {Pichot}, {Piergiovanni}, {Pillant},
  {Pinard}, {Pirello}, {Pitkin}, {Poggiani}, {Pong}, {Ponrathnam}, {Popolizio},
  {Porter}, {Powell}, {Prajapati}, {Prasad}, {Prasai}, {Prasanna}, {Pratten},
  {Prestegard}, {Privitera}, {Prodi}, {Prokhorov}, {Puncken}, {Punturo},
  {Puppo}, {P{\"u}rrer}, {Qi}, {Quetschke}, {Quinonez}, {Quintero},
  {Quitzow-James}, {Raab}, {Radkins}, {Radulescu}, {Raffai}, {Raja}, {Rajan},
  {Rajbhandari}, {Rakhmanov}, {Ramirez}, {Ramos-Buades}, {Rana}, {Rao},
  {Rapagnani}, {Raymond}, {Razzano}, {Read}, {Regimbau}, {Rei}, {Reid},
  {Reitze}, {Ren}, {Ricci}, {Richardson}, {Richardson}, {Ricker}, {Riles},
  {Rizzo}, {Robertson}, {Robie}, {Robinet}, {Rocchi}, {Rolland}, {Rollins},
  {Roma}, {Romanelli}, {Romano}, {Romel}, {Romie}, {Rose}, {Rosi{\'n}ska},
  {Rosofsky}, {Ross}, {Rowan}, {R{\"u}diger}, {Ruggi}, {Rutins}, {Ryan},
  {Sachdev}, {Sadecki}, {Sakellariadou}, {Salconi}, {Saleem}, {Samajdar},
  {Sammut}, {Sanchez}, {Sanchez}, {Sanchis-Gual}, {Sandberg}, {Sanders},
  {Santiago}, {Sarin}, {Sassolas}, {Sathyaprakash}, {Saulson}, {Sauter},
  {Savage}, {Schale}, {Scheel}, {Scheuer}, {Schmidt}, {Schnabel}, {Schofield},
  {Sch{\"o}nbeck}, {Schreiber}, {Schulte}, {Schutz}, {Schwalbe}, {Scott},
  {Scott}, {Seidel}, {Sellers}, {Sengupta}, {Sennett}, {Sentenac}, {Sequino},
  {Sergeev}, {Setyawati}, {Shaddock}, {Shaffer}, {Shahriar}, {Shaner}, {Shao},
  {Sharma}, {Shawhan}, {Shen}, {Shink}, {Shoemaker}, {Shoemaker},
  {ShyamSundar}, {Siellez}, {Sieniawska}, {Sigg}, {Silva}, {Singer}, {Singh},
  {Singhal}, {Sintes}, {Sitmukhambetov}, {Skliris}, {Slagmolen},
  {Slaven-Blair}, {Smith}, {Smith}, {Somala}, {Son}, {Sorazu}, {Sorrentino},
  {Souradeep}, {Sowell}, {Spencer}, {Spera}, {Srivastava}, {Srivastava},
  {Staats}, {Stachie}, {Standke}, {Steer}, {Steinke}, {Steinlechner},
  {Steinlechner}, {Steinmeyer}, {Stevenson}, {Stocks}, {Stone}, {Stops},
  {Strain}, {Stratta}, {Strigin}, {Strunk}, {Sturani}, {Stuver}, {Sudhir},
  {Summerscales}, {Sun}, {Sunil}, {Suresh}, {Sutton}, {Swinkels},
  {Szczepa{\'n}czyk}, {Tacca}, {Tait}, {Talbot}, {Talukder}, {Tanner},
  {T{\'a}pai}, {Taracchini}, {Tasson}, {Taylor}, {Thies}, {Thomas}, {Thomas},
  {Thondapu}, {Thorne}, {Thrane}, {Tiwari}, {Tiwari}, {Tiwari}, {Toland},
  {Tonelli}, {Tornasi}, {Torres-Forn{\'e}}, {Torrie}, {T{\"o}yr{\"a}},
  {Travasso}, {Traylor}, {Tringali}, {Trovato}, {Trozzo}, {Trudeau}, {Tsang},
  {Tse}, {Tso}, {Tsukada}, {Tsuna}, {Tuyenbayev}, {Ueno}, {Ugolini},
  {Unnikrishnan}, {Urban}, {Usman}, {Vahlbruch}, {Vajente}, {Valdes}, {van
  Bakel}, {van Beuzekom}, {van den Brand}, {Van Den Broeck}, {Vander-Hyde},
  {van der Schaaf}, {van Heijningen}, {van Veggel}, {Vardaro}, {Varma}, {Vass},
  {Vas{\'u}th}, {Vecchio}, {Vedovato}, {Veitch}, {Veitch}, {Venkateswara},
  {Venugopalan}, {Verkindt}, {Vetrano}, {Vicer{\'e}}, {Viets}, {Vine}, {Vinet},
  {Vitale}, {Vo}, {Vocca}, {Vorvick}, {Vyatchanin}, {Wade}, {Wade}, {Wade},
  {Walet}, {Walker}, {Wallace}, {Walsh}, {Wang}, {Wang}, {Wang}, {Wang},
  {Wang}, {Ward}, {Warden}, {Warner}, {Was}, {Watchi}, {Weaver}, {Wei},
  {Weinert}, {Weinstein}, {Weiss}, {Wellmann}, {Wen}, {Wessel}, {We{\ss}els},
  {Westhouse}, {Wette}, {Whelan}, {Whiting}, {Whittle}, {Wilken}, {Williams},
  {Williamson}, {Willis}, {Willke}, {Wimmer}, {Winkler}, {Wipf}, {Wittel},
  {Woan}, {Woehler}, {Wofford}, {Worden}, {Wright}, {Wu}, {Wysocki}, {Xiao},
  {Yamamoto}, {Yancey}, {Yang}, {Yap}, {Yazback}, {Yeeles}, {Yu}, {Yu}, {Yuen},
  {Yvert}, {Zadro{\.z}ny}, {Zanolin}, {Zelenova}, {Zendri}, {Zevin}, {Zhang},
  {Zhang}, {Zhang}, {Zhao}, {Zhou}, {Zhou}, {Zhu}, {Zimmerman}, {Zlochower},
  {Zucker}, {Zweizig}, {LIGO Scientific Collaboration}, \& {Virgo
  Collaboration}}]{2019ApJ...882L..24A}
---. 2019, \apjl, 882, L24, \dodoi{10.3847/2041-8213/ab3800}

\bibitem[{{Abbott} {et~al.}(2020{\natexlab{a}}){Abbott}, {Abbott}, {Abraham},
  {Acernese}, {Ackley}, {Adams}, {Adams}, {Adhikari}, {Adya}, {Affeldt},
  {Agathos}, {Agatsuma}, {Aggarwal}, {Aguiar}, {Aiello}, {Ain}, {Ajith},
  {Akcay}, {Allen}, {Allocca}, {Altin}, {Amato}, {Anand}, {Ananyeva},
  {Anderson}, {Anderson}, {Angelova}, {Ansoldi}, {Antelis}, {Antier}, {Appert},
  {Arai}, {Araya}, {Areeda}, {Ar{\`e}ne}, {Arnaud}, {Aronson}, {Arun}, {Asali},
  {Ascenzi}, {Ashton}, {Aston}, {Astone}, {Aubin}, {Aufmuth}, {AultONeal},
  {Austin}, {Avendano}, {Babak}, {Badaracco}, {Bader}, {Bae}, {Baer},
  {Bagnasco}, {Baird}, {Ball}, {Ballardin}, {Ballmer}, {Bals}, {Balsamo},
  {Baltus}, {Banagiri}, {Bankar}, {Bankar}, {Barayoga}, {Barbieri}, {Barish},
  {Barker}, {Barneo}, {Barnum}, {Barone}, {Barr}, {Barsotti}, {Barsuglia},
  {Barta}, {Bartlett}, {Bartos}, {Bassiri}, {Basti}, {Bawaj}, {Bayley},
  {Bazzan}, {Becher}, {B{\'e}csy}, {Bedakihale}, {Bejger}, {Belahcene},
  {Beniwal}, {Benjamin}, {Bennett}, {Bentley}, {Bergamin}, {Berger},
  {Bergmann}, {Bernuzzi}, {Berry}, {Bersanetti}, {Bertolini}, {Betzwieser},
  {Bhand are}, {Bhandari}, {Bhattacharjee}, {Bidler}, {Bilenko}, {Billingsley},
  {Birney}, {Birnholtz}, {Biscans}, {Bischi}, {Biscoveanu}, {Bisht}, {Bitossi},
  {Bizouard}, {Blackburn}, {Blackman}, {Blair}, {Blair}, {Blair}, {Blanch},
  {Bobba}, {Bode}, {Boer}, {Boetzel}, {Bogaert}, {Boldrini}, {Bondu},
  {Bonnand}, {Bonilla}, {Booker}, {Boom}, {Bork}, {Boschi}, {Bose},
  {Bossilkov}, {Boudart}, {Bouffanais}, {Bozzi}, {Bradaschia}, {Brady},
  {Bramley}, {Branchesi}, {Brau}, {Breschi}, {Briant}, {Briggs}, {Brighenti},
  {Brillet}, {Brinkmann}, {Brockill}, {Brooks}, {Brooks}, {Brown}, {Brunett},
  {Bruno}, {Bruntz}, {Buikema}, {Bulik}, {Bulten}, {Buonanno}, {Buscicchio},
  {Buskulic}, {Byer}, {Cabero}, {Cadonati}, {Caesar}, {Cagnoli}, {Cahillane},
  {Calder{\'o}n Bustillo}, {Callaghan}, {Callister}, {Calloni}, {Camp},
  {Canepa}, {Cannon}, {Cao}, {Cao}, {Carapella}, {Carbognani}, {Carney},
  {Carpinelli}, {Carullo}, {Carver}, {Casanueva Diaz}, {Casentini}, {Caudill},
  {Cavagli{\`a}}, {Cavalier}, {Cavalieri}, {Cella}, {Cerd{\'a}-Dur{\'a}n},
  {Cesarini}, {Chaibi}, {Chakravarti}, {Chan}, {Chan}, {Chandra}, {Chanial},
  {Chao}, {Charlton}, {Chase}, {Chassande-Mottin}, {Chatterjee},
  {Chattopadhyay}, {Chaturvedi}, {Chatziioannou}, {Chen}, {Chen}, {Chen},
  {Chen}, {Cheng}, {Cheong}, {Chia}, {Chiadini}, {Chierici}, {Chincarini},
  {Chiummo}, {Cho}, {Cho}, {Cho}, {Choate}, {Christensen}, {Chu}, {Chua},
  {Chung}, {Chung}, {Ciani}, {Ciecielag}, {Cie{\'s}lar}, {Cifaldi}, {Ciobanu},
  {Ciolfi}, {Cipriano}, {Cirone}, {Clara}, {Clark}, {Clark}, {Clarke},
  {Clearwater}, {Clesse}, {Cleva}, {Coccia}, {Cohadon}, {Cohen}, {Colleoni},
  {Collette}, {Collins}, {Colpi}, {Constancio}, {Conti}, {Cooper}, {Corban},
  {Corbitt}, {Cordero-Carri{\'o}n}, {Corezzi}, {Corley}, {Cornish}, {Corre},
  {Corsi}, {Cortese}, {Costa}, {Cotesta}, {Coughlin}, {Coughlin}, {Coulon},
  {Countryman}, {Cousins}, {Couvares}, {Covas}, {Coward}, {Cowart}, {Coyne},
  {Coyne}, {Creighton}, {Creighton}, {Croquette}, {Crowder}, {Cudell},
  {Cullen}, {Cumming}, {Cummings}, {Cunningham}, {Cuoco}, {Curylo}, {Dal
  Canton}, {D{\'a}lya}, {Dana}, {DaneshgaranBajastani}, {D'Angelo}, {Danila},
  {Danilishin}, {D'Antonio}, {Danzmann}, {Darsow-Fromm}, {Dasgupta}, {Datrier},
  {Dattilo}, {Dave}, {Davier}, {Davies}, {Davis}, {Daw}, {Dean}, {DeBra},
  {Deenadayalan}, {Degallaix}, {De Laurentis}, {Del{\'e}glise}, {Del Favero},
  {De Lillo}, {De Lillo}, {Del Pozzo}, {DeMarchi}, {De Matteis}, {D'Emilio},
  {Demos}, {Denker}, {Dent}, {Depasse}, {De Pietri}, {De Rosa}, {De Rossi},
  {DeSalvo}, {de Varona}, {Dhurand har}, {D{\'\i}az}, {Diaz-Ortiz}, {Didio},
  {Dietrich}, {Di Fiore}, {DiFronzo}, {Di Giorgio}, {Di Giovanni}, {Di
  Giovanni}, {Di Girolamo}, {Di Lieto}, {Ding}, {Di Pace}, {Di Palma}, {Di
  Renzo}, {Divakarla}, {Dmitriev}, {Doctor}, {D'Onofrio}, {Donovan}, {Dooley},
  {Doravari}, {Dorrington}, {Downes}, {Drago}, {Driggers}, {Du}, {Ducoin},
  {Dupej}, {Durante}, {D'Urso}, {Duverne}, {Dwyer}, {Easter}, {Eddolls},
  {Edelman}, {Edo}, {Edy}, {Effler}, {Eichholz}, {Eikenberry}, {Eisenmann},
  {Eisenstein}, {Ejlli}, {Errico}, {Essick}, {Estell{\'e}s}, {Estevez},
  {Etienne}, {Etzel}, {Evans}, {Evans}, {Ewing}, {Fafone}, {Fair}, {Fairhurst},
  {Fan}, {Farah}, {Farinon}, {Farr}, {Farr}, {Fauchon-Jones}, {Favata}, {Fays},
  {Fazio}, {Feicht}, {Fejer}, {Feng}, {Fenyvesi}, {Ferguson},
  {Fernandez-Galiana}, {Ferrante}, {Ferreira}, {Fidecaro}, {Figura}, {Fiori},
  {Fiorucci}, {Fishbach}, {Fisher}, {Fishner}, {Fittipaldi}, {Fitz-Axen},
  {Fiumara}, {Flaminio}, {Floden}, {Flynn}, {Fong}, {Font}, {Forsyth},
  {Fournier}, {Frasca}, {Frasconi}, {Frei}, {Freise}, {Frey}, {Frey},
  {Fritschel}, {Frolov}, {Fronz{\'e}}, {Fulda}, {Fyffe}, {Gabbard}, {Gadre},
  {Gaebel}, {Gair}, {Gais}, {Galaudage}, {Gamba}, {Ganapathy}, {Ganguly},
  {Gaonkar}, {Garaventa}, {Garc{\'\i}a-Quir{\'o}s}, {Garufi}, {Gateley},
  {Gaudio}, {Gayathri}, {Gemme}, {Gennai}, {George}, {George}, {George},
  {Gergely}, {Ghonge}, {Ghosh}, {Ghosh}, {Ghosh}, {Giacomazzo}, {Giacoppo},
  {Giaime}, {Giardina}, {Gibson}, {Gier}, {Gill}, {Giri}, {Glanzer}, {Gleckl},
  {Godwin}, {Goetz}, {Goetz}, {Gohlke}, {Goncharov}, {Gonz{\'a}lez},
  {Gopakumar}, {Gossan}, {Gosselin}, {Gouaty}, {Grace}, {Grado}, {Granata},
  {Granata}, {Grant}, {Gras}, {Grassia}, {Gray}, {Gray}, {Greco}, {Green},
  {Green}, {Gretarsson}, {Griggs}, {Grignani}, {Grimaldi}, {Grimes}, {Grimm},
  {Grote}, {Grunewald}, {Gruning}, {Guerrero}, {Guidi}, {Guimaraes},
  {Guix{\'e}}, {Gulati}, {Guo}, {Gupta}, {Gupta}, {Gupta}, {Gustafson},
  {Gustafson}, {Guzman}, {Haegel}, {Halim}, {Hall}, {Hamilton}, {Hammond},
  {Haney}, {Hanke}, {Hanks}, {Hanna}, {Hannam}, {Hannuksela}, {Hannuksela},
  {Hansen}, {Hansen}, {Hanson}, {Harder}, {Hardwick}, {Haris}, {Harms},
  {Harry}, {Harry}, {Hartwig}, {Hasskew}, {Haster}, {Haughian}, {Hayes},
  {Healy}, {Heidmann}, {Heintze}, {Heinze}, {Heinzel}, {Heitmann}, {Hellman},
  {Hello}, {Helmling-Cornell}, {Hemming}, {Hendry}, {Heng}, {Hennes}, {Hennig},
  {Hennig}, {Hernandez Vivanco}, {Heurs}, {Hild}, {Hill}, {Hines}, {Hochheim},
  {Hofgard}, {Hofman}, {Hohmann}, {Holgado}, {Holland}, {Hollows}, {Holmes},
  {Holt}, {Holz}, {Hopkins}, {Horst}, {Hough}, {Howell}, {Hoy}, {Hoyland},
  {Huang}, {H{\"u}bner}, {Huddart}, {Huerta}, {Hughey}, {Hui}, {Husa},
  {Huttner}, {Hutzler}, {Huxford}, {Huynh-Dinh}, {Idzkowski}, {Iess},
  {Imperato}, {Inchauspe}, {Ingram}, {Intini}, {Isi}, {Iyer},
  {JaberianHamedan}, {Jacqmin}, {Jadhav}, {Jadhav}, {James}, {Jani},
  {Janssens}, {Janthalur}, {Jaranowski}, {Jariwala}, {Jaume}, {Jenkins},
  {Jeunon}, {Jiang}, {Johns}, {Johnson-McDaniel}, {Jones}, {Jones}, {Jones},
  {Jones}, {Jones}, {Jonker}, {Ju}, {Junker}, {Kalaghatgi}, {Kalogera},
  {Kamai}, {Kand hasamy}, {Kang}, {Kanner}, {Kapadia}, {Kapasi},
  {Karathanasis}, {Karki}, {Kashyap}, {Kasprzack}, {Kastaun}, {Katsanevas},
  {Katsavounidis}, {Katzman}, {Kawabe}, {K{\'e}f{\'e}lian}, {Keitel}, {Key},
  {Khadka}, {Khalili}, {Khan}, {Khan}, {Khazanov}, {Khetan}, {Khursheed},
  {Kijbunchoo}, {Kim}, {Kim}, {Kim}, {Kim}, {Kim}, {Kim}, {Kimball}, {King},
  {Kinley-Hanlon}, {Kirchhoff}, {Kissel}, {Kleybolte}, {Klimenko}, {Knowles},
  {Knyazev}, {Koch}, {Koehlenbeck}, {Koekoek}, {Koley}, {Kolstein}, {Komori},
  {Kondrashov}, {Kontos}, {Koper}, {Korobko}, {Korth}, {Kovalam}, {Kozak},
  {Kr{\"a}mer}, {Kringel}, {Krishnendu}, {Kr{\'o}lak}, {Kuehn}, {Kumar},
  {Kumar}, {Kumar}, {Kumar}, {Kuns}, {Kwang}, {Lackey}, {Laghi}, {Laland e},
  {Lam}, {Lamberts}, {Landry}, {Lane}, {Lang}, {Lange}, {Lantz}, {Lanza}, {La
  Rosa}, {Lartaux-Vollard}, {Lasky}, {Laxen}, {Lazzarini}, {Lazzaro}, {Leaci},
  {Leavey}, {Lecoeuche}, {Lee}, {Lee}, {Lee}, {Lee}, {Lehmann}, {Leon},
  {Leroy}, {Letendre}, {Levin}, {Li}, {Li}, {Li}, {Li}, {Li}, {Linde},
  {Linker}, {Linley}, {Littenberg}, {Liu}, {Liu}, {Llorens-Monteagudo}, {Lo},
  {Lockwood}, {London}, {Longo}, {Lorenzini}, {Loriette}, {Lormand}, {Losurdo},
  {Lough}, {Lousto}, {Lovelace}, {L{\"u}ck}, {Lumaca}, {Lundgren}, {Ma},
  {Macas}, {MacInnis}, {Macleod}, {MacMillan}, {Macquet}, {Maga{\~n}a
  Hernandez}, {Maga{\~n}a-Sandoval}, {Magazz{\`u}}, {Magee}, {Majorana},
  {Maksimovic}, {Maliakal}, {Malik}, {Man}, {Mandic}, {Mangano}, {Mansell},
  {Manske}, {Mantovani}, {Mapelli}, {Marchesoni}, {Marion}, {M{\'a}rka},
  {M{\'a}rka}, {Markakis}, {Markosyan}, {Markowitz}, {Maros}, {Marquina},
  {Marsat}, {Martelli}, {Martin}, {Martin}, {Martinez}, {Martinez}, {Martynov},
  {Masalehdan}, {Mason}, {Massera}, {Masserot}, {Massinger}, {Masso-Reid},
  {Mastrogiovanni}, {Matas}, {Mateu-Lucena}, {Matichard}, {Matiushechkina},
  {Mavalvala}, {Maynard}, {McCann}, {McCarthy}, {McClelland}, {McCormick},
  {McCuller}, {McGuire}, {McIsaac}, {McIver}, {McManus}, {McRae}, {McWilliams},
  {Meacher}, {Meadors}, {Mehmet}, {Mehta}, {Melatos}, {Melchor}, {Mendell},
  {Menendez-Vazquez}, {Mercer}, {Mereni}, {Merfeld}, {Merilh}, {Merritt},
  {Merzougui}, {Meshkov}, {Messenger}, {Messick}, {Metzdorff}, {Meyers},
  {Meylahn}, {Mhaske}, {Miani}, {Miao}, {Michaloliakos}, {Michel}, {Middleton},
  {Milano}, {Miller}, {Millhouse}, {Mills}, {Milotti}, {Milovich-Goff},
  {Minazzoli}, {Minenkov}, {Mir}, {Mishkin}, {Mishra}, {Mistry}, {Mitra},
  {Mitrofanov}, {Mitselmakher}, {Mittleman}, {Mo}, {Mogushi}, {Mohapatra},
  {Mohite}, {Molina}, {Molina-Ruiz}, {Mondin}, {Montani}, {Moore}, {Moraru},
  {Morawski}, {Moreno}, {Morisaki}, {Mours}, {Mow-Lowry}, {Mozzon},
  {Muciaccia}, {Mukherjee}, {Mukherjee}, {Mukherjee}, {Mukherjee}, {Mukund},
  {Mullavey}, {Munch}, {Mu{\~n}iz}, {Murray}, {Nadji}, {Nagar}, {Nardecchia},
  {Naticchioni}, {Nayak}, {Neil}, {Neilson}, {Nelemans}, {Nelson}, {Nery},
  {Neunzert}, {Nitz}, {Ng}, {Ng}, {Nguyen}, {Nguyen}, {Nguyen}, {Nichols},
  {Nissanke}, {Nocera}, {Noh}, {North}, {Nothard}, {Nuttall}, {Oberling},
  {O'Brien}, {O'Dell}, {Oganesyan}, {Ogin}, {Oh}, {Oh}, {Ohme}, {Ohta},
  {Okada}, {Olivetto}, {Oppermann}, {Oram}, {O'Reilly}, {Ormiston}, {Ortega},
  {O'Shaughnessy}, {Ossokine}, {Osthelder}, {Ottaway}, {Overmier}, {Owen},
  {Pace}, {Pagano}, {Page}, {Pagliaroli}, {Pai}, {Pai}, {Palamos}, {Palashov},
  {Palomba}, {Pan}, {Panda}, {Pang}, {Pankow}, {Pannarale}, {Pant}, {Paoletti},
  {Paoli}, {Paolone}, {Parker}, {Pascucci}, {Pasqualetti}, {Passaquieti},
  {Passuello}, {Patel}, {Patricelli}, {Payne}, {Pechsiri}, {Pedraza},
  {Pegoraro}, {Pele}, {Penn}, {Perego}, {Perez}, {P{\'e}rigois}, {Perreca},
  {Perri{\`e}s}, {Petermann}, {Petterson}, {Pfeiffer}, {Pham}, {Phukon},
  {Piccinni}, {Pichot}, {Piendibene}, {Piergiovanni}, {Pierini}, {Pierro},
  {Pillant}, {Pilo}, {Pinard}, {Pinto}, {Piotrzkowski}, {Pirello}, {Pitkin},
  {Placidi}, {Plastino}, {Pluchar}, {Poggiani}, {Polini}, {Pong}, {Ponrathnam},
  {Popolizio}, {Porter}, {Poverman}, {Powell}, {Pracchia}, {Prajapati},
  {Prasai}, {Prasanna}, {Pratten}, {Prestegard}, {Principe}, {Prodi},
  {Prokhorov}, {Prosposito}, {Prudenzi}, {Puecher}, {Punturo}, {Puosi},
  {Puppo}, {P{\"u}rrer}, {Qi}, {Quetschke}, {Quinonez}, {Quitzow-James},
  {Raab}, {Raaijmakers}, {Radkins}, {Radulesco}, {Raffai}, {Rafferty}, {Rail},
  {Raja}, {Rajan}, {Rajbhandari}, {Rakhmanov}, {Ramirez}, {Ramirez},
  {Ramos-Buades}, {Rana}, {Rao}, {Rapagnani}, {Rapol}, {Ratto}, {Raymond},
  {Razzano}, {Read}, {Regimbau}, {Rei}, {Reid}, {Reitze}, {Rettegno}, {Ricci},
  {Richardson}, {Richardson}, {Richardson}, {Ricker}, {Riemenschneider},
  {Riles}, {Rizzo}, {Robertson}, {Robinet}, {Rocchi}, {Rocha}, {Rodriguez},
  {Rodriguez-Soto}, {Rolland}, {Rollins}, {Roma}, {Romanelli}, {Romano},
  {Romel}, {Romero}, {Romero-Shaw}, {Romie}, {Ronchini}, {Rose}, {Rose},
  {Rose}, {Rosell}, {Rosi{\'n}ska}, {Rosofsky}, {Ross}, {Rowan}, {Rowlinson},
  {Roy}, {Roy}, {Ruggi}, {Ryan}, {Sachdev}, {Sadecki}, {Sadiq},
  {Sakellariadou}, {Salafia}, {Salconi}, {Saleem}, {Samajdar}, {Sanchez},
  {Sanchez}, {Sanchez}, {Sanchis-Gual}, {Sanders}, {Sandles}, {Santiago},
  {Santos}, {Saravanan}, {Sarin}, {Sassolas}, {Sathyaprakash}, {Sauter},
  {Savage}, {Savant}, {Sawant}, {Sayah}, {Schaetzl}, {Schale}, {Scheel},
  {Scheuer}, {Schindler-Tyka}, {Schmidt}, {Schnabel}, {Schofield},
  {Sch{\"o}nbeck}, {Schreiber}, {Schulte}, {Schutz}, {Schwarm}, {Schwartz},
  {Scott}, {Scott}, {Seglar-Arroyo}, {Seidel}, {Sellers}, {Sengupta},
  {Sennett}, {Sentenac}, {Sequino}, {Sergeev}, {Setyawati}, {Shaffer},
  {Shahriar}, {Sharifi}, {Sharma}, {Sharma}, {Shawhan}, {Shen}, {Shikauchi},
  {Shink}, {Shoemaker}, {Shoemaker}, {Shukla}, {ShyamSundar}, {Sieniawska},
  {Sigg}, {Singer}, {Singh}, {Singh}, {Singha}, {Singhal}, {Sintes}, {Sipala},
  {Skliris}, {Slagmolen}, {Slaven-Blair}, {Smetana}, {Smith}, {Smith},
  {Somala}, {Son}, {Soni}, {Soni}, {Sorazu}, {Sordini}, {Sorrentino},
  {Sorrentino}, {Soulard}, {Souradeep}, {Sowell}, {Spencer}, {Spera},
  {Srivastava}, {Srivastava}, {Staats}, {Stachie}, {Steer}, {Steinhoff},
  {Steinke}, {Steinlechner}, {Steinlechner}, {Steinmeyer}, {Stevenson},
  {Stolle-McAllister}, {Stops}, {Stover}, {Strain}, {Stratta}, {Strunk},
  {Sturani}, {Stuver}, {S{\"u}dbeck}, {Sudhagar}, {Sudhir}, {Suh},
  {Summerscales}, {Sun}, {Sun}, {Sunil}, {Sur}, {Suresh}, {Sutton}, {Swinkels},
  {Szczepa{\'n}czyk}, {Tacca}, {Tait}, {Talbot}, {Tanasijczuk}, {Tanner},
  {Tao}, {Tapia}, {Tapia San Martin}, {Tasson}, {Taylor}, {Tenorio},
  {Terkowski}, {Thirugnanasambandam}, {Thomas}, {Thomas}, {Thomas}, {Thompson},
  {Thondapu}, {Thorne}, {Thrane}, {Tiwari}, {Tiwari}, {Tiwari}, {Toland},
  {Tolley}, {Tonelli}, {Tornasi}, {Torres-Forn{\'e}}, {Torrie}, {Melo},
  {T{\"o}yr{\"a}}, {Tran}, {Trapananti}, {Travasso}, {Traylor}, {Tringali},
  {Tripathee}, {Trovato}, {Trudeau}, {Tsai}, {Tsang}, {Tse}, {Tso}, {Tsukada},
  {Tsuna}, {Tsutsui}, {Turconi}, {Ubhi}, {Udall}, {Ueno}, {Ugolini},
  {Unnikrishnan}, {Urban}, {Usman}, {Utina}, {Vahlbruch}, {Vajente}, {Vajpeyi},
  {Valdes}, {Valentini}, {Valsan}, {van Bakel}, {van Beuzekom}, {van den Brand
  }, {Van Den Broeck}, {Vander-Hyde}, {van der Schaaf}, {van Heijningen},
  {Vardaro}, {Vargas}, {Varma}, {Vass}, {Vas{\'u}th}, {Vecchio}, {Vedovato},
  {Veitch}, {Veitch}, {Venkateswara}, {Venneberg}, {Venugopalan}, {Verkindt},
  {Verma}, {Veske}, {Vetrano}, {Vicer{\'e}}, {Viets}, {Vijaykumar},
  {Villa-Ortega}, {Vinet}, {Vitale}, {Vo}, {Vocca}, {Vorvick}, {Vyatchanin},
  {Wade}, {Wade}, {Wade}, {Walet}, {Walker}, {Wallace}, {Wallace}, {Walsh},
  {Wang}, {Wang}, {Wang}, {Wang}, {Ward}, {Warner}, {Was}, {Washington},
  {Watchi}, {Weaver}, {Wei}, {Weinert}, {Weinstein}, {Weiss}, {Wellmann},
  {Wen}, {We{\ss}els}, {Westhouse}, {Wette}, {Whelan}, {White}, {White},
  {Whiting}, {Whittle}, {Wilken}, {Williams}, {Williams}, {Williamson},
  {Willis}, {Willke}, {Wilson}, {Wimmer}, {Winkler}, {Wipf}, {Woan}, {Woehler},
  {Wofford}, {Wong}, {Wrangel}, {Wright}, {Wu}, {Wysocki}, {Xiao}, {Yamamoto},
  {Yang}, {Yang}, {Yang}, {Yap}, {Yeeles}, {Yoon}, {Yu}, {Yu}, {Yuen},
  {Zadro{\.z}ny}, {Zanolin}, {Zelenova}, {Zendri}, {Zevin}, {Zhang}, {Zhang},
  {Zhang}, {Zhang}, {Zhao}, {Zhao}, {Zheng}, {Zhou}, {Zhou}, {Zhu},
  {Zimmerman}, {Zlochower}, {Zucker}, \& {Zweizig}}]{2020arXiv201014527A}
{Abbott}, R., {Abbott}, T.~D., {Abraham}, S., {et~al.} 2020{\natexlab{a}},
  arXiv e-prints, arXiv:2010.14527.
\newblock \doarXiv{2010.14527}

\bibitem[{{Abbott} {et~al.}(2020{\natexlab{b}}){Abbott}, {Abbott}, {Abraham},
  {Acernese}, {Ackley}, {Adams}, {Adhikari}, {Adya}, {Affeldt}, {Agathos},
  {Agatsuma}, {Aggarwal}, {Aguiar}, {Aich}, {Aiello}, {Ain}, {Ajith}, {Akcay},
  {Allen}, {Allocca}, {Altin}, {Amato}, {Anand}, {Ananyeva}, {Anderson},
  {Anderson}, {Angelova}, {Ansoldi}, {Antier}, {Appert}, {Arai}, {Araya},
  {Areeda}, {Ar{\`e}ne}, {Arnaud}, {Aronson}, {Arun}, {Asali}, {Ascenzi},
  {Ashton}, {Aston}, {Astone}, {Aubin}, {Aufmuth}, {AultONeal}, {Austin},
  {Avendano}, {Babak}, {Bacon}, {Badaracco}, {Bader}, {Bae}, {Baer}, {Baird},
  {Baldaccini}, {Ballardin}, {Ballmer}, {Bals}, {Balsamo}, {Baltus},
  {Banagiri}, {Bankar}, {Bankar}, {Barayoga}, {Barbieri}, {Barish}, {Barker},
  {Barkett}, {Barneo}, {Barone}, {Barr}, {Barsotti}, {Barsuglia}, {Barta},
  {Bartlett}, {Bartos}, {Bassiri}, {Basti}, {Bawaj}, {Bayley}, {Bazzan},
  {B{\'e}csy}, {Bejger}, {Belahcene}, {Bell}, {Beniwal}, {Benjamin}, {Bentley},
  {Bergamin}, {Berger}, {Bergmann}, {Bernuzzi}, {Berry}, {Bersanetti},
  {Bertolini}, {Betzwieser}, {Bhand are}, {Bhandari}, {Bidler}, {Biggs},
  {Bilenko}, {Billingsley}, {Birney}, {Birnholtz}, {Biscans}, {Bischi},
  {Biscoveanu}, {Bisht}, {Bissenbayeva}, {Bitossi}, {Bizouard}, {Blackburn},
  {Blackman}, {Blair}, {Blair}, {Blair}, {Bobba}, {Bode}, {Boer}, {Boetzel},
  {Bogaert}, {Bondu}, {Bonilla}, {Bonnand}, {Booker}, {Boom}, {Bork}, {Boschi},
  {Bose}, {Bossilkov}, {Bosveld}, {Bouffanais}, {Bozzi}, {Bradaschia}, {Brady},
  {Bramley}, {Branchesi}, {Brau}, {Breschi}, {Briant}, {Briggs}, {Brighenti},
  {Brillet}, {Brinkmann}, {Brockill}, {Brooks}, {Brooks}, {Brown}, {Brunett},
  {Bruno}, {Bruntz}, {Buikema}, {Bulik}, {Bulten}, {Buonanno}, {Buscicchio},
  {Buskulic}, {Byer}, {Cabero}, {Cadonati}, {Cagnoli}, {Cahillane},
  {Calder{\'o}n Bustillo}, {Callaghan}, {Callister}, {Calloni}, {Camp},
  {Canepa}, {Cannon}, {Cao}, {Cao}, {Carapella}, {Carbognani}, {Caride},
  {Carney}, {Carullo}, {Casanueva Diaz}, {Casentini}, {Casta{\~n}eda},
  {Caudill}, {Cavagli{\`a}}, {Cavalier}, {Cavalieri}, {Cella},
  {Cerd{\'a}-Dur{\'a}n}, {Cesarini}, {Chaibi}, {Chakravarti}, {Chan}, {Chan},
  {Chandra}, {Chao}, {Charlton}, {Chase}, {Chassande-Mottin}, {Chatterjee},
  {Chaturvedi}, {Chatziioannou}, {Chen}, {Chen}, {Chen}, {Cheng}, {Cheong},
  {Chia}, {Chiadini}, {Chierici}, {Chincarini}, {Chiummo}, {Cho}, {Cho}, {Cho},
  {Christensen}, {Chu}, {Chua}, {Chung}, {Chung}, {Ciani}, {Ciecielag},
  {Cie{\'s}lar}, {Ciobanu}, {Ciolfi}, {Cipriano}, {Cirone}, {Clara}, {Clark},
  {Clearwater}, {Clesse}, {Cleva}, {Coccia}, {Cohadon}, {Cohen}, {Colleoni},
  {Collette}, {Collins}, {Colpi}, {Constancio}, {Conti}, {Cooper}, {Corban},
  {Corbitt}, {Cordero-Carri{\'o}n}, {Corezzi}, {Corley}, {Cornish}, {Corre},
  {Corsi}, {Cortese}, {Costa}, {Cotesta}, {Coughlin}, {Coughlin}, {Coulon},
  {Countryman}, {Couvares}, {Covas}, {Coward}, {Cowart}, {Coyne}, {Coyne},
  {Creighton}, {Creighton}, {Cripe}, {Croquette}, {Crowder}, {Cudell},
  {Cullen}, {Cumming}, {Cummings}, {Cunningham}, {Cuoco}, {Curylo}, {Canton},
  {D{\'a}lya}, {Dana}, {Daneshgaran-Bajastani}, {D'Angelo}, {Danilishin},
  {D'Antonio}, {Danzmann}, {Darsow-Fromm}, {Dasgupta}, {Datrier}, {Dattilo},
  {Dave}, {Davier}, {Davies}, {Davis}, {Daw}, {DeBra}, {Deenadayalan},
  {Degallaix}, {De Laurentis}, {Del{\'e}glise}, {Delfavero}, {De Lillo}, {Del
  Pozzo}, {DeMarchi}, {D'Emilio}, {Demos}, {Dent}, {De Pietri}, {De Rosa}, {De
  Rossi}, {DeSalvo}, {de Varona}, {Dhurand har}, {D{\'\i}az}, {Diaz-Ortiz},
  {Dietrich}, {Di Fiore}, {Di Fronzo}, {Di Giorgio}, {Di Giovanni}, {Di
  Giovanni}, {Di Girolamo}, {Di Lieto}, {Ding}, {Di Pace}, {Di Palma}, {Di
  Renzo}, {Divakarla}, {Dmitriev}, {Doctor}, {Donovan}, {Dooley}, {Doravari},
  {Dorrington}, {Downes}, {Drago}, {Driggers}, {Du}, {Ducoin}, {Dupej},
  {Durante}, {D'Urso}, {Dwyer}, {Easter}, {Eddolls}, {Edelman}, {Edo}, {Edy},
  {Effler}, {Ehrens}, {Eichholz}, {Eikenberry}, {Eisenmann}, {Eisenstein},
  {Ejlli}, {Errico}, {Essick}, {Estelles}, {Estevez}, {Etienne}, {Etzel},
  {Evans}, {Evans}, {Ewing}, {Fafone}, {Fairhurst}, {Fan}, {Farinon}, {Farr},
  {Farr}, {Fauchon-Jones}, {Favata}, {Fays}, {Fazio}, {Feicht}, {Fejer},
  {Feng}, {Fenyvesi}, {Ferguson}, {Fernandez-Galiana}, {Ferrante}, {Ferreira},
  {Ferreira}, {Fidecaro}, {Fiori}, {Fiorucci}, {Fishbach}, {Fisher},
  {Fittipaldi}, {Fitz-Axen}, {Fiumara}, {Flaminio}, {Floden}, {Flynn}, {Fong},
  {Font}, {Forsyth}, {Fournier}, {Frasca}, {Frasconi}, {Frei}, {Freise},
  {Frey}, {Frey}, {Fritschel}, {Frolov}, {Fronz{\`e}}, {Fulda}, {Fyffe},
  {Gabbard}, {Gadre}, {Gaebel}, {Gair}, {Galaudage}, {Ganapathy}, {Ganguly},
  {Gaonkar}, {Garc{\'\i}a-Quir{\'o}s}, {Garufi}, {Gateley}, {Gaudio},
  {Gayathri}, {Gemme}, {Genin}, {Gennai}, {George}, {George}, {Gergely},
  {Ghonge}, {Ghosh}, {Ghosh}, {Ghosh}, {Giacomazzo}, {Giaime}, {Giardina},
  {Gibson}, {Gier}, {Gill}, {Glanzer}, {Gniesmer}, {Godwin}, {Goetz}, {Goetz},
  {Gohlke}, {Goncharov}, {Gonz{\'a}lez}, {Gopakumar}, {Gossan}, {Gosselin},
  {Gouaty}, {Grace}, {Grado}, {Granata}, {Grant}, {Gras}, {Grassia}, {Gray},
  {Gray}, {Greco}, {Green}, {Green}, {Gretarsson}, {Griggs}, {Grignani},
  {Grimaldi}, {Grimm}, {Grote}, {Grunewald}, {Gruning}, {Guidi}, {Guimaraes},
  {Guix{\'e}}, {Gulati}, {Guo}, {Gupta}, {Gupta}, {Gupta}, {Gustafson},
  {Gustafson}, {Haegel}, {Halim}, {Hall}, {Hamilton}, {Hammond}, {Haney},
  {Hanke}, {Hanks}, {Hanna}, {Hannam}, {Hannuksela}, {Hansen}, {Hanson},
  {Harder}, {Hardwick}, {Haris}, {Harms}, {Harry}, {Harry}, {Hasskew},
  {Haster}, {Haughian}, {Hayes}, {Healy}, {Heidmann}, {Heintze}, {Heinze},
  {Heitmann}, {Hellman}, {Hello}, {Hemming}, {Hendry}, {Heng}, {Hennes},
  {Hennig}, {Heurs}, {Hild}, {Hinderer}, {Hoback}, {Hochheim}, {Hofgard},
  {Hofman}, {Holgado}, {Holland}, {Holt}, {Holz}, {Hopkins}, {Horst}, {Hough},
  {Howell}, {Hoy}, {Huang}, {H{\"u}bner}, {Huerta}, {Huet}, {Hughey}, {Hui},
  {Husa}, {Huttner}, {Huxford}, {Huynh-Dinh}, {Idzkowski}, {Iess}, {Inchauspe},
  {Ingram}, {Intini}, {Isac}, {Isi}, {Iyer}, {Jacqmin}, {Jadhav}, {Jadhav},
  {James}, {Jani}, {Janthalur}, {Jaranowski}, {Jariwala}, {Jaume}, {Jenkins},
  {Jiang}, {Johns}, {Johnson-McDaniel}, {Jones}, {Jones}, {Jones}, {Jones},
  {Jones}, {Jonker}, {Ju}, {Junker}, {Kalaghatgi}, {Kalogera}, {Kamai},
  {Kandhasamy}, {Kang}, {Kanner}, {Kapadia}, {Karki}, {Kashyap}, {Kasprzack},
  {Kastaun}, {Katsanevas}, {Katsavounidis}, {Katzman}, {Kaufer}, {Kawabe},
  {K{\'e}f{\'e}lian}, {Keitel}, {Keivani}, {Kennedy}, {Key}, {Khadka},
  {Khalili}, {Khan}, {Khan}, {Khan}, {Khazanov}, {Khetan}, {Khursheed},
  {Kijbunchoo}, {Kim}, {Kim}, {Kim}, {Kim}, {Kim}, {Kim}, {Kim}, {Kimball},
  {King}, {Kinley-Hanlon}, {Kirchhoff}, {Kissel}, {Kleybolte}, {Klimenko},
  {Knowles}, {Knyazev}, {Koch}, {Koehlenbeck}, {Koekoek}, {Koley},
  {Kondrashov}, {Kontos}, {Koper}, {Korobko}, {Korth}, {Kovalam}, {Kozak},
  {Kringel}, {Krishnendu}, {Kr{\'o}lak}, {Krupinski}, {Kuehn}, {Kumar},
  {Kumar}, {Kumar}, {Kumar}, {Kumar}, {Kuo}, {Kutynia}, {Lackey}, {Laghi},
  {Lalande}, {Lam}, {Lamberts}, {Landry}, {Lane}, {Lang}, {Lange}, {Lantz},
  {Lanza}, {La Rosa}, {Lartaux-Vollard}, {Lasky}, {Laxen}, {Lazzarini},
  {Lazzaro}, {Leaci}, {Leavey}, {Lecoeuche}, {Lee}, {Lee}, {Lee}, {Lee}, {Lee},
  {Lehmann}, {Leroy}, {Letendre}, {Levin}, {Li}, {Li}, {li}, {Li}, {Li},
  {Linde}, {Linker}, {Linley}, {Littenberg}, {Liu}, {Liu},
  {Llorens-Monteagudo}, {Lo}, {Lockwood}, {London}, {Longo}, {Lorenzini},
  {Loriette}, {Lormand}, {Losurdo}, {Lough}, {Lousto}, {Lovelace}, {L{\"u}ck},
  {Lumaca}, {Lundgren}, {Ma}, {Macas}, {Macfoy}, {MacInnis}, {Macleod},
  {MacMillan}, {Macquet}, {Maga{\~n}a Hernandez}, {Maga{\~n}a-Sandoval},
  {Magee}, {Majorana}, {Maksimovic}, {Malik}, {Man}, {Mandic}, {Mangano},
  {Mansell}, {Manske}, {Mantovani}, {Mapelli}, {Marchesoni}, {Marion},
  {M{\'a}rka}, {M{\'a}rka}, {Markakis}, {Markosyan}, {Markowitz}, {Maros},
  {Marquina}, {Marsat}, {Martelli}, {Martin}, {Martin}, {Martinez}, {Martynov},
  {Masalehdan}, {Mason}, {Massera}, {Masserot}, {Massinger}, {Masso-Reid},
  {Mastrogiovanni}, {Matas}, {Matichard}, {Mavalvala}, {Maynard}, {McCann},
  {McCarthy}, {McClelland}, {McCormick}, {McCuller}, {McGuire}, {McIsaac},
  {McIver}, {McManus}, {McRae}, {McWilliams}, {Meacher}, {Meadors}, {Mehmet},
  {Mehta}, {Mejuto Villa}, {Melatos}, {Mendell}, {Mercer}, {Mereni}, {Merfeld},
  {Merilh}, {Merritt}, {Merzougui}, {Meshkov}, {Messenger}, {Messick},
  {Metzdorff}, {Meyers}, {Meylahn}, {Mhaske}, {Miani}, {Miao}, {Michaloliakos},
  {Michel}, {Middleton}, {Milano}, {Miller}, {Millhouse}, {Mills}, {Milotti},
  {Milovich-Goff}, {Minazzoli}, {Minenkov}, {Mishkin}, {Mishra}, {Mistry},
  {Mitra}, {Mitrofanov}, {Mitselmakher}, {Mittleman}, {Mo}, {Mogushi},
  {Mohapatra}, {Mohite}, {Molina-Ruiz}, {Mondin}, {Montani}, {Moore}, {Moraru},
  {Morawski}, {Moreno}, {Morisaki}, {Mours}, {Mow-Lowry}, {Mozzon},
  {Muciaccia}, {Mukherjee}, {Mukherjee}, {Mukherjee}, {Mukherjee}, {Mukund},
  {Mullavey}, {Munch}, {Mu{\~n}iz}, {Murray}, {Nagar}, {Nardecchia},
  {Naticchioni}, {Nayak}, {Neil}, {Neilson}, {Nelemans}, {Nelson}, {Nery},
  {Neunzert}, {Ng}, {Ng}, {Nguyen}, {Nguyen}, {Nichols}, {Nichols}, {Nissanke},
  {Nitz}, {Nocera}, {Noh}, {North}, {Nothard}, {Nuttall}, {Oberling},
  {O'Brien}, {Oganesyan}, {Ogin}, {Oh}, {Oh}, {Ohme}, {Ohta}, {Okada},
  {Oliver}, {Olivetto}, {Oppermann}, {Oram}, {O'Reilly}, {Ormiston}, {Ortega},
  {O'Shaughnessy}, {Ossokine}, {Osthelder}, {Ottaway}, {Overmier}, {Owen},
  {Pace}, {Pagano}, {Page}, {Pagliaroli}, {Pai}, {Pai}, {Palamos}, {Palashov},
  {Palomba}, {Pan}, {Panda}, {Pang}, {Pankow}, {Pannarale}, {Pant}, {Paoletti},
  {Paoli}, {Parida}, {Parker}, {Pascucci}, {Pasqualetti}, {Passaquieti},
  {Passuello}, {Patricelli}, {Payne}, {Pearlstone}, {Pechsiri}, {Pedersen},
  {Pedraza}, {Pele}, {Penn}, {Perego}, {Perez}, {P{\'e}rigois}, {Perreca},
  {Perri{\`e}s}, {Petermann}, {Pfeiffer}, {Phelps}, {Phukon}, {Piccinni},
  {Pichot}, {Piendibene}, {Piergiovanni}, {Pierro}, {Pillant}, {Pinard},
  {Pinto}, {Piotrzkowski}, {Pirello}, {Pitkin}, {Plastino}, {Poggiani}, {Pong},
  {Ponrathnam}, {Popolizio}, {Porter}, {Powell}, {Prajapati}, {Prasai},
  {Prasanna}, {Pratten}, {Prestegard}, {Principe}, {Prodi}, {Prokhorov},
  {Punturo}, {Puppo}, {P{\"u}rrer}, {Qi}, {Quetschke}, {Quinonez}, {Raab},
  {Raaijmakers}, {Radkins}, {Radulesco}, {Raffai}, {Rafferty}, {Raja}, {Rajan},
  {Rajbhandari}, {Rakhmanov}, {Ramirez}, {Ramos-Buades}, {Rana}, {Rao},
  {Rapagnani}, {Raymond}, {Razzano}, {Read}, {Regimbau}, {Rei}, {Reid},
  {Reitze}, {Rettegno}, {Ricci}, {Richardson}, {Richardson}, {Ricker},
  {Riemenschneider}, {Riles}, {Rizzo}, {Robertson}, {Robinet}, {Rocchi},
  {Rodriguez-Soto}, {Rolland}, {Rollins}, {Roma}, {Romanelli}, {Romano},
  {Romel}, {Romero-Shaw}, {Romie}, {Rose}, {Rose}, {Rose}, {Rosi{\'n}ska},
  {Rosofsky}, {Ross}, {Rowan}, {Rowlinson}, {Roy}, {Roy}, {Roy}, {Ruggi},
  {Rutins}, {Ryan}, {Sachdev}, {Sadecki}, {Sakellariadou}, {Salafia},
  {Salconi}, {Saleem}, {Salemi}, {Samajdar}, {Sanchez}, {Sanchez},
  {Sanchis-Gual}, {Sanders}, {Santiago}, {Santos}, {Sarin}, {Sassolas},
  {Sathyaprakash}, {Sauter}, {Savage}, {Savant}, {Sawant}, {Sayah}, {Schaetzl},
  {Schale}, {Scheel}, {Scheuer}, {Schmidt}, {Schnabel}, {Schofield},
  {Sch{\"o}nbeck}, {Schreiber}, {Schulte}, {Schutz}, {Schwarm}, {Schwartz},
  {Scott}, {Scott}, {Seidel}, {Sellers}, {Sengupta}, {Sennett}, {Sentenac},
  {Sequino}, {Sergeev}, {Setyawati}, {Shaddock}, {Shaffer}, {Sharifi},
  {Shahriar}, {Sharma}, {Sharma}, {Shawhan}, {Shen}, {Shikauchi}, {Shink},
  {Shoemaker}, {Shoemaker}, {Shukla}, {ShyamSundar}, {Siellez}, {Sieniawska},
  {Sigg}, {Singer}, {Singh}, {Singh}, {Singha}, {Singhal}, {Sintes}, {Sipala},
  {Skliris}, {Slagmolen}, {Slaven-Blair}, {Smetana}, {Smith}, {Smith},
  {Somala}, {Son}, {Soni}, {Sorazu}, {Sordini}, {Sorrentino}, {Souradeep},
  {Sowell}, {Spencer}, {Spera}, {Srivastava}, {Srivastava}, {Staats},
  {Stachie}, {Standke}, {Steer}, {Steinke}, {Steinlechner}, {Steinlechner},
  {Steinmeyer}, {Stevenson}, {Stocks}, {Stops}, {Stover}, {Strain}, {Stratta},
  {Strunk}, {Sturani}, {Stuver}, {Sudhagar}, {Sudhir}, {Summerscales}, {Sun},
  {Sunil}, {Sur}, {Suresh}, {Sutton}, {Swinkels}, {Szczepa{\'n}czyk}, {Tacca},
  {Tait}, {Talbot}, {Tanasijczuk}, {Tanner}, {Tao}, {T{\'a}pai}, {Tapia},
  {Tapia San Martin}, {Tasson}, {Taylor}, {Tenorio}, {Terkowski},
  {Thirugnanasambandam}, {Thomas}, {Thomas}, {Thompson}, {Thondapu}, {Thorne},
  {Thrane}, {Tinsman}, {Saravanan}, {Tiwari}, {Tiwari}, {Tiwari}, {Toland},
  {Tonelli}, {Tornasi}, {Torres-Forn{\'e}}, {Torrie}, {Tosta e Melo},
  {T{\"o}yr{\"a}}, {Travasso}, {Traylor}, {Tringali}, {Tripathee}, {Trovato},
  {Trudeau}, {Tsang}, {Tse}, {Tso}, {Tsukada}, {Tsuna}, {Tsutsui}, {Turconi},
  {Ubhi}, {Udall}, {Ueno}, {Ugolini}, {Unnikrishnan}, {Urban}, {Usman},
  {Utina}, {Vahlbruch}, {Vajente}, {Valdes}, {Valentini}, {van Bakel}, {van
  Beuzekom}, {van den Brand}, {Van Den Broeck}, {Vander-Hyde}, {van der
  Schaaf}, {Van Heijningen}, {van Veggel}, {Vardaro}, {Varma}, {Vass},
  {Vas{\'u}th}, {Vecchio}, {Vedovato}, {Veitch}, {Veitch}, {Venkateswara},
  {Venugopalan}, {Verkindt}, {Veske}, {Vetrano}, {Vicer{\'e}}, {Viets},
  {Vinciguerra}, {Vine}, {Vinet}, {Vitale}, {Vivanco}, {Vo}, {Vocca},
  {Vorvick}, {Vyatchanin}, {Wade}, {Wade}, {Wade}, {Walet}, {Walker},
  {Wallace}, {Wallace}, {Walsh}, {Wang}, {Wang}, {Wang}, {Ward}, {Warden},
  {Warner}, {Was}, {Watchi}, {Weaver}, {Wei}, {Weinert}, {Weinstein}, {Weiss},
  {Wellmann}, {Wen}, {We{\ss}els}, {Westhouse}, {Wette}, {Whelan}, {Whiting},
  {Whittle}, {Wilken}, {Williams}, {Willis}, {Willke}, {Winkler}, {Wipf},
  {Wittel}, {Woan}, {Woehler}, {Wofford}, {Wong}, {Wright}, {Wu}, {Wysocki},
  {Xiao}, {Yamamoto}, {Yang}, {Yang}, {Yang}, {Yap}, {Yazback}, {Yeeles}, {Yu},
  {Yu}, {Yuen}, {Zadro{\.Z}ny}, {Zadro{\.Z}ny}, {Zanolin}, {Zelenova},
  {Zendri}, {Zevin}, {Zhang}, {Zhang}, {Zhang}, {Zhao}, {Zhao}, {Zhou}, {Zhou},
  {Zhu}, {Zimmerman}, {Zucker}, {Zweizig}, {LIGO Scientific Collaboration}, \&
  {Virgo Collaboration}}]{2020PhRvL.125j1102A}
---. 2020{\natexlab{b}}, \prl, 125, 101102,
  \dodoi{10.1103/PhysRevLett.125.101102}

\bibitem[{{Acernese} {et~al.}(2015){Acernese}, {Agathos}, {Agatsuma}, {Aisa},
  {Allemandou}, {Allocca}, {Amarni}, {Astone}, {Balestri}, {Ballardin}, \&
  et~al.}]{2015CQGra..32b4001A}
{Acernese}, F., {Agathos}, M., {Agatsuma}, K., {et~al.} 2015, Classical and
  Quantum Gravity, 32, 024001, \dodoi{10.1088/0264-9381/32/2/024001}

\bibitem[{{Adams} {et~al.}(2017){Adams}, {Kochanek}, {Gerke}, {Stanek}, \&
  {Dai}}]{2017MNRAS.468.4968A}
{Adams}, S.~M., {Kochanek}, C.~S., {Gerke}, J.~R., {Stanek}, K.~Z., \& {Dai},
  X. 2017, \mnras, 468, 4968, \dodoi{10.1093/mnras/stx816}

\bibitem[{Anderson \& Darling(1954)}]{andersondarling}
Anderson, T.~W., \& Darling, D.~A. 1954, Journal of the American Statistical
  Association, 49, 765, \dodoi{10.1080/01621459.1954.10501232}

\bibitem[{{Atri} {et~al.}(2020){Atri}, {Miller-Jones}, {Bahramian}, {Plotkin},
  {Deller}, {Jonker}, {Maccarone}, {Sivakoff}, {Soria}, {Altamirano},
  {Belloni}, {Fender}, {Koerding}, {Maitra}, {Markoff}, {Migliari}, {Russell},
  {Russell}, {Sarazin}, {Tetarenko}, \& {Tudose}}]{atri2020}
{Atri}, P., {Miller-Jones}, J.~C.~A., {Bahramian}, A., {et~al.} 2020, \mnras,
  493, L81, \dodoi{10.1093/mnrasl/slaa010}

\bibitem[{{Bailyn} {et~al.}(1998){Bailyn}, {Jain}, {Coppi}, \&
  {Orosz}}]{1998ApJ...499..367B}
{Bailyn}, C.~D., {Jain}, R.~K., {Coppi}, P., \& {Orosz}, J.~A. 1998, \apj, 499,
  367, \dodoi{10.1086/305614}

\bibitem[{{Baumgardt} {et~al.}(2019){Baumgardt}, {Hilker}, {Sollima}, \&
  {Bellini}}]{2019MNRAS.482.5138B}
{Baumgardt}, H., {Hilker}, M., {Sollima}, A., \& {Bellini}, A. 2019, \mnras,
  482, 5138, \dodoi{10.1093/mnras/sty2997}

\bibitem[{{Belokurov} {et~al.}(2020){Belokurov}, {Penoyre}, {Oh}, {Iorio},
  {Hodgkin}, {Evans}, {Everall}, {Koposov}, {Tout}, {Izzard}, {Clarke}, \&
  {Brown}}]{2020MNRAS.496.1922B}
{Belokurov}, V., {Penoyre}, Z., {Oh}, S., {et~al.} 2020, \mnras, 496, 1922,
  \dodoi{10.1093/mnras/staa1522}

\bibitem[{{Blaauw}(1961)}]{1961BAN....15..265B}
{Blaauw}, A. 1961, \bain, 15, 265

\bibitem[{{Breivik} {et~al.}(2017){Breivik}, {Chatterjee}, \&
  {Larson}}]{2017ApJ...850L..13B}
{Breivik}, K., {Chatterjee}, S., \& {Larson}, S.~L. 2017, \apjl, 850, L13,
  \dodoi{10.3847/2041-8213/aa97d5}

\bibitem[{{Burrows} \& {Vartanyan}(2021)}]{2021Natur.589...29B}
{Burrows}, A., \& {Vartanyan}, D. 2021, \nat, 589, 29,
  \dodoi{10.1038/s41586-020-03059-w}

\bibitem[{{Casares}(2015)}]{2015ApJ...808...80C}
{Casares}, J. 2015, \apj, 808, 80, \dodoi{10.1088/0004-637X/808/1/80}

\bibitem[{{Casares}(2016)}]{2016ApJ...822...99C}
---. 2016, \apj, 822, 99, \dodoi{10.3847/0004-637X/822/2/99}

\bibitem[{{Casares}(2018)}]{2018MNRAS.473.5195C}
---. 2018, \mnras, 473, 5195, \dodoi{10.1093/mnras/stx2690}

\bibitem[{{Casares} \& {Jonker}(2014)}]{2014SSRv..183..223C}
{Casares}, J., \& {Jonker}, P.~G. 2014, \ssr, 183, 223,
  \dodoi{10.1007/s11214-013-0030-6}

\bibitem[{{Casares} {et~al.}(2014){Casares}, {Negueruela}, {Rib{\'o}}, {Ribas},
  {Paredes}, {Herrero}, \& {Sim{\'o}n-D{\'\i}az}}]{2014Natur.505..378C}
{Casares}, J., {Negueruela}, I., {Rib{\'o}}, M., {et~al.} 2014, \nat, 505, 378,
  \dodoi{10.1038/nature12916}

\bibitem[{{Casares} \& {Torres}(2018)}]{2018MNRAS.481.4372C}
{Casares}, J., \& {Torres}, M. A.~P. 2018, \mnras, 481, 4372,
  \dodoi{10.1093/mnras/sty2570}

\bibitem[{{Casares} {et~al.}(2009){Casares}, {Orosz}, {Zurita}, {Shahbaz},
  {Corral-Santana}, {McClintock}, {Garcia}, {Mart{\'{\i}}nez-Pais}, {Charles},
  {Fender}, \& {Remillard}}]{2009ApJS..181..238C}
{Casares}, J., {Orosz}, J.~A., {Zurita}, C., {et~al.} 2009, \apjs, 181, 238,
  \dodoi{10.1088/0067-0049/181/1/238}

\bibitem[{Chan {et~al.}(2018){Chan}, {Mueller}, {Heger}, {Pakmor}, \&
  {Springel}}]{Chan_2018}
{Chan}, C., {Mueller}, B., {Heger}, A., {Pakmor}, R., \& {Springel}, V. 2018, \apj, 
852, L19, \dodoi{10.3847/2041-8213/aaa28c}

\bibitem[{{Chaty} {et~al.}(2002){Chaty}, {Mirabel}, {Goldoni}, {Mereghetti},
  {Duc}, {Mart{\'\i}}, \& {Mignani}}]{2002MNRAS.331.1065C}
{Chaty}, S., {Mirabel}, I.~F., {Goldoni}, P., {et~al.} 2002, \mnras, 331, 1065,
  \dodoi{10.1046/j.1365-8711.2002.05267.x}

\bibitem[{{Chauhan} {et~al.}(2020){Chauhan}, {Miller-Jones}, {Raja}, {Allison},
  {Jacob}, {Anderson}, {Carotenuto}, {Corbel}, {Fender}, {Hotan}, {Whiting},
  {Woudt}, {Koribalski}, \& {Mahony}}]{2020MNRAS.tmpL.236C}
{Chauhan}, J., {Miller-Jones}, J.~C.~A., {Raja}, W., {et~al.} 2020, \mnras,
  \dodoi{10.1093/mnrasl/slaa195}

\bibitem[{{Chevalier}(1989)}]{1989ApJ...346..847C}
{Chevalier}, R.~A. 1989, \apj, 346, 847, \dodoi{10.1086/168066}

\bibitem[{{Chomiuk} {et~al.}(2013){Chomiuk}, {Strader}, {Maccarone},
  {Miller-Jones}, {Heinke}, {Noyola}, {Seth}, \&
  {Ransom}}]{2013ApJ...777...69C}
{Chomiuk}, L., {Strader}, J., {Maccarone}, T.~J., {et~al.} 2013, \apj, 777, 69,
  \dodoi{10.1088/0004-637X/777/1/69}

\bibitem[{{Chrimes} {et~al.}(2021){Chrimes}, {Levan}, {Groot}, {Lyman}, \&
  {Nelemans}}]{2021arXiv210504549C}
{Chrimes}, A.~A., {Levan}, A.~J., {Groot}, P.~J., {Lyman}, J.~D., \&
  {Nelemans}, G. 2021, arXiv e-prints, arXiv:2105.04549.
\newblock \doarXiv{2105.04549}

\bibitem[{{Corral-Santana} {et~al.}(2016){Corral-Santana}, {Casares},
  {Mu{\~n}oz-Darias}, {Bauer}, {Mart{\'\i}nez-Pais}, \& {Russell}}]{blackcat}
{Corral-Santana}, J.~M., {Casares}, J., {Mu{\~n}oz-Darias}, T., {et~al.} 2016,
  \aap, 587, A61, \dodoi{10.1051/0004-6361/201527130}

\bibitem[{{Corral-Santana} {et~al.}(2011){Corral-Santana}, {Casares},
  {Shahbaz}, {Zurita}, {Mart{\'{\i}}nez-Pais}, \&
  {Rodr{\'{\i}}guez-Gil}}]{2011MNRAS.413L..15C}
{Corral-Santana}, J.~M., {Casares}, J., {Shahbaz}, T., {et~al.} 2011, \mnras,
  413, L15, \dodoi{10.1111/j.1745-3933.2011.01022.x}

\bibitem[{{Debattista} {et~al.}(2017){Debattista}, {Ness}, {Gonzalez},
  {Freeman}, {Zoccali}, \& {Minniti}}]{2017MNRAS.469.1587D}
{Debattista}, V.~P., {Ness}, M., {Gonzalez}, O.~A., {et~al.} 2017, \mnras, 469,
  1587, \dodoi{10.1093/mnras/stx947}

\bibitem[{{Deegan} {et~al.}(2009){Deegan}, {Combet}, \&
  {Wynn}}]{2009MNRAS.400.1337D}
{Deegan}, P., {Combet}, C., \& {Wynn}, G.~A. 2009, \mnras, 400, 1337,
  \dodoi{10.1111/j.1365-2966.2009.15573.x}

\bibitem[{{Dubus} {et~al.}(2001){Dubus}, {Hameury}, \&
  {Lasota}}]{2001A&A...373..251D}
{Dubus}, G., {Hameury}, J.~M., \& {Lasota}, J.~P. 2001, \aap, 373, 251,
  \dodoi{10.1051/0004-6361:20010632}

\bibitem[{{Eggleton} \& {Verbunt}(1986)}]{1986MNRAS.220P..13E}
{Eggleton}, P.~P., \& {Verbunt}, F. 1986, \mnras, 220, 13P,
  \dodoi{10.1093/mnras/220.1.13P}

\bibitem[{{Farr} {et~al.}(2011){Farr}, {Sravan}, {Cantrell}, {Kreidberg},
  {Bailyn}, {Mandel}, \& {Kalogera}}]{2011ApJ...741..103F}
{Farr}, W.~M., {Sravan}, N., {Cantrell}, A., {et~al.} 2011, \apj, 741, 103,
  \dodoi{10.1088/0004-637X/741/2/103}

\bibitem[{{Filippenko} {et~al.}(1999){Filippenko}, {Leonard}, {Matheson}, {Li},
  {Moran}, \& {Riess}}]{1999PASP..111..969F}
{Filippenko}, A.~V., {Leonard}, D.~C., {Matheson}, T., {et~al.} 1999, \pasp,
  111, 969, \dodoi{10.1086/316413}

\bibitem[{{Frontera} {et~al.}(2001){Frontera}, {Zdziarski}, {Amati},
  {Miko{\l}ajewska}, {Belloni}, {Del Sordo}, {Haardt}, {Kuulkers}, {Masetti},
  {Orlandini}, {Palazzi}, {Parmar}, {Remillard}, {Santangelo}, \&
  {Stella}}]{2001ApJ...561.1006F}
{Frontera}, F., {Zdziarski}, A.~A., {Amati}, L., {et~al.} 2001, \apj, 561,
  1006, \dodoi{10.1086/323258}

\bibitem[{{Fryer}(1999)}]{Fryer99}
{Fryer}, C.~L. 1999, \apj, 522, 413, \dodoi{10.1086/307647}

\bibitem[{{Fryer} {et~al.}(2012){Fryer}, {Belczynski}, {Wiktorowicz},
  {Dominik}, {Kalogera}, \& {Holz}}]{2012ApJ...749...91F}
{Fryer}, C.~L., {Belczynski}, K., {Wiktorowicz}, G., {et~al.} 2012, \apj, 749,
  91, \dodoi{10.1088/0004-637X/749/1/91}

\bibitem[{{Fryer} \& {Kalogera}(2001)}]{2001ApJ...554..548F}
{Fryer}, C.~L., \& {Kalogera}, V. 2001, \apj, 554, 548, \dodoi{10.1086/321359}

\bibitem[{{Gandhi} {et~al.}(2020{\natexlab{a}}){Gandhi}, {Buckley}, {Charles},
  {Hodgkin}, {Scaringi}, {Knigge}, \& {Rao}}]{2020arXiv200907277G}
{Gandhi}, P., {Buckley}, D. A.~H., {Charles}, P., {et~al.} 2020{\natexlab{a}},
  arXiv e-prints, arXiv:2009.07277.
\newblock \doarXiv{2009.07277}

\bibitem[{{Gandhi} {et~al.}(2020{\natexlab{b}}){Gandhi}, {Rao}, {Charles},
  {Belczynski}, {Maccarone}, {Arur}, \& {Corral-Santana}}]{poshak2020}
{Gandhi}, P., {Rao}, A., {Charles}, P.~A., {et~al.} 2020{\natexlab{b}}, \mnras,
  496, L22, \dodoi{10.1093/mnrasl/slaa081}

\bibitem[{{Gandhi} {et~al.}(2019){Gandhi}, {Rao}, {Johnson}, {Paice}, \&
  {Maccarone}}]{poshak2019}
{Gandhi}, P., {Rao}, A., {Johnson}, M. A.~C., {Paice}, J.~A., \& {Maccarone},
  T.~J. 2019, \mnras, 485, 2642, \dodoi{10.1093/mnras/stz438}

\bibitem[{{Garcia} {et~al.}(2001){Garcia}, {McClintock}, {Narayan}, {Callanan},
  {Barret}, \& {Murray}}]{2001ApJ...553L..47G}
{Garcia}, M.~R., {McClintock}, J.~E., {Narayan}, R., {et~al.} 2001, \apjl, 553,
  L47, \dodoi{10.1086/320494}

\bibitem[{{Gelino} \& {Harrison}(2003)}]{2003ApJ...599.1254G}
{Gelino}, D.~M., \& {Harrison}, T.~E. 2003, \apj, 599, 1254,
  \dodoi{10.1086/379311}

\bibitem[{{Giesers} {et~al.}(2018){Giesers}, {Dreizler}, {Husser}, {Kamann},
  {Anglada Escud{\'e}}, {Brinchmann}, {Carollo}, {Roth}, {Weilbacher}, \&
  {Wisotzki}}]{2018MNRAS.475L..15G}
{Giesers}, B., {Dreizler}, S., {Husser}, T.-O., {et~al.} 2018, \mnras, 475,
  L15, \dodoi{10.1093/mnrasl/slx203}

\bibitem[{{Giesers} {et~al.}(2019){Giesers}, {Kamann}, {Dreizler}, {Husser},
  {Askar}, {G{\"o}ttgens}, {Brinchmann}, {Latour}, {Weilbacher}, {Wendt}, \&
  {Roth}}]{2019A&A...632A...3G}
{Giesers}, B., {Kamann}, S., {Dreizler}, S., {et~al.} 2019, \aap, 632, A3,
  \dodoi{10.1051/0004-6361/201936203}

\bibitem[{{Giesler} {et~al.}(2018){Giesler}, {Clausen}, \&
  {Ott}}]{2018MNRAS.477.1853G}
{Giesler}, M., {Clausen}, D., \& {Ott}, C.~D. 2018, \mnras, 477, 1853,
  \dodoi{10.1093/mnras/sty659}

\bibitem[{{Gonz{\'a}lez Hern{\'a}ndez} {et~al.}(2006){Gonz{\'a}lez
  Hern{\'a}ndez}, {Rebolo}, {Israelian}, {Harlaftis}, {Filippenko}, \&
  {Chornock}}]{2006ApJ...644L..49G}
{Gonz{\'a}lez Hern{\'a}ndez}, J.~I., {Rebolo}, R., {Israelian}, G., {et~al.}
  2006, \apjl, 644, L49, \dodoi{10.1086/505391}

\bibitem[{{Gould} \& {Salim}(2002)}]{2002ApJ...572..944G}
{Gould}, A., \& {Salim}, S. 2002, \apj, 572, 944, \dodoi{10.1086/340435}

\bibitem[{{Harris}(1996)}]{1996AJ....112.1487H}
{Harris}, W.~E. 1996, \aj, 112, 1487, \dodoi{10.1086/118116}

\bibitem[{{Heida} {et~al.}(2017){Heida}, {Jonker}, {Torres}, \&
  {Chiavassa}}]{2017ApJ...846..132H}
{Heida}, M., {Jonker}, P.~G., {Torres}, M.~A.~P., \& {Chiavassa}, A. 2017,
  \apj, 846, 132, \dodoi{10.3847/1538-4357/aa85df}

\bibitem[{{Hjellming} \& {Rupen}(1995)}]{1995Natur.375..464H}
{Hjellming}, R.~M., \& {Rupen}, M.~P. 1995, \nat, 375, 464,
  \dodoi{10.1038/375464a0}

\bibitem[{{Homan} {et~al.}(2013){Homan}, {Fridriksson}, {Jonker}, {Russell},
  {Gallo}, {Kuulkers}, {Rea}, \& {Altamirano}}]{2013ApJ...775....9H}
{Homan}, J., {Fridriksson}, J.~K., {Jonker}, P.~G., {et~al.} 2013, \apj, 775,
  9, \dodoi{10.1088/0004-637X/775/1/9}

\bibitem[{{Homan} {et~al.}(2006){Homan}, {Wijnands}, {Kong}, {Miller}, {Rossi},
  {Belloni}, \& {Lewin}}]{2006MNRAS.366..235H}
{Homan}, J., {Wijnands}, R., {Kong}, A., {et~al.} 2006, \mnras, 366, 235,
  \dodoi{10.1111/j.1365-2966.2005.09843.x}

\bibitem[{{Hynes} {et~al.}(2002){Hynes}, {Haswell}, {Chaty}, {Shrader}, \&
  {Cui}}]{2002MNRAS.331..169H}
{Hynes}, R.~I., {Haswell}, C.~A., {Chaty}, S., {Shrader}, C.~R., \& {Cui}, W.
  2002, \mnras, 331, 169, \dodoi{10.1046/j.1365-8711.2002.05175.x}

\bibitem[{{Igoshev} \& {Perets}(2019)}]{2019MNRAS.486.4098I}
{Igoshev}, A.~P., \& {Perets}, H.~B. 2019, \mnras, 486, 4098,
  \dodoi{10.1093/mnras/stz1024}

\bibitem[{{in't Zand} {et~al.}(2004){in't Zand}, {Verbunt}, {Heise}, {Bazzano},
  {Cocchi}, {Cornelisse}, {Kuulkers}, {Natalucci}, \&
  {Ubertini}}]{2004NuPhS.132..486I}
{in't Zand}, J., {Verbunt}, F., {Heise}, J., {et~al.} 2004, Nuclear Physics B
  Proceedings Supplements, 132, 486, \dodoi{10.1016/j.nuclphysbps.2004.04.083}

\bibitem[{{Janka}(2013)}]{2013MNRAS.434.1355J}
{Janka}, H.-T. 2013, \mnras, 434, 1355, \dodoi{10.1093/mnras/stt1106}

\bibitem[{{Jonker} {et~al.}(2012){Jonker}, {Miller-Jones}, {Homan}, {Tomsick},
  {Fender}, {Kaaret}, {Markoff}, \& {Gallo}}]{jonker2012}
{Jonker}, P.~G., {Miller-Jones}, J.~C.~A., {Homan}, J., {et~al.} 2012, \mnras,
  423, 3308, \dodoi{10.1111/j.1365-2966.2012.21116.x}

\bibitem[{{Jonker} \& {Nelemans}(2004)}]{2004MNRAS.354..355J}
{Jonker}, P.~G., \& {Nelemans}, G. 2004, \mnras, 354, 355,
  \dodoi{10.1111/j.1365-2966.2004.08193.x}

\bibitem[{{Jonker} {et~al.}(2011){Jonker}, {Bassa}, {Nelemans}, {Steeghs},
  {Torres}, {Maccarone}, {Hynes}, {Greiss}, {Clem}, {Dieball}, {Mikles},
  {Britt}, {Gossen}, {Collazzi}, {Wijnands}, {In't Zand}, {M{\'e}ndez}, {Rea},
  {Kuulkers}, {Ratti}, {van Haaften}, {Heinke}, {{\"O}zel}, {Groot}, \&
  {Verbunt}}]{2011ApJS..194...18J}
{Jonker}, P.~G., {Bassa}, C.~G., {Nelemans}, G., {et~al.} 2011, \apjs, 194, 18,
  \dodoi{10.1088/0067-0049/194/2/18}

\bibitem[{{Jonker} {et~al.}(2014){Jonker}, {Torres}, {Hynes}, {Maccarone},
  {Steeghs}, {Greiss}, {Britt}, {Wu}, {Johnson}, {Nelemans}, \&
  {Heinke}}]{2014ApJS..210...18J}
{Jonker}, P.~G., {Torres}, M. A.~P., {Hynes}, R.~I., {et~al.} 2014, \apjs, 210,
  18, \dodoi{10.1088/0067-0049/210/2/18}

\bibitem[{{Khargharia} {et~al.}(2010){Khargharia}, {Froning}, \&
  {Robinson}}]{2010ApJ...716.1105K}
{Khargharia}, J., {Froning}, C.~S., \& {Robinson}, E.~L. 2010, \apj, 716, 1105,
  \dodoi{10.1088/0004-637X/716/2/1105}

\bibitem[{{Khargharia} {et~al.}(2013){Khargharia}, {Froning}, {Robinson}, \&
  {Gelino}}]{2013AJ....145...21K}
{Khargharia}, J., {Froning}, C.~S., {Robinson}, E.~L., \& {Gelino}, D.~M. 2013,
  \aj, 145, 21, \dodoi{10.1088/0004-6256/145/1/21}

\bibitem[{{King} {et~al.}(1997){King}, {Frank}, {Kolb}, \&
  {Ritter}}]{1997ApJ...484..844K}
{King}, A.~R., {Frank}, J., {Kolb}, U., \& {Ritter}, H. 1997, \apj, 484, 844,
  \dodoi{10.1086/304383}

\bibitem[{{Kreidberg} {et~al.}(2012){Kreidberg}, {Bailyn}, {Farr}, \&
  {Kalogera}}]{2012ApJ...757...36K}
{Kreidberg}, L., {Bailyn}, C.~D., {Farr}, W.~M., \& {Kalogera}, V. 2012, \apj,
  757, 36, \dodoi{10.1088/0004-637X/757/1/36}

\bibitem[{{Kremer} {et~al.}(2018){Kremer}, {Chatterjee}, {Rodriguez}, \&
  {Rasio}}]{2018ApJ...852...29K}
{Kremer}, K., {Chatterjee}, S., {Rodriguez}, C.~L., \& {Rasio}, F.~A. 2018,
  \apj, 852, 29, \dodoi{10.3847/1538-4357/aa99df}

\bibitem[{Krimm {et~al.}(2013)Krimm, Holland, Corbet, Pearlman, Romano, Kennea,
  Bloom, Barthelmy, Baumgartner, Cummings, Gehrels, Lien, Markwardt, Palmer,
  Sakamoto, Stamatikos, \& Ukwatta}]{Krimm_2013}
Krimm, H.~A., Holland, S.~T., Corbet, R. H.~D., {et~al.} 2013, The
  Astrophysical Journal Supplement Series, 209, 14,
  \dodoi{10.1088/0067-0049/209/1/14}

\bibitem[{{Kulkarni} {et~al.}(1993){Kulkarni}, {Hut}, \&
  {McMillan}}]{1993Natur.364..421K}
{Kulkarni}, S.~R., {Hut}, P., \& {McMillan}, S. 1993, \nat, 364, 421,
  \dodoi{10.1038/364421a0}

\bibitem[{{Kuulkers} {et~al.}(2007){Kuulkers}, {Shaw}, {Paizis}, {Chenevez},
  {Brandt}, {Courvoisier}, {Domingo}, {Ebisawa}, {Kretschmar}, {Markwardt},
  {Mowlavi}, {Oosterbroek}, {Orr}, {R{\'\i}squez}, {Sanchez-Fernandez}, \&
  {Wijnands}}]{2007A&A...466..595K}
{Kuulkers}, E., {Shaw}, S.~E., {Paizis}, A., {et~al.} 2007, \aap, 466, 595,
  \dodoi{10.1051/0004-6361:20066651}

\bibitem[{{Kuulkers} {et~al.}(2013){Kuulkers}, {Kouveliotou}, {Belloni},
  {Cadolle Bel}, {Chenevez}, {D{\'\i}az Trigo}, {Homan}, {Ibarra}, {Kennea},
  {Mu{\~n}oz-Darias}, {Ness}, {Parmar}, {Pollock}, {van den Heuvel}, \& {van
  der Horst}}]{2013A&A...552A..32K}
{Kuulkers}, E., {Kouveliotou}, C., {Belloni}, T., {et~al.} 2013, \aap, 552,
  A32, \dodoi{10.1051/0004-6361/201219447}

\bibitem[{{Levine} {et~al.}(1996){Levine}, {Bradt}, {Cui}, {Jernigan},
  {Morgan}, {Remillard}, {Shirey}, \& {Smith}}]{1996ApJ...469L..33L}
{Levine}, A.~M., {Bradt}, H., {Cui}, W., {et~al.} 1996, \apjl, 469, L33,
  \dodoi{10.1086/310260}

\bibitem[{{L{\'o}pez} {et~al.}(2019){L{\'o}pez}, {Jonker}, {Torres}, {Heida},
  {Rau}, \& {Steeghs}}]{2019MNRAS.482.2149L}
{L{\'o}pez}, K.~M., {Jonker}, P.~G., {Torres}, M.~A.~P., {et~al.} 2019, \mnras,
  482, 2149, \dodoi{10.1093/mnras/sty2793}

\bibitem[{{Maccarone}(2003)}]{2003A&A...409..697M}
{Maccarone}, T.~J. 2003, \aap, 409, 697, \dodoi{10.1051/0004-6361:20031146}

\bibitem[{{Maccarone} {et~al.}(2007){Maccarone}, {Kundu}, {Zepf}, \&
  {Rhode}}]{2007Natur.445..183M}
{Maccarone}, T.~J., {Kundu}, A., {Zepf}, S.~E., \& {Rhode}, K.~L. 2007, \nat,
  445, 183, \dodoi{10.1038/nature05434}

\bibitem[{{MacDonald} {et~al.}(2014){MacDonald}, {Bailyn}, {Buxton},
  {Cantrell}, {Chatterjee}, {Kennedy-Shaffer}, {Orosz}, {Markwardt}, \&
  {Swank}}]{2014ApJ...784....2M}
{MacDonald}, R.~K.~D., {Bailyn}, C.~D., {Buxton}, M., {et~al.} 2014, \apj, 784,
  2, \dodoi{10.1088/0004-637X/784/1/2}

\bibitem[{Mann \& Whitney(1947)}]{mann1947}
Mann, H.~B., \& Whitney, D.~R. 1947, Ann. Math. Statist., 18, 50,
  \dodoi{10.1214/aoms/1177730491}

\bibitem[{{Mata S{\'a}nchez} {et~al.}(2015){Mata S{\'a}nchez},
  {Mu{\~n}oz-Darias}, {Casares}, {Corral-Santana}, \&
  {Shahbaz}}]{2015MNRAS.454.2199M}
{Mata S{\'a}nchez}, D., {Mu{\~n}oz-Darias}, T., {Casares}, J.,
  {Corral-Santana}, J.~M., \& {Shahbaz}, T. 2015, \mnras, 454, 2199,
  \dodoi{10.1093/mnras/stv2111}

\bibitem[{{Mata S{\'a}nchez} {et~al.}(2021){Mata S{\'a}nchez}, {Rau},
  {{\'A}lvarez Hern{\'a}ndez}, {van Grunsven}, {Torres}, \&
  {Jonker}}]{2021MNRAS.tmp.1510M}
{Mata S{\'a}nchez}, D., {Rau}, A., {{\'A}lvarez Hern{\'a}ndez}, A., {et~al.}
  2021, \mnras, \dodoi{10.1093/mnras/stab1714}

\bibitem[{Matsuoka {et~al.}(2009)Matsuoka, Kawasaki, Ueno, Tomida, Kohama,
  Suzuki, Adachi, Ishikawa, Mihara, Sugizaki, Isobe, Nakagawa, Tsunemi, Miyata,
  Kawai, Kataoka, Morii, Yoshida, Negoro, Nakajima, Ueda, Chujo, Yamaoka,
  Yamazaki, Nakahira, You, Ishiwata, Miyoshi, Eguchi, Hiroi, Katayama, \&
  Ebisawa}]{10.1093/pasj/61.5.999}
Matsuoka, M., Kawasaki, K., Ueno, S., {et~al.} 2009, Publications of the
  Astronomical Society of Japan, 61, 999, \dodoi{10.1093/pasj/61.5.999}

\bibitem[{{McClintock} \& {Remillard}(2006)}]{2006csxs.book..157M}
{McClintock}, J.~E., \& {Remillard}, R.~A. 2006, {Black hole binaries},
  Vol.~39, 157--213

\bibitem[{{Meyer-Hofmeister}(2004)}]{2004A&A...423..321M}
{Meyer-Hofmeister}, E. 2004, \aap, 423, 321, \dodoi{10.1051/0004-6361:20040369}

\bibitem[{{Mignard}(2000)}]{2000A&A...354..522M}
{Mignard}, F. 2000, \aap, 354, 522

\bibitem[{{Miller-Jones} {et~al.}(2009{\natexlab{a}}){Miller-Jones}, {Jonker},
  {Dhawan}, {Brisken}, {Rupen}, {Nelemans}, \& {Gallo}}]{2009ApJ...706L.230M}
{Miller-Jones}, J.~C.~A., {Jonker}, P.~G., {Dhawan}, V., {et~al.}
  2009{\natexlab{a}}, \apjl, 706, L230, \dodoi{10.1088/0004-637X/706/2/L230}

\bibitem[{{Miller-Jones} {et~al.}(2009{\natexlab{b}}){Miller-Jones}, {Jonker},
  {Nelemans}, {Portegies Zwart}, {Dhawan}, {Brisken}, {Gallo}, \&
  {Rupen}}]{2009MNRAS.394.1440M}
{Miller-Jones}, J.~C.~A., {Jonker}, P.~G., {Nelemans}, G., {et~al.}
  2009{\natexlab{b}}, \mnras, 394, 1440,
  \dodoi{10.1111/j.1365-2966.2009.14404.x}

\bibitem[{{Miller-Jones} {et~al.}(2015){Miller-Jones}, {Strader}, {Heinke},
  {Maccarone}, {van den Berg}, {Knigge}, {Chomiuk}, {Noyola}, {Russell},
  {Seth}, \& {Sivakoff}}]{2015MNRAS.453.3918M}
{Miller-Jones}, J.~C.~A., {Strader}, J., {Heinke}, C.~O., {et~al.} 2015,
  \mnras, 453, 3918, \dodoi{10.1093/mnras/stv1869}

\bibitem[{{Miller-Jones} {et~al.}(2021){Miller-Jones}, {Bahramian}, {Orosz},
  {Mandel}, {Gou}, {Maccarone}, {Neijssel}, {Zhao}, {Zi{\'o}{\l}kowski},
  {Reid}, {Uttley}, {Zheng}, {Byun}, {Dodson}, {Grinberg}, {Jung}, {Kim},
  {Marcote}, {Markoff}, {Rioja}, {Rushton}, {Russell}, {Sivakoff}, {Tetarenko},
  {Tudose}, \& {Wilms}}]{2021arXiv210209091M}
{Miller-Jones}, J. C.~A., {Bahramian}, A., {Orosz}, J.~A., {et~al.} 2021, arXiv
  e-prints, arXiv:2102.09091.
\newblock \doarXiv{2102.09091}

\bibitem[{{Mirabel} \& {Rodrigues}(2003)}]{2003Sci...300.1119M}
{Mirabel}, I.~F., \& {Rodrigues}, I. 2003, Science, 300, 1119,
  \dodoi{10.1126/science.1083451}

\bibitem[{{Moe} \& {Di Stefano}(2017)}]{MoeDiStefano17}
{Moe}, M., \& {Di Stefano}, R. 2017, \apjs, 230, 15,
  \dodoi{10.3847/1538-4365/aa6fb6}

\bibitem[{{Moon} {et~al.}(2009){Moon}, {Koo}, {Lee}, {Matthews}, {Lee}, {Pyo},
  {Seok}, \& {Hayashi}}]{2009ApJ...703L..81M}
{Moon}, D.-S., {Koo}, B.-C., {Lee}, H.-G., {et~al.} 2009, \apjl, 703, L81,
  \dodoi{10.1088/0004-637X/703/1/L81}

\bibitem[{{Mr{\'o}z} {et~al.}(2019){Mr{\'o}z}, {Udalski}, {Skowron},
  {Szyma{\'n}ski}, {Soszy{\'n}ski}, {Wyrzykowski}, {Pietrukowicz},
  {Koz{\l}owski}, {Poleski}, {Ulaczyk}, {Rybicki}, \&
  {Iwanek}}]{2019ApJS..244...29M}
{Mr{\'o}z}, P., {Udalski}, A., {Skowron}, J., {et~al.} 2019, \apjs, 244, 29,
  \dodoi{10.3847/1538-4365/ab426b}

\bibitem[{{Orosz} {et~al.}(1998){Orosz}, {Jain}, {Bailyn}, {McClintock}, \&
  {Remillard}}]{1998ApJ...499..375O}
{Orosz}, J.~A., {Jain}, R.~K., {Bailyn}, C.~D., {McClintock}, J.~E., \&
  {Remillard}, R.~A. 1998, \apj, 499, 375, \dodoi{10.1086/305620}

\bibitem[{{Orosz} {et~al.}(2004){Orosz}, {McClintock}, {Remillard}, \&
  {Corbel}}]{2004ApJ...616..376O}
{Orosz}, J.~A., {McClintock}, J.~E., {Remillard}, R.~A., \& {Corbel}, S. 2004,
  \apj, 616, 376, \dodoi{10.1086/424892}

\bibitem[{{Orosz} {et~al.}(2011){Orosz}, {Steiner}, {McClintock}, {Torres},
  {Remillard}, {Bailyn}, \& {Miller}}]{2011ApJ...730...75O}
{Orosz}, J.~A., {Steiner}, J.~F., {McClintock}, J.~E., {et~al.} 2011, \apj,
  730, 75, \dodoi{10.1088/0004-637X/730/2/75}

\bibitem[{{{\"O}zel} {et~al.}(2010){{\"O}zel}, {Psaltis}, {Narayan}, \&
  {McClintock}}]{2010ApJ...725.1918O}
{{\"O}zel}, F., {Psaltis}, D., {Narayan}, R., \& {McClintock}, J.~E. 2010,
  \apj, 725, 1918, \dodoi{10.1088/0004-637X/725/2/1918}

\bibitem[{{Perna} {et~al.}(2019){Perna}, {Wang}, {Farr}, {Leigh}, \&
  {Cantiello}}]{2019ApJ...878L...1P}
{Perna}, R., {Wang}, Y.-H., {Farr}, W.~M., {Leigh}, N., \& {Cantiello}, M.
  2019, \apjl, 878, L1, \dodoi{10.3847/2041-8213/ab2336}

\bibitem[{{Plotkin} {et~al.}(2013){Plotkin}, {Gallo}, \&
  {Jonker}}]{2013ApJ...773...59P}
{Plotkin}, R.~M., {Gallo}, E., \& {Jonker}, P.~G. 2013, \apj, 773, 59,
  \dodoi{10.1088/0004-637X/773/1/59}

\bibitem[{{Portail} {et~al.}(2015){Portail}, {Wegg}, {Gerhard}, \&
  {Martinez-Valpuesta}}]{2015MNRAS.448..713P}
{Portail}, M., {Wegg}, C., {Gerhard}, O., \& {Martinez-Valpuesta}, I. 2015,
  \mnras, 448, 713, \dodoi{10.1093/mnras/stv058}

\bibitem[{{Reid} {et~al.}(2011){Reid}, {McClintock}, {Narayan}, {Gou},
  {Remillard}, \& {Orosz}}]{2011ApJ...742...83R}
{Reid}, M.~J., {McClintock}, J.~E., {Narayan}, R., {et~al.} 2011, \apj, 742,
  83, \dodoi{10.1088/0004-637X/742/2/83}

\bibitem[{{Reid} {et~al.}(2014){Reid}, {McClintock}, {Steiner}, {Steeghs},
  {Remillard}, {Dhawan}, \& {Narayan}}]{2014ApJ...796....2R}
{Reid}, M.~J., {McClintock}, J.~E., {Steiner}, J.~F., {et~al.} 2014, \apj, 796,
  2, \dodoi{10.1088/0004-637X/796/1/2}

\bibitem[{{Reid} {et~al.}(2019){Reid}, {Menten}, {Brunthaler}, {Zheng}, {Dame},
  {Xu}, {Li}, {Sakai}, {Wu}, {Immer}, {Zhang}, {Sanna}, {Moscadelli}, {Rygl},
  {Bartkiewicz}, {Hu}, {Quiroga-Nu{\~n}ez}, \& {van Langevelde}}]{reid2019}
{Reid}, M.~J., {Menten}, K.~M., {Brunthaler}, A., {et~al.} 2019, \apj, 885,
  131, \dodoi{10.3847/1538-4357/ab4a11}

\bibitem[{{Reynolds} {et~al.}(2015){Reynolds}, {Fraser}, \&
  {Gilmore}}]{2015MNRAS.453.2885R}
{Reynolds}, T.~M., {Fraser}, M., \& {Gilmore}, G. 2015, \mnras, 453, 2885,
  \dodoi{10.1093/mnras/stv1809}

\bibitem[{{Sana} {et~al.}(2012){Sana}, {de Mink}, {de Koter}, {Langer},
  {Evans}, {Gieles}, {Gosset}, {Izzard}, {Le Bouquin}, \&
  {Schneider}}]{2012Sci...337..444S}
{Sana}, H., {de Mink}, S.~E., {de Koter}, A., {et~al.} 2012, Science, 337, 444,
  \dodoi{10.1126/science.1223344}

\bibitem[{{Schneider} {et~al.}(2021){Schneider}, {Podsiadlowski}, \&
  {M{\"u}ller}}]{2021A&A...645A...5S}
{Schneider}, F.~R.~N., {Podsiadlowski}, P., \& {M{\"u}ller}, B. 2021, \aap,
  645, A5, \dodoi{10.1051/0004-6361/202039219}

\bibitem[{{Shahaf} {et~al.}(2019){Shahaf}, {Mazeh}, {Faigler}, \&
  {Holl}}]{2019MNRAS.487.5610S}
{Shahaf}, S., {Mazeh}, T., {Faigler}, S., \& {Holl}, B. 2019, \mnras, 487,
  5610, \dodoi{10.1093/mnras/stz1636}

\bibitem[{{Shahbaz}(2003)}]{2003MNRAS.339.1031S}
{Shahbaz}, T. 2003, \mnras, 339, 1031, \dodoi{10.1046/j.1365-8711.2003.06258.x}

\bibitem[{{Shen} \& {Zheng}(2020)}]{2020RAA....20..159S}
{Shen}, J., \& {Zheng}, X.-W. 2020, Research in Astronomy and Astrophysics, 20,
  159, \dodoi{10.1088/1674-4527/20/10/159}

\bibitem[{{Shikauchi} {et~al.}(2020){Shikauchi}, {Kumamoto}, {Tanikawa}, \&
  {Fujii}}]{2020PASJ...72...45S}
{Shikauchi}, M., {Kumamoto}, J., {Tanikawa}, A., \& {Fujii}, M.~S. 2020, \pasj,
  72, 45, \dodoi{10.1093/pasj/psaa030}

\bibitem[{{Shishkovsky} {et~al.}(2018){Shishkovsky}, {Strader}, {Chomiuk},
  {Bahramian}, {Tremou}, {Li}, {Salinas}, {Tudor}, {Miller-Jones}, {Maccarone},
  {Heinke}, \& {Sivakoff}}]{2018ApJ...855...55S}
{Shishkovsky}, L., {Strader}, J., {Chomiuk}, L., {et~al.} 2018, \apj, 855, 55,
  \dodoi{10.3847/1538-4357/aaadb1}

\bibitem[{{Simionescu} {et~al.}(2019){Simionescu}, {Nakashima}, {Yamaguchi},
  {Matsushita}, {Mernier}, {Werner}, {Tamura}, {Nomoto}, {de Plaa}, {Leung},
  {Bamba}, {Bulbul}, {Eckart}, {Ezoe}, {Fabian}, {Fukazawa}, {Gu}, {Ichinohe},
  {Ishigaki}, {Kaastra}, {Kilbourne}, {Kitayama}, {Leutenegger}, {Loewenstein},
  {Maeda}, {Miller}, {Mushotzky}, {Noda}, {Pinto}, {Porter}, {Safi-Harb},
  {Sato}, {Takahashi}, {Ueda}, \& {Zha}}]{2019MNRAS.483.1701S}
{Simionescu}, A., {Nakashima}, S., {Yamaguchi}, H., {et~al.} 2019, \mnras, 483,
  1701, \dodoi{10.1093/mnras/sty3220}

\bibitem[{Smirnov(1948)}]{smirnov1948}
Smirnov, N. 1948, Ann. Math. Statist., 19, 279, \dodoi{10.1214/aoms/1177730256}

\bibitem[{{Steeghs} {et~al.}(2013){Steeghs}, {McClintock}, {Parsons}, {Reid},
  {Littlefair}, \& {Dhillon}}]{2013ApJ...768..185S}
{Steeghs}, D., {McClintock}, J.~E., {Parsons}, S.~G., {et~al.} 2013, \apj, 768,
  185, \dodoi{10.1088/0004-637X/768/2/185}

\bibitem[{{Strader} {et~al.}(2012){Strader}, {Chomiuk}, {Maccarone},
  {Miller-Jones}, \& {Seth}}]{2012Natur.490...71S}
{Strader}, J., {Chomiuk}, L., {Maccarone}, T.~J., {Miller-Jones}, J. C.~A., \&
  {Seth}, A.~C. 2012, \nat, 490, 71, \dodoi{10.1038/nature11490}

\bibitem[{{Tauris} \& {van den Heuvel}(2006)}]{2006csxs.book..623T}
{Tauris}, T.~M., \& {van den Heuvel}, E.~P.~J. 2006, {Formation and evolution
  of compact stellar X-ray sources}, Vol.~39, 623--665

\bibitem[{{Torres} {et~al.}(2020{\natexlab{a}}){Torres}, {Casares},
  {Jim{\'e}nez-Ibarra}, {{\'A}lvarez-Hern{\'a}ndez}, {Mu{\~n}oz-Darias}, {Armas
  Padilla}, {Jonker}, \& {Heida}}]{torres-1820-2020}
{Torres}, M.~A.~P., {Casares}, J., {Jim{\'e}nez-Ibarra}, F., {et~al.}
  2020{\natexlab{a}}, \apjl, 893, L37, \dodoi{10.3847/2041-8213/ab863a}

\bibitem[{{Torres} {et~al.}(2019){Torres}, {Casares}, {Jim{\'e}nez-Ibarra},
  {Mu{\~n}oz-Darias}, {Armas Padilla}, {Jonker}, \& {Heida}}]{torres-1820-2019}
---. 2019, \apjl, 882, L21, \dodoi{10.3847/2041-8213/ab39df}

\bibitem[{{Torres} {et~al.}(2020{\natexlab{b}}){Torres}, {Jonker}, {Casares},
  {Miller-Jones}, \& {Steeghs}}]{torres-1659-2020}
{Torres}, M.~A.~P., {Jonker}, P.~G., {Casares}, J., {Miller-Jones}, J.~C.~A.,
  \& {Steeghs}, D. 2020{\natexlab{b}}, arXiv e-prints, arXiv:2011.02383.
\newblock \doarXiv{2011.02383}

\bibitem[{{Ubertini} {et~al.}(2003){Ubertini}, {Lebrun}, {Di Cocco}, {Bazzano},
  {Bird}, {Broenstad}, {Goldwurm}, {La Rosa}, {Labanti}, {Laurent}, {Mirabel},
  {Quadrini}, {Ramsey}, {Reglero}, {Sabau}, {Sacco}, {Staubert}, {Vigroux},
  {Weisskopf}, \& {Zdziarski}}]{2003A&A...411L.131U}
{Ubertini}, P., {Lebrun}, F., {Di Cocco}, G., {et~al.} 2003, \aap, 411, L131,
  \dodoi{10.1051/0004-6361:20031224}

\bibitem[{{Urquhart} {et~al.}(2014){Urquhart}, {Figura}, {Moore}, {Hoare},
  {Lumsden}, {Mottram}, {Thompson}, \& {Oudmaijer}}]{urquhart2014}
{Urquhart}, J.~S., {Figura}, C.~C., {Moore}, T.~J.~T., {et~al.} 2014, \mnras,
  437, 1791, \dodoi{10.1093/mnras/stt2006}

\bibitem[{{Urquhart} {et~al.}(2020){Urquhart}, {Bahramian}, {Strader},
  {Chomiuk}, {Ransom}, {Wang}, {Heinke}, {Tudor}, {Miller-Jones}, {Tetarenko},
  {Maccarone}, {Sivakoff}, {Shishkovsky}, {Swihart}, \&
  {Tremou}}]{2020ApJ...904..147U}
{Urquhart}, R., {Bahramian}, A., {Strader}, J., {et~al.} 2020, \apj, 904, 147,
  \dodoi{10.3847/1538-4357/abb6fc}

\bibitem[{{van Grunsven} {et~al.}(2017){van Grunsven}, {Jonker}, {Verbunt}, \&
  {Robinson}}]{2017MNRAS.472.1907V}
{van Grunsven}, T.~F.~J., {Jonker}, P.~G., {Verbunt}, F.~W.~M., \& {Robinson},
  E.~L. 2017, \mnras, 472, 1907, \dodoi{10.1093/mnras/stx2071}

\bibitem[{{Verbunt} {et~al.}(2017){Verbunt}, {Igoshev}, \&
  {Cator}}]{2017A&A...608A..57V}
{Verbunt}, F., {Igoshev}, A., \& {Cator}, E. 2017, \aap, 608, A57,
  \dodoi{10.1051/0004-6361/201731518}

\bibitem[{{Vink} {et~al.}(2020){Vink}, {Higgins}, {Sander}, \&
  {Sabhahit}}]{2020arXiv201011730V}
{Vink}, J.~S., {Higgins}, E.~R., {Sander}, A. A.~C., \& {Sabhahit}, G.~N. 2020,
  arXiv e-prints, arXiv:2010.11730.
\newblock \doarXiv{2010.11730}

\bibitem[{{Wegg} {et~al.}(2015){Wegg}, {Gerhard}, \&
  {Portail}}]{2015MNRAS.450.4050W}
{Wegg}, C., {Gerhard}, O., \& {Portail}, M. 2015, \mnras, 450, 4050,
  \dodoi{10.1093/mnras/stv745}

\bibitem[{{Wielen}(1977)}]{1977A&A....60..263W}
{Wielen}, R. 1977, \aap, 60, 263

\bibitem[{{Wijnands} {et~al.}(2006){Wijnands}, {in't Zand}, {Rupen},
  {Maccarone}, {Homan}, {Cornelisse}, {Fender}, {Grindlay}, {van der Klis},
  {Kuulkers}, {Markwardt}, {Miller-Jones}, \& {Wang}}]{2006A&A...449.1117W}
{Wijnands}, R., {in't Zand}, J.~J.~M., {Rupen}, M., {et~al.} 2006, \aap, 449,
  1117, \dodoi{10.1051/0004-6361:20054129}

\bibitem[{{Wiktorowicz} {et~al.}(2020){Wiktorowicz}, {Lu}, {Wyrzykowski},
  {Zhang}, {Liu}, {Justham}, \& {Belczynski}}]{2020ApJ...905..134W}
{Wiktorowicz}, G., {Lu}, Y., {Wyrzykowski}, {\L}., {et~al.} 2020, \apj, 905,
  134, \dodoi{10.3847/1538-4357/abc699}

\bibitem[{{Winkler} {et~al.}(2003){Winkler}, {Courvoisier}, {Di Cocco},
  {Gehrels}, {Gim{\'e}nez}, {Grebenev}, {Hermsen}, {Mas-Hesse}, {Lebrun},
  {Lund}, {Palumbo}, {Paul}, {Roques}, {Schnopper}, {Sch{\"o}nfelder},
  {Sunyaev}, {Teegarden}, {Ubertini}, {Vedrenne}, \&
  {Dean}}]{2003A&A...411L...1W}
{Winkler}, C., {Courvoisier}, T.~J.~L., {Di Cocco}, G., {et~al.} 2003, \aap,
  411, L1, \dodoi{10.1051/0004-6361:20031288}

\bibitem[{{Wong} {et~al.}(2014){Wong}, {Fryer}, {Ellinger}, {Rockefeller}, \&
  {Kalogera}}]{2014arXiv1401.3032W}
{Wong}, T.-W., {Fryer}, C.~L., {Ellinger}, C.~I., {Rockefeller}, G., \&
  {Kalogera}, V. 2014, arXiv e-prints, arXiv:1401.3032.
\newblock \doarXiv{1401.3032}

\bibitem[{{Wu} {et~al.}(2016){Wu}, {Orosz}, {McClintock}, {Hasan}, {Bailyn},
  {Gou}, \& {Chen}}]{2016ApJ...825...46W}
{Wu}, J., {Orosz}, J.~A., {McClintock}, J.~E., {et~al.} 2016, \apj, 825, 46,
  \dodoi{10.3847/0004-637X/825/1/46}

\bibitem[{{Wyrzykowski} \& {Mandel}(2020)}]{2020A&A...636A..20W}
{Wyrzykowski}, {\L}., \& {Mandel}, I. 2020, \aap, 636, A20,
  \dodoi{10.1051/0004-6361/201935842}

\bibitem[{{Wyrzykowski} {et~al.}(2016){Wyrzykowski}, {Kostrzewa-Rutkowska},
  {Skowron}, {Rybicki}, {Mr{\'o}z}, {Koz{\l}owski}, {Udalski}, {Szyma{\'n}ski},
  {Pietrzy{\'n}ski}, {Soszy{\'n}ski}, {Ulaczyk}, {Pietrukowicz}, {Poleski},
  {Pawlak}, {I{\l}kiewicz}, \& {Rattenbury}}]{2016MNRAS.458.3012W}
{Wyrzykowski}, {\L}., {Kostrzewa-Rutkowska}, Z., {Skowron}, J., {et~al.} 2016,
  \mnras, 458, 3012, \dodoi{10.1093/mnras/stw426}

\bibitem[{{Yamaguchi} {et~al.}(2018){Yamaguchi}, {Kawanaka}, {Bulik}, \&
  {Piran}}]{2018ApJ...861...21Y}
{Yamaguchi}, M.~S., {Kawanaka}, N., {Bulik}, T., \& {Piran}, T. 2018, \apj,
  861, 21, \dodoi{10.3847/1538-4357/aac5ec}

\end{thebibliography}


\begin{appendix}

\begin{deluxetable}{lcccccc}
\caption{The name, the Galacto-centric distance $d\rm _{GC}$, the associated uncertainties, and the absolute distance to the Galactic plane $|z|$ and its associated uncertainties for the sample of Galactic candidate BH LMXBs. 
  }
\label{tab:candi}
\tablehead{\colhead{Name} & \colhead{D$_{GC}$} & \colhead{-Err-d$_{GC}$} & \colhead{+Err-d$_{GC}$} & \colhead{$|z|$} & \colhead{Err-$z_{min}$} & \colhead{Err-$z_{max}$}\\ \colhead{ } & \colhead{$\mathrm{ kpc}$} & \colhead{ kpc} & \colhead{ kpc} & \colhead{$\mathrm{pc}$} & \colhead{$\mathrm{pc}$} & \colhead{$\mathrm{pc}$}}
\startdata
SRGAJ043520.9+552226 & 15.63 & 2.88 & 2.92 & 749.13 & 278.86 & 278.86 \\
MAXIJ0637-430 & 13.0 & 2.25 & 2.51 & 2817.41 & 1058.59 & 1058.59 \\
MAXIJ1348-630 & 6.97 & 0.23 & 0.27 & 37.79 & 11.81 & 11.81 \\
MAXIJ1631-479 & 3.32 & 0.78 & 0.72 & 44.19 & 14.51 & 14.51 \\
MAXIJ1813-095 & 2.87 & 1.01 & 0.53 & 551.3 & 204.68 & 204.68 \\
SwiftJ1658.2-4242 & 2.36 & 1.3 & 0.31 & 7.8 & 0.86 & 0.86 \\
MAXIJ1535-571 & 5.03 & 0.04 & 1.48 & 156.46 & 60.73 & 60.73 \\
IGRJ17454-2919 & 0.16 & 2.69 & 0.3 & 24.53 & 11.26 & 11.26 \\
IGRJ17451-3022 & 0.25 & 2.61 & 0.29 & 91.76 & 36.47 & 36.47 \\
MAXIJ1828-249 & 1.47 & 1.85 & 0.03 & 911.65 & 343.93 & 343.93 \\
SwiftJ1753.7-2544 & 0.54 & 2.38 & 0.26 & 14.57 & 3.4 & 3.4 \\
SwiftJ174510.8-262411 & 0.39 & 2.49 & 0.28 & 196.02 & 71.45 & 71.45 \\
SwiftJ1910.2-0546 & 4.26 & 0.35 & 1.14 & 952.48 & 359.24 & 359.24 \\
MAXIJ1836-194 & 2.1 & 1.46 & 0.2 & 746.23 & 281.9 & 281.9 \\
MAXIJ1543-564 & 4.85 & 0.11 & 1.4 & 155.46 & 60.36 & 60.36 \\
SwiftJ1357.2-0933 & 7.67 & 0.84 & 1.76 & 6131.28 & 2297.17 & 2297.17 \\
XTEJ1752-223 & 0.96 & 2.1 & 0.18 & 295.29 & 108.67 & 108.67 \\
XTEJ1652-453 & 2.74 & 1.08 & 0.47 & 109.44 & 43.1 & 43.1 \\
SwiftJ1539.2-6227 & 5.44 & 0.11 & 1.54 & 785.28 & 296.54 & 296.54 \\
SwiftJ1842.5-1124 & 3.08 & 0.9 & 0.62 & 443.16 & 168.25 & 168.25 \\
SwiftJ174540.2-290005 & 0.15 & 2.7 & 0.3 & 5.92 & 4.28 & 4.28 \\
IGRJ17497-2821 & 0.21 & 2.64 & 0.3 & 63.11 & 25.73 & 25.73 \\
XTEJ1817-330 & 1.14 & 2.0 & 0.13 & 1112.6 & 419.29 & 419.29 \\
XTEJ1726-476 & 2.68 & 1.12 & 0.44 & 963.88 & 363.52 & 363.52 \\
XTEJ1818-245 & 1.21 & 1.97 & 0.11 & 584.55 & 221.27 & 221.27 \\
SwiftJ1753.5-0127 & 3.85 & 0.53 & 0.96 & 1689.41 & 631.47 & 631.47 \\
IGRJ17098-3628 & 1.51 & 1.84 & 0.02 & 289.79 & 106.61 & 106.61 \\
IGRJ17091-3624 & 1.51 & 1.84 & 0.01 & 309.07 & 113.84 & 113.84 \\
XTEJ1720-318 & 0.89 & 2.14 & 0.2 & 432.95 & 160.29 & 160.29 \\
XTEJ1908+094 & 5.96 & 0.29 & 1.59 & 62.18 & 21.25 & 21.25 \\
SAXJ1711.6-3808 & 1.64 & 1.77 & 0.03 & 111.74 & 39.84 & 39.84 \\
XTEJ2012+381 & 9.88 & 1.46 & 2.05 & 317.8 & 117.11 & 117.11 \\
XTEJ1748-288 & 0.18 & 2.67 & 0.3 & 30.89 & 13.65 & 13.65 \\
XTEJ1755-324 & 0.6 & 2.33 & 0.25 & 506.58 & 192.03 & 192.03 \\
GRS1737-31 & 0.37 & 2.51 & 0.28 & 13.77 & 7.23 & 7.23 \\
GRS1739-278 & 0.24 & 2.62 & 0.29 & 164.26 & 59.53 & 59.53 \\
XTEJ1856+053 & 5.3 & 0.06 & 1.53 & 175.61 & 63.79 & 63.79 \\
GRS1730-312 & 0.51 & 2.4 & 0.27 & 140.62 & 50.67 & 50.67 \\
GRS1716-249 & 1.0 & 2.08 & 0.17 & 973.83 & 363.13 & 363.13 \\
GS1734-275 & 0.4 & 2.49 & 0.28 & 361.7 & 133.57 & 133.57 \\
EXO1846-031 & 4.18 & 0.38 & 1.1 & 127.42 & 49.84 & 49.84 \\
SLX1746-331 & 0.63 & 2.31 & 0.25 & 414.89 & 157.64 & 157.64 \\
H1743-322 & 0.49 & 2.41 & 0.27 & 255.77 & 97.98 & 97.98 \\
3A1524-617 & 5.51 & 0.14 & 1.55 & 616.13 & 233.11 & 233.11 \\
4U1755-338 & 0.8 & 2.19 & 0.21 & 679.36 & 256.82 & 256.82 \\
4U1630-472 & 3.24 & 0.82 & 0.69 & 35.48 & 11.24 & 11.24 \\
CenX-2 & 6.81 & 0.57 & 1.67 & 402.74 & 153.09 & 153.09\\
\enddata
\end{deluxetable}
\end{appendix}

\end{document}